\documentclass[aps,pra,floatfix,twocolumn,superscriptaddress]{revtex4-2}

\usepackage{amsmath}
\usepackage{amsfonts}
\usepackage{amssymb}
\usepackage{graphicx}
\usepackage{parskip}
\usepackage{braket}
\usepackage{esdiff}
\usepackage{lettrine}
\usepackage{accents}
\usepackage{mathtools}
\usepackage[T1]{fontenc}
\usepackage{xcolor}

\DeclarePairedDelimiter\floor{\lfloor}{\rfloor}

\newcommand{\ubar}[1]{\underaccent{\bar}{#1}}

\begin{document}

\title{Third quantization with Hartree approximation for open-system bosonic transport}

\author{Fernando Espinoza-Ortiz}
\affiliation{Department of Physics, University of California, Merced, CA 95343, USA}

\author{Chih-Chun Chien}
\email{cchien5@ucmerced.edu}
\affiliation{Department of Physics, University of California, Merced, CA 95343, USA}

\begin{abstract}
The third quantization (3rd Q) for bosons provides the exact steady-state solution of the Lindblad equation with quadratic Hamiltonians. By decomposing the interaction of the Bose Hubbard model (BHM) according to Hartree approximation, we present a self-consistent formalism for solving the open-system bosonic Lindblad equation with weak interactions in the steady state. The 3rd Q with Hartree approximation takes into account the infinite Fock space of bosons while its demand of resource scales polynomially with the system size. We examine the method by analyzing three examples of the BHM, including the uniform chain, interaction induced diode effect, and Su-Schrieffer-Heeger (SSH) Hubbard model. When compared with the simulations with capped boson numbers for small systems, the 3rd Q with Hartree approximation captures the qualitative behavior and suggests an upper bound of the steady-state value. Finite-size scaling confirms the results from the 3rd Q with Hartree approximation converge towards the thermodynamic limit. Thus, the manageable method allows us to characterize and predict large-system behavior of quantum transport in interacting bosonic systems relevant to cold-atom experiments. 
\end{abstract}

\maketitle

\section{Introduction}
The emergence of ultracold-atoms as a versatile platform for studying quantum physics in and out of equilibrium has brought many interesting simulations and analogues of condensed matter physics~\cite{pethick2002bose,ueda2010fundamentals,cosco2018bosehubbard,maciej2007ultracold}. For example, quantum transport has been an important subject in electronic systems~\cite{di2008electrical,Nazarov_book}, and it has received attention in atomic systems as well~\cite{ChienNatPhys}. While bosonic and fermionic atoms have been used in experiments, bosonic physics has less counterparts in conventional condensed matter physics and poses challenges for theoretical understanding because of the infinite-dimensional Fock space allowed by the Bose-Einstein statistics. Therefore, some methods developed for electronic transport~\cite{Hirose_book,Waintal24} may not apply to bosonic systems directly.

On the other hand, the general concept of modeling quantum transport as quantum dynamics of a system driven by external reservoirs applies to both fermionic and bosonic systems. In the open quantum system framework, dynamics of the system may be described by various master equation~\cite{breuer2002theory,weiss2012quantum,Li2018markov,Minganti24} following different assumptions and approximations. This approach has been applied to bosonic atomtronic analogs of diodes, transistors, and batteries~\cite{pepino2009atomtronic,pepino2010open,ghosh2021fast,Amico2022atomtronic.94.041001}, some of which have been realized~\cite{caliga2017experimental,caliga2016transport,gati2006realization}. These atomtronic devices consist of ultracold atoms loaded into optical lattices with the system usually modeled by the Bose Hubbard model (BHM). 

The infinite Fock space of bosonic systems makes it challenging to obtain the exact solution of the BHM driven by external reservoirs in general. To include interaction effects, many available methods restrict the number of bosons to reduce the problem to one with a finite Fock space. For example, the exact solution of three site and four bosons has been studied in Ref.~\cite{gangat2018symmetry} and up to 128 sites with 3 Fock states on each site in Ref.~\cite{cosco2018bosehubbard}. One may use further truncation to study systems with more bosons or larger lattices~\cite{Lai2016}.
However, experiments have shown approximately 150 atoms in a single site~\cite{caliga2016transport} and can load $10^4$ to $10^7$ atoms into relative large lattices~\cite{gati2006realization,vochezer2018light,yang2020observation,aidelsburger2011experimental,tao2024high,gyger2024continuous}. Ref.~\cite{PhysRevA.105.013307} used a semi-classical approach by considering the effective distribution of bosons in the system to obtain steady-state currents. However, an approximate method for qualitatively describing interacting bosonic systems in an open quantum system setting is desirable from both theoretical and experimental points of view.

If the interactions between the bosons are excluded, numerical simulations for a forty-one site lattice with twenty-one bosons have been performed~\cite{trivedi2023filling}, and analytic results from analogues of the Landauer formula have been derived~\cite{PhysRevA.98.043623,PhysRevE.94.062143}. Meanwhile, an analytical formalism to include the entire Fock space for quadratic bosonic Hamiltonians, called third quantization (3rd Q) for bosons~\cite{prosen2010quantization}, provides the exact steady-state solution of the Lindblad equation, which is a quantum master equation commonly found in the literature~\cite{lindblad1976semigroup,breuer2002theory,Schlosshauer_book,weiss2012quantum,Li2018markov,Minganti24}. Refs.~\cite{PhysRevLett.129.120401,Barthel22} generalized the Lindblad operators in the 3rd Q, Ref.~\cite{He_2022} applied the 3rd Q to classify open bosonic systems, and Ref.~\cite{PhysRevA.95.052107} used the 3rd Q to study transport induced by artificial gauge fields. 

Nevertheless, interactions in the system result in many interesting transport phenomena, including diode effect~\cite{pepino2009atomtronic,pepino2010open,PRXQuantum.5.010341} and conducting-non conducting transition~\cite{chien2013interaction}. Tensor-based methods have been a useful tool for studying 1D interacting systems, including up to 200 sites with 4 bosonic states per site of the 1D BHM~\cite{Ramana2009signature} and up to 96 sites and 6 bosons per site of a two-species Bose-Hubbard ring~\cite{contessi2021collisionless}. A combination of the mean-field theory and density-matrix renormalization group method allows the study of the correlation functions for large (up to 3700-site) systems with a cap of three bosons per site \cite{cmf2022gaude}. However, the Fock space of bosons has been truncated to render the calculations manageable.
Here we introduce the Hartree approximation~\cite{fetter1971many,chern2014dynamically,metcalf2016matterwave,parajuli2019mass,parajuli2023atomic} to the third quantization of bosons without truncating the bosonic Fock space by approximating the interaction term of the BHM with quadratic terms. 
Utilizing the analytic framework of third quantization provides an efficient and scalable method to find the steady-state solution of the driven BHM, described by the Lindblad equation, without truncating the infinite Fock space of bosons. We mention that multi-configurational Hartree (or Hartree-Fock) approximations have been implemented in many-body calculations~\cite{PhysRevA.77.033613,10.1063/1.4821350,PhysRevA.93.063601,10.1007/978-3-319-47066-5_6,e23040392,Molignini24}, and there are other approaches, such as phase-space methods using the Wigner function~\cite{anton2013bosonic,kordas2015nonequilibrium,LEE1995147,LajosDiosi_2002,POLKOVNIKOV20101790,10.1063/1.5046663,10.1002/qute.202100016}, to study interacting systems in or out of equilibrium.

We will examine three examples of the BHM: the uniform chain, interaction-induced diode effect, and Su-Schrieffer-Heeger (SSH) Hubbard model. The first serves as a calibration of the results, the second has been studied in the context of atomtronics~\cite{pepino2009atomtronic,pepino2010open}, and the third is based on the SSH model with alternating hopping coefficients~\cite{SSH,Asboth2016}, which has illustrated interaction effects in topological systems~\cite{diliberto2017twobound,azcona2021doublons,Le2020,PhysRevB.106.L081114}. By comparing the results of small systems with the exact interaction term but in finite truncated Fock space, the 3rd Q with Hartree approximation correctly captures the qualitative features in all cases. However, it over-estimates the steady-state current due to the mean-field treatment of the interactions. Nevertheless, we are able to study the steady states of bosonic systems having thousands of lattice sites without truncating the Fock space, allowing an interpolation between small and large systems. The 3rd Q with Hartree approximation thus serves as a fast and scalable estimation of the upper-bound of bosonic transport for large systems with many bosons in the weak interaction regime and complements other available numerical methods.

The rest of the paper is organized as follows. Sec.~\ref{Sec:Theory} reviews the BHM, Lindblad equation, simulations with capped boson numbers, and the third quantization for bosons. We then introduce the 3rd Q with Hartree approximation and the iteration method for obtaining the self-consistent solution in the steady state. Sec.~\ref{Sec:Example} presents three examples of the BHM: the uniform chain, interaction induced diode effect, and SSH-Hubbard model with alternating hopping coefficients. We compare the 3rd Q with Hartree approximation with simulations with capped boson numbers and examine the scaling with system size and interactions. Sec.~\ref{Sec:Discussion} discusses implications from the 3rd Q with Hartree approximation and their experimental relevance. Finally, Sec.~\ref{Sec:Conclusion} concludes our work.

\section{Theory and Approximation}\label{Sec:Theory}
\subsection{Lindblad equation of Bose Hubbard model}
We consider the 1D Bose Hubbard model (BHM) of length $L$ with the second quantized Hamiltonian 
\begin{equation}
    \hat{H} = -\sum\limits_{\braket{i,j}}^{L-1}t_{ij}(\hat{a}^\dagger_i\hat{a}_j + \hat{a}^\dagger_j\hat{a}_i) + \frac{1}{2}\sum\limits_{i}^{L}U_i\hat{a}^\dagger_i\hat{a}_i\hat{a}^\dagger_i\hat{a}_i,
    \label{eq:H}
\end{equation}
where $\hat{a}^\dagger$ and $\hat{a}$ are the bosonic creation and annihilation operators acting on site $i$ with onsite coupling constant $U_i$ and nearest neighbor hopping coefficient $t_{ij}$ between nearest neighbors $\braket{i,j}$. The system becomes more complex as we introduce couplings to the environment, but if the environment is sufficiently larger and returns to equilibrium faster than the system, only the system dynamics is of interest~\cite{Li2018markov} and may be formulated by examining the equation of motion for its density matrix.

To model the dynamics of the 1D BHM driven by external reservoirs, we follow the open-system approach using the Lindblad master equation (LME)~\cite{breuer2002theory,Li2018markov}
\begin{equation}
    \diff{\hat{\rho}}{t} = \hat{\mathfrak{L}}\hat{\rho} = -i[\hat{H},\hat{\rho}] + \sum\limits_{\mu}(2\hat{L}_{\mu}\hat{\rho}\hat{L}_{\mu}^{\dagger}-\{\hat{L}_{\mu}^{\dagger}\hat{L}_{\mu},\hat{\rho}\}),
    \label{eq:lme}
\end{equation}
where $\hat{\rho}$ is the density matrix of the system and $\hat{L}_\mu$ are Lindblad operators that couple the system to the environment~\cite{breuer2002theory}. The LME assumes separate time scales of the system and reservoirs as well as Markovian approximation~\cite{wichterich2007modeling,purkayastha2016out,lidar2019lecture}.  
We consider two reservoirs connected to the two ends of the 1D BHM and exchange particles with the system. The Lindblad operators on the left side are 
\begin{equation}  
    \hat{L}_1 = \sqrt{\frac{\gamma_\mathcal{L} N_\mathcal{L}}{2}}\hat{a}^\dagger_1, \quad \hat{L}_2 = \sqrt{\frac{\gamma_\mathcal{L} (N_\mathcal{L}+1)}{2}}\hat{a}_1.
    \label{eq:leftcouple}
\end{equation}
Here $N_\mathcal{L}$ is the average particle number of the left reservoir, which is assumed to follow the Bose-Einstein distribution. The choice of the coefficients guarantees thermal equilibrium if only the left reservoir is present.
Similarly, the Lindblad operators at the right end are described by
\begin{equation}
    \hat{L}_3 = \sqrt{\frac{\gamma_\mathcal{R} N_\mathcal{R}}{2}}\hat{a}^\dagger_L, \quad \hat{L}_4 = \sqrt{\frac{\gamma_\mathcal{R} (N_\mathcal{R}+1)}{2}}\hat{a}_L
    \label{eq:rightcouple}.
\end{equation}
Here $N_\mathcal{R}$ is the average particle number of the right reservoir. The choice of the coefficients again ensures thermal equilibrium if only the right reservoir is present.

The expectation value of a physical observable $\hat{A}$ is 
\begin{equation}
    \braket{\hat{A}} = Tr(\hat{\rho}\hat{A}).
    \label{eq:expect}
\end{equation}
In the following, we will focus on the steady-state solutions with $\diff{\hat{\rho}}{t}=0$ in the long-time limit, which are more relevant in experiments. Numerically, the steady state $\hat{\rho}_{ss}$ may be obtained by either solving the LME (\ref{eq:lme}) over a sufficiently long period such that $\hat{\rho}$ no longer varies with time or solving the equation $\hat{\mathfrak{L}}\hat{\rho}_{ss}=0$ algebraically. 
The current operator between sites $i$ and $j$ is \cite{chien2013interaction,lai2018tunable,palak2022geometry}
\begin{equation}
    \hat{J}_{ij} = -i(t_{ij}\hat{a}_{i}^{\dagger}\hat{a}_{j}-t_{ji}\hat{a}_{j}^{\dagger}\hat{a}_{i})
    , \label{eq:curr2nd}
\end{equation}
and the number operator on site $i$ is $\hat{n}_{i} = \hat{a}_{i}\hat{a}^{\dagger}_{i}$. Their steady-state expectation values can be evaluated accordingly. For a system in the steady state, the incoming current should be the same as the outgoing current through any part of the system if there is no source or sink within the system.

\subsection{Solutions of LME with truncated Fock space}
The Lindblad equation of a 1D $L$-site BHM is a differential equation, and it is tempting to solve the equation by integration. It is convenient to choose our basis as a tensor product of Fock states $\ket{b} =  \ket{n_1}\otimes\ket{n_2}\otimes\ket{n_3}\otimes\cdots\ket{n_L}$, where $n_j$ is the number of bosons on site $j$. However, a challenge of bosonic systems immediately surfaces because the Bose-Einstein distribution allows an infinite number of bosons on each site, thereby an exact solution becomes impractical on available computers. Typically, the Fock space is truncated by imposing $\ket{n_j} \in \{\ket{0},\ket{1},\cdots,\ket{M-1},\ket{M}\}$~\cite{gangat2018symmetry,trivedi2023filling,pepino2009atomtronic,pepino2010open}.

We solve the LME (\ref{eq:lme}) numerically with the python library QuTiP~\cite{johansson2013qutip} by capping the number of bosons on each site by $M$. 
Once the creation and annihilation operators on each site are defined, the Hamiltonian of Eq.~(\ref{eq:H}) and Lindblad operators shown in Eqs.~(\ref{eq:leftcouple}) and (\ref{eq:rightcouple}) can be explicitly constructed. The Hamiltonian and Lindblad operators are then passed as arguments to the LME solver with a given initial density matrix and a time interval large enough to reach the steady state, outputting the time-evolved density matrices which are used to obtain the steady-state expectation values by using Eq. (\ref{eq:expect}). In our calculations, we have checked that the steady state is insensitive to various initial density matrix for the solver.

However, the LME solver is limited to small numbers of bosons and system size due to the heavy usage of computer memory. In our simulations, we do not exceed four sites with six bosons maximal per site or three sites with twelve bosons maximal per site. It is known that the number of allowed states scales factorially according to~\cite{maciej2007ultracold} 
    $\sum_{j=0}^{\floor*{\frac{N}{M+1}}}(-1)^{j}\binom{N+L-1-j(M+1)}{L-1}\binom{L}{j}$,
where $N$ is the total number of bosons, $M$ is the maximum number of bosons per site, and $\floor{\:}$ is the floor function. As the number of bosons or lattice sites increases, the problem soon becomes too resource-demanding to solve numerically. Therefore, for a proper description of open-system bosonic transport we look towards a scalable method that does not restrict the number of bosons and system size so severely.

\subsection{Third quantization of bosons}
Ref.~\cite{prosen2010quantization} introduces a method called the third quantization (3rd Q) for finding the exact solution of the Lindblad equation with a quadratic Hamiltonian and linear Lindblad operators. This includes the case with $U_i=0$ for all $i$ of the BHM, which has been discussed previously~\cite{palak2022geometry,palak2020geometry,pivzorn2013one}. Importantly, the infinite Fock space of bosons is included in the formalism. 
We briefly review the method, which begins with a pair of vector spaces $\mathcal{K}$ and $\mathcal{K}^{'}$, where $\mathcal{K}$ contains trace class operators (i.e., density matrices) and $\mathcal{K}^{'}$  physical observables. Here $\ket{\rho}$ denotes an element of $\mathcal{K}$ and $(A|$ an element of $\mathcal{K}^{'}$ such that the inner product is the expectation value of $A$ with respect to the state $\rho$. Explicitly,
    $(A\ket{\rho} = Tr(A\rho)$.
Left and right multiplications are introduced for both spaces as $ \hat{a}^{L}\ket{\rho} = \ket{a\rho}$, $\hat{a}^{R}\ket{\rho} = \ket{\rho a}$, $(A|\hat{a}^{L} = (aA|$, and $(A|\hat{a}^{R} = (Aa|$.
These left and right multiplications are used to define $4L$ maps of 3rd Q: $\hat{a}_{0,j}=\hat{a}_{j}^{L}$, $\hat{a}_{0,j}^{'}=\hat{a}_{j}^{\dagger L}-\hat{a}_{j}^{\dagger R}$, $\hat{a}_{1,j}=\hat{a}_{j}^{\dagger R}$, and $\hat{a}_{1,j}^{'}=\hat{a}_{j}^{R}-\hat{a}_{j}^{L}$.
The Liouvillean operator $\hat{\mathfrak{L}}$ in Eq.~\eqref{eq:lme} is rewritten in terms of the $4L$ maps and a vector $\ubar{\hat{b}} = (\ubar{\hat{a}}_0,\ubar{\hat{a}}_1,
\ubar{\hat{a}}_0^{'},\ubar{\hat{a}}_1^{'})^{T}$ as
\begin{equation}
    \hat{\mathfrak{L}} = \ubar{\hat{b}}\cdot \mathcal{S} \ubar{\hat{b}} - S_0\mathbb{I},
    \label{eq:3superop}
\end{equation}
where $S_0$ is a scalar, and $\mathcal{S}$ is a complex symmetric $4L\times4L$ matrix. Onward variables underlined by a bar, such as $\ubar{\hat{a}}_0$, will denote vectors.

For the 1D BHM, the matrix $\mathcal{S}$ takes the following form:
\begin{gather}
    \mathcal{S} = \left[ {\begin{array}{cc}
                    0 & -X \\
                    -X^{T} & Y \\
                            \end{array} } \right],\label{eq:3qs}\\
    {X} = \frac{1}{2} \left[ {\begin{array}{cc}
                   \Tilde{X} & 0 \\
                    0 & \Tilde{X}^{*} \\
                            \end{array} } \right],                            
    {Y} = \frac{1}{2} \left[ {\begin{array}{cc}
                    0 & \Tilde{Y} \\
                    \Tilde{Y} &  0\\
                            \end{array} } \right].
    \label{eq:3XYbuff}
\end{gather}
Here $\Tilde{X} = i\textbf{H}-\textbf{N}+\textbf{M}$, $\Tilde{Y} = 2\textbf{N}$, and \textbf{H}, \textbf{N}, \textbf{M} are $L\times L$ matrices with entries from the coefficients from the Hamiltonian and Lindblad operators. Moreover, $S_0=Tr(X)$. The matrix elements of \textbf{H} are obtained by reformulating the quadratic Hamiltonian from Eq.~\eqref{eq:H} with $U_i=0$ as a matrix equation of the form $\hat{H} = \ubar{\hat{a}}^{\dagger}\cdot \textbf{H} \ubar{\hat{a}}$. Here $\ubar{\hat{a}}^{\dagger}$ and $\ubar{\hat{a}}$ are column vectors of the creation and annihilation operators. Explicitly, 
\begin{equation}
    \textbf{H} = \left[{\begin{smallmatrix}
                    0 & -t_{12} & & & \\
                    -t_{12} & 0 &-t_{23} & & \\
                     & -t_{23}& \ddots& \ddots& \\
                    &  &  \ddots& \ddots& -t_{(L-1)L}\\
                    & & & -t_{(L-1)L} & 0\\
                            \end{smallmatrix}} \right]\\.
    \label{eq:3qH}
\end{equation}
The matrices \textbf{N}, \textbf{M}, \textbf{L} are obtained from the tensor product $\otimes$\ of the coefficient vectors of our Lindblad operators ($\ubar{l}_{\mu}$, $\ubar{k}_{\mu}$) as
\begin{eqnarray}
    \hat{L}_{\mu} &=& \ubar{l}_{\mu}\cdot\ubar{\hat{a}} + \ubar{k}_{\mu}\cdot\ubar{\hat{a}}^{\dagger},\\
    \textbf{M}&=&\sum\limits_{\mu}\ubar{l}_{\mu}\otimes\ubar{l}_{\mu}^{*}, \textbf{N}=\sum\limits_{\mu}\ubar{k}_{\mu}\otimes\ubar{k}_{\mu}^{*},    \textbf{L}=\sum\limits_{\mu}\ubar{l}_{\mu}\otimes\ubar{k}_{\mu}^{*}. \nonumber
    \label{eq:outerNML}
\end{eqnarray}
From the Lindblad operators $\hat{L}_{\mu}$ of Eqs.~(\ref{eq:leftcouple}) and (\ref{eq:rightcouple}), the matrix \textbf{L} is a null $L\times L$ matrix because of no simultaneous creation and annihilation operators in the Lindblad operators. The matrix \textbf{N} and \textbf{M} are diagonal matrices with the diagonal elements given by $(\frac{\gamma_\mathcal{L} N_\mathcal{L}}{2}, 0, \cdots, 0, \frac{\gamma_\mathcal{R} N_\mathcal{R}}{2})$ and $(\frac{\gamma_\mathcal{L} (N_\mathcal{L}+1)}{2}, 0, \cdots, 0, \frac{\gamma_\mathcal{R} (N_\mathcal{R}+1)}{2})$, respectively.

In the steady state, $d\rho/dt = \hat{\mathfrak{L}} \hat{\rho}=0$. One can diagonalize $\hat{\mathfrak{L}}$ of Eq.~\eqref{eq:3superop} and find that its eigenvalues come in complex conjugate pairs $\beta_i,\beta_i^*$. Furthermore, the covariance matrix is given by a $2L\times2L$ complex symmetric matrix of the form  $Z =\left[ {\begin{array}{cc}
                    0 & \Tilde{Z} \\
                    \Tilde{Z}^T &  0\\
                            \end{array} } \right]$. Here the $L \times L$ Hermitian matrix $\Tilde{Z}$ has elements $\Tilde{Z}_{ij}= \braket{\hat{a}^{\dagger}_i\hat{a}_j}$.
Importantly, $Z$ is a solution of the Lyapunov equation $X^TZ + ZX = Y$, which reduces to a Sylvester equation of $L\times L$ matrices~\cite{pivzorn2013one}. Explicitly, 
\begin{equation}
    \Tilde{X} \Tilde{Z} + \Tilde{Z}\Tilde{X}^{*} = \Tilde{Y}.
    \label{eq:slyfast}
\end{equation}

Our task is to solve  Eq.~\eqref{eq:slyfast}. After obtaining the matrix  $\Tilde{Z}$, the steady-state current is given by
\begin{equation}
    J_{ij}=\braket{\hat{J}_{ij}} = -i(t_{ij}\Tilde{Z}_{ij} - t_{ji}\Tilde{Z}_{ji}).
    \label{eq:3qcurr}
\end{equation}
Moreover, the steady-sate number density on site $i$ can also be obtained from $\langle \hat{n}_i \rangle=\Tilde{Z}_{ii}$. 
The resource needed for obtaining the steady-state solution from the 3rd Q method now scales only polynomially with the number of sites.
We emphasize that the 3rd Q method gives the exact steady-state solutions in the infinite Fock space of bosons with the limitation of dealing with only quadratic Hamiltonians.

\subsection{Self-consistent Hartree approximation for interacting bosonic systems}
Although 3rd Q solves the Lindblad equation with the complete Fock space of bosons, it requires quadratic Hamiltonians with linear Lindblad operators. Thus, only the $U_i=0$ case of the BHM can be studied. In order to incorporate the interactions, we will apply the Hartree approximation~\cite{fetter1971many} and approximate the interaction term in Eq.~\eqref{eq:H} by a combination of quadratic terms. Let $n_{i} = \braket{\hat{a}_i^\dagger\hat{a}_i}$ be the density on site $i$. The Hartree approximation for bosons decomposes the interaction term in the following form: 
\begin{align}
    \frac{1}{2}U_i\hat{a}_i^\dagger\hat{a}_i\hat{a}^\dagger_i\hat{a}_i &\approx \frac{1}{2}U_i(\braket{\hat{a}_i^\dagger\hat{a}_i}\hat{a}_i^\dagger\hat{a}_i + \hat{a}_i^\dagger\hat{a}_i\braket{\hat{a}_i^\dagger\hat{a}_i} ). 
    \label{eq:Hartree}
\end{align}
Here a scalar term proportional to $n_i^2$ has been dropped since it commutes with any operators. The BHM Hamiltonian~\eqref{eq:H} after the approximation then has the following quadratic form:
\begin{equation}
    \hat{H}_{H} = -\sum\limits_{\braket{i,j}}^{L-1}t_{ij}(\hat{a}^\dagger_i\hat{a}_j + \hat{a}^\dagger_j\hat{a}_i) + \sum\limits_{i}^{L}U_{i}n_{i}\hat{a}_i^\dagger\hat{a}_i.
    \label{eq:quadH}
\end{equation}
Therefore, the matrix $\mathbf{H}$ of Eq.~\eqref{eq:3qH} has additional diagonal terms $(U_1 n_1, \cdots, U_L n_L)$. 
The Hartree approximation is expected to work when the interaction strength $U_i\leq1$, so we refrain from exceeding the limit unless we want to investigate if the approximation exhibits any breakdown in the strongly interacting regime.

We then use a self-consistent iteration method to find the steady-state solution of the Lindblad equation of the 1D BHM in the 3rd Q formalism with Hartree approximation. The iteration method is similar to the Bogoliubov-de Gennes (BdG) equation of inhomogeneous superconductors~\cite{gennes1966superconductivity,zhu2016bogoliubov}, but the Hartree approximation studied here has no order parameter.
In the iteration method, the boson densities $n_{i}$ for all $i$ in $\mathbf{H}$ and $\Tilde{X}$ need to be determined self-consistently. We being with an initial guess of $n_{i}$ and obtain the steady-state solution from the 3rd Q formalism. The new densities $n^{new}_{i}$ can then be obtained from $\Tilde{Z}$ in Eq.~\eqref{eq:slyfast}, which will be used to construct $\mathbf{H}$ in the next iteration. The iteration terminates when the following convergence condition is met: 
    $\sum\limits_{i}\sqrt{(n^{old}_{i} - n^{new}_{i})^{2}} < \epsilon$.
The condition implies that a self-consistent density profile has been obtained for the steady-state solution. In the following, we will examine some examples to see the results from the 3rd Q with Hartree approximation.

We remark that with the rapid development of quantum simulators of many-body physics~\cite{pethick2002bose,ueda2010fundamentals,cosco2018bosehubbard,maciej2007ultracold,Amico2022atomtronic.94.041001,Polo_2024}, it may become feasible to engineer a system with density-dependent onsite interactions described exactly by Eq.~\eqref{eq:quadH} instead of treating it as Hartree approximation of the BHM. Then our results following Eq.~\eqref{eq:quadH} will faithfully depict the behavior of the realization of such a system rather than confining it within the BHM framework.

\section{Examples}\label{Sec:Example}
Since quantum transport typically deals with a flowing current in a selected direction through a system, hereinafter we focus on 1D systems and examine the Lindblad equation of the 1D BHM with symmetric system-reservoir couplings $\gamma_\mathcal{L}=\gamma_\mathcal{R}=\gamma$. Previous studies~\cite{palak2022geometry} have shown that asymmetric system-reservoir couplings only introduce quantitative differences in the steady state. The energy and time units are $t_{12}$ (abbreviated as $t_{1}$ or $t$ whenever applicable) and $T_{0}=\frac{\hbar}{t_{1}}$, and we will set $\hbar=1$ in the following. In the steady state, the current is the same through the system without any sink or source. We have verified this property in our simulations and will present $J = \braket{\hat{J}_{23}}$ whenever applicable.

\subsection{Uniform chain}
We begin with the uniform 1D BHM with $t_{ij} = t$ and on-site interaction $U_i = U$ in the BHM Hamiltonian shown in Eq.~\eqref{eq:H}. To verify the validity of the 3rd Q with Hartree approximation, we compare its results of small systems ($L\le 4$) with the steady-state currents from the QuTiP simulations with at most $M$ bosons per site because the latter is limited to small systems. As summarized in Appendix~\ref{app:a}, the 3rd Q with Hartree approximation captures the qualitative features of interacting bosonic systems while possibly over-estimates the steady-state current in the comparison. 
Therefore, one may treat the 3rd Q with Hartree approximation for interacting bosonic systems as an upper bound of the steady-state values without any truncation of the bosonic Fock space. 

\begin{figure}[t]
    \centering
    \includegraphics[width=0.9\columnwidth]{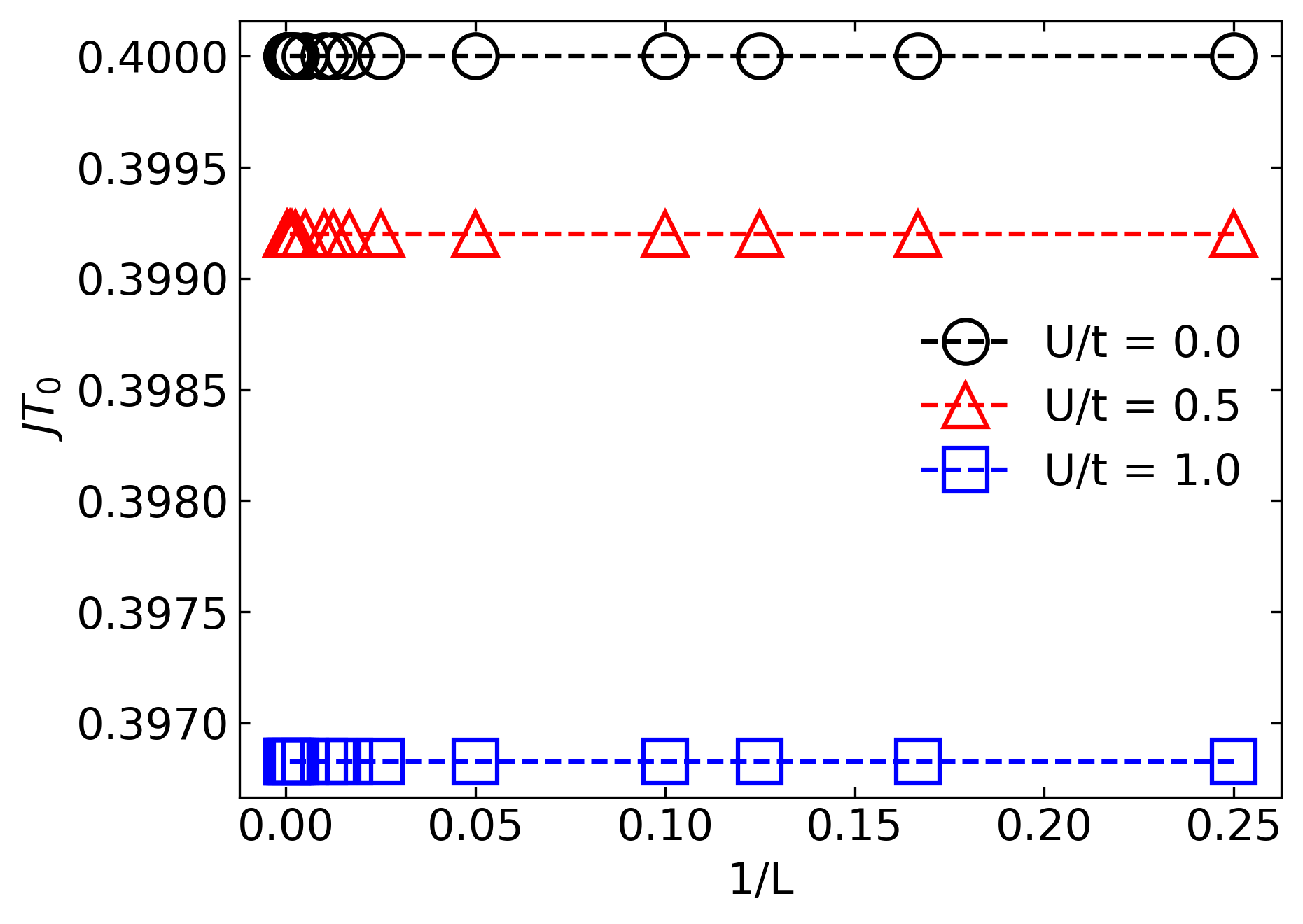}
    \caption{Finite-size scaling of the steady-state currents of a uniform BHM from the 3rd Q with Hartree approximation. Here $\gamma/t=1$, $N_\mathcal{L}=0$, and $N_\mathcal{R}=1$. 
    }
    \label{fig:steady_vbose_FS}
\end{figure}

One possible reason of the overestimate of the 3rd Q with Hartree approximation is that the onsite interaction may also suppress the system-reservoir couplings $\gamma_{\mathcal{L,R}}$, which further reduces the current.
After showing the favorable comparisons in small systems in Appendix~\ref{app:a}, however, we apply the 3rd Q with Hartree approximation to large ($L\ge 1000$) systems to demonstrate its advantage without constraining the boson numbers on each site  as $L$ increases. In contrast, the QuTiP simulation of the exact solution soon becomes too demanding. Fig.~\ref{fig:steady_vbose_FS} shows the finite-size scaling of the steady-state current from the 3rd Q with Hartree approximation for the uniform BHM with selected values of $U/t$. One can see that the results stably converge to the thermodynamic limit ($L\rightarrow\infty$), thereby demonstrating the applicability of the method for large systems. Moreover, the steady-state current decreases with $U$ because of the scattering effects it introduces.

The driven BHM system provides interesting bosonic analogues of electronic devices. For example, the uniform chain discussed above may be viewed as an atomtronic version of an electronic wire~\cite{pepino2009atomtronic,pepino2010open}. The bosonic current is driven by enforcing $N_\mathcal{L} \neq N_\mathcal{R}$ and causing a potential difference, akin to a battery. The manageable calculations of the 3rd Q with Hartree approximation allows investigations of the bulk behavior of large systems in the steady state that otherwise would be computationally expensive or even impossible due to the infinite Fock space of bosonic systems.

\subsection{Interaction induced rectification}
Another electronic device of interest is a diode which allows current to flow when applying a voltage bias and shuts current when the voltage bias is reversed. It has been shown that the BHM with inhomogeneous interactions may exhibit an analogous diode effect with rectification of the current~\cite{pepino2009atomtronic,pepino2010open}. Previously, the diode effect was shown for a bosonic system with $M=2$ and two sites $L=2$ and later to slightly larger $M$ and $L$. 

\begin{figure}[t]
    \centering
    \includegraphics[width=0.9\columnwidth]{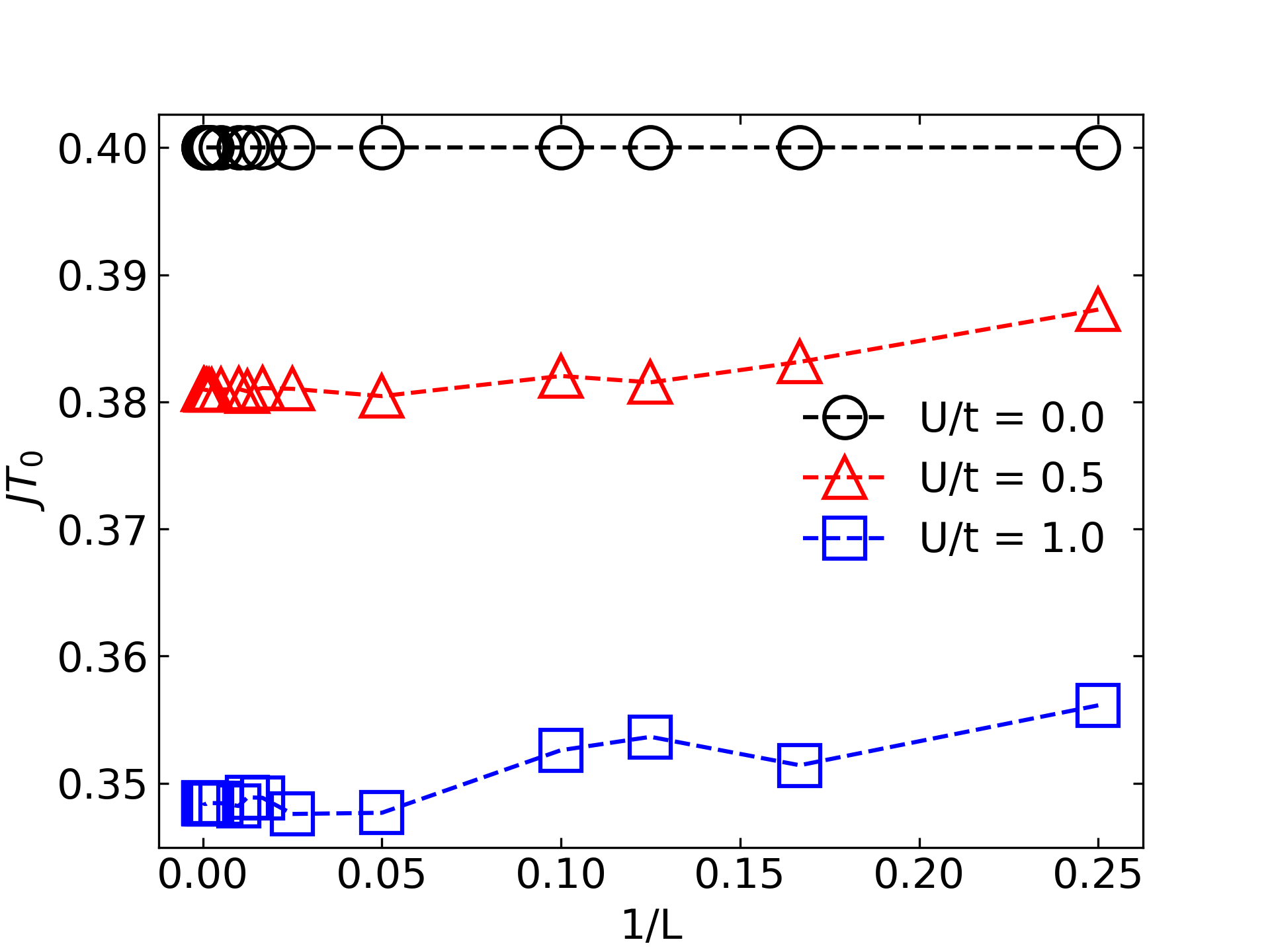}
    \caption{Finite-size scaling of the steady-state current in the presence of the inhomogeneous interaction profile for the diode effect up to $L=5000$ with $\gamma_\mathcal{L}=\gamma_\mathcal{R}=\gamma=t$, $N_\mathcal{L}=0$, and $N_\mathcal{R}=1$. Here $U$ is the coupling constant on half of the system with interactions. 
    }
    \label{fig:diode_finite}
\end{figure}

By using the 3rd Q with Hartree approximation, we are able to investigate the diode effect of the BHM in an unrestricted Fock space and up to $L=5000$. 
Following the configurations of Refs.~\cite{pepino2009atomtronic,pepino2010open}, we consider a 1D BHM split into two sections in real space: half uniformly interacting with $U>0$ and the other noninteracting with $U=0$. However, the hopping coefficient $t$ is assumed to be uniform. Therefore, the BHM Hamiltonian in Eq.~\eqref{eq:H} has $t_i=t$ for all $i$ while $U_i=0$ for $0\le i< L/2$ (or $L/2\le i< L$)  and $U_i=U$ for $L/2\le i< L$ (or $0\le i< L/2$). Explicitly, $H_{diode}=-t\sum_{i=1}^{L-1}c^{\dagger}_{i}c_{i+1}+H.c.+U\sum_{half}n_i n_i$. The reverse bias is induced by swapping the reservoirs, or swapping $N_\mathcal{L}\leftrightarrow N_\mathcal{R}$ in the simulation. We will present the results with $N_\mathcal{L}=0$ and $N_\mathcal{R}=1$ ($N_\mathcal{L}=1$ and $N_\mathcal{R}=0$) for the forward (reverse) case.

Equivalently, one can reverse the inhomogeneous interaction profile of the BHM without swapping the reservoirs.
Given an interaction profile of $[0,0,\cdots U/t,U/t,\cdots]$, the reverse configuration is $[U/t,U/t,\cdots,0,0,\cdots]$. We denote the reverse configuration by a superscript $R$. In our convention, the forward configuration has $U > 0$ in contact with the reservoir with the larger number density. Therefore, $J_{ij}^{R}$ is greater than $J_{ij}$ because it is easier for the bosons to enter the region with lower onsite interactions due to low scattering among the bosons. Fig.~\ref{fig:diode_finite} shows the finite-size scaling of the steady-state currents in the forward configuration. As $L$ increases, the steady-state current approaches the thermodynamic-limit ($L\rightarrow \infty$) results according to the 3rd Q with Hartree approximation. The steady-state currents of the reverse configuration show similar finite-size scaling.

\begin{figure}[t]
    \centering
    \includegraphics[width=0.9\columnwidth]{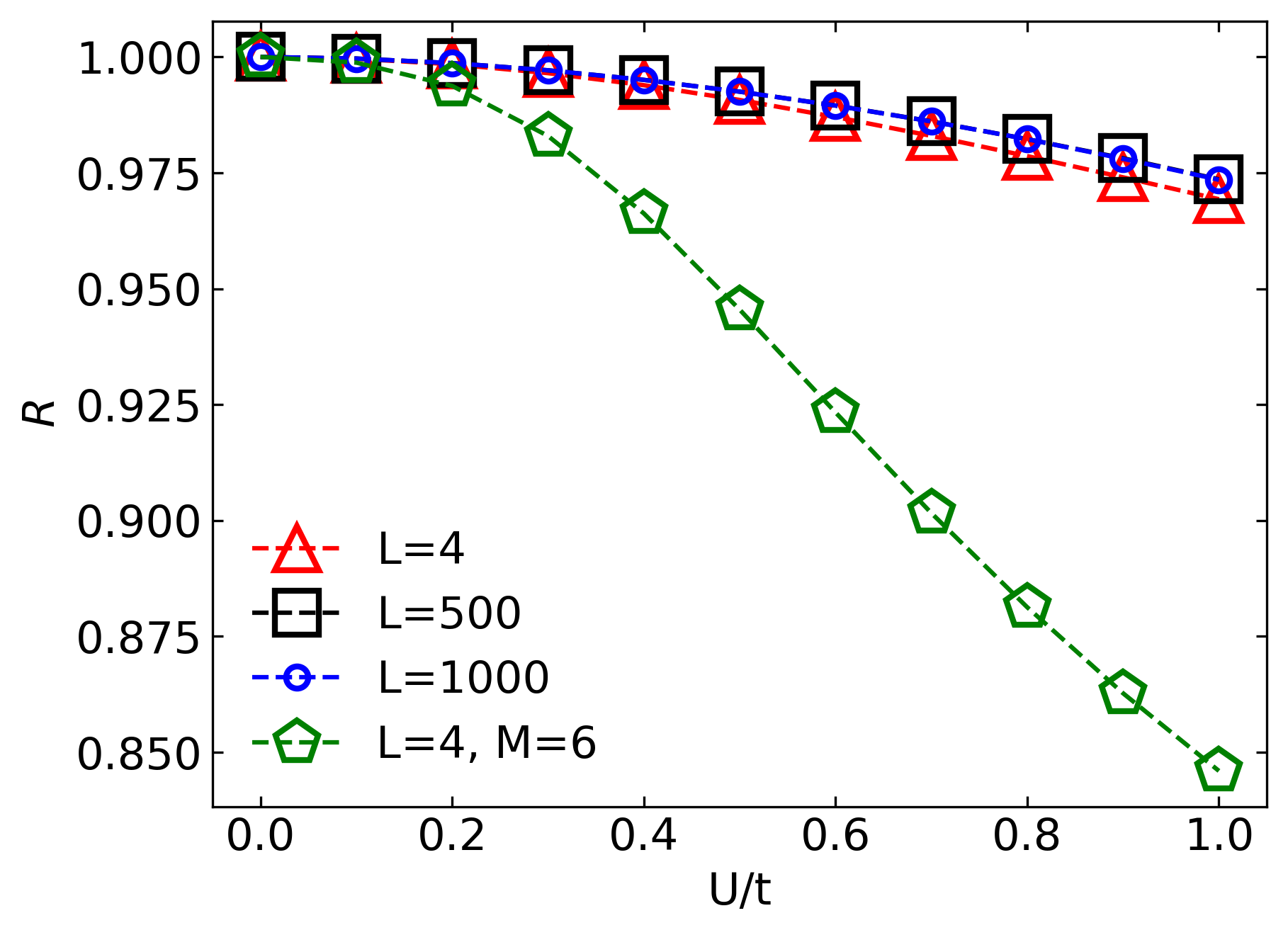}
    \caption{Interaction induced rectification of the inhomogeneous BHM described by the Lindblad equation. The 3rd Q with Hartree approximation with different values of $L$ shows similar decay of the rectification ratio $R$ with $U/t$. Meanwhile, the QuTiP simulation with $L=4$ and $M=6$ exhibits a steeper decline of $R$ as $U$ increases. Here $\gamma_\mathcal{L}=\gamma_\mathcal{R}=\gamma=t$, $N_\mathcal{L}=0$, and $N_\mathcal{R}=1$.}
    \label{fig:rectvinter_sites}
\end{figure}

We define the degree of the diode effect by means of the rectification ratio, or rectification for short, as in Refs.~\cite{mendoza2024giant,capozzi2015single,kornilovitch2002current}. The rectification $R_{ij}$ is the current ratio of the forward and reversed configurations. Explicitly,
\begin{equation}
    R_{ij} = \frac{\braket{J_{ij}}}{\braket{J^{R}_{ij}}}.
    \label{eq:rect}
\end{equation}
Since $J_{ij}^{R} \geq J_{ij}$ following our convention, the rectification ranges from 0 to 1. $R_{ij}=0$ corresponds to perfect diode effect and $R_{ij}=1$ shows no diode effect. In the steady state, $R_{ij}$ is independent of the location, so we will present $R=R_{23}$ in the following. We remark that rectification due to a swap of the reservoirs is only present for interacting systems as confirmed by our numerical simulations, as shown in Fig.~\ref{fig:rectvinter_sites}. The rectification from the 3rd Q with Hartree approximation decreases with $U/t$ and increases slightly as $L$ increases. As a comparison, we also show the result from the QuTiP simulation with $L=4$ and $M=6$. The increasing deviations between the $L=4$ results from the 3rd Q with Hartree approximation and QuTiP simulation again suggest that while the former method captures the qualitative behavior, its values may serve as an upper bound for the exact solution with $M\rightarrow\infty$.

\begin{figure}[t]
    \centering
    \includegraphics[width=0.9\columnwidth]{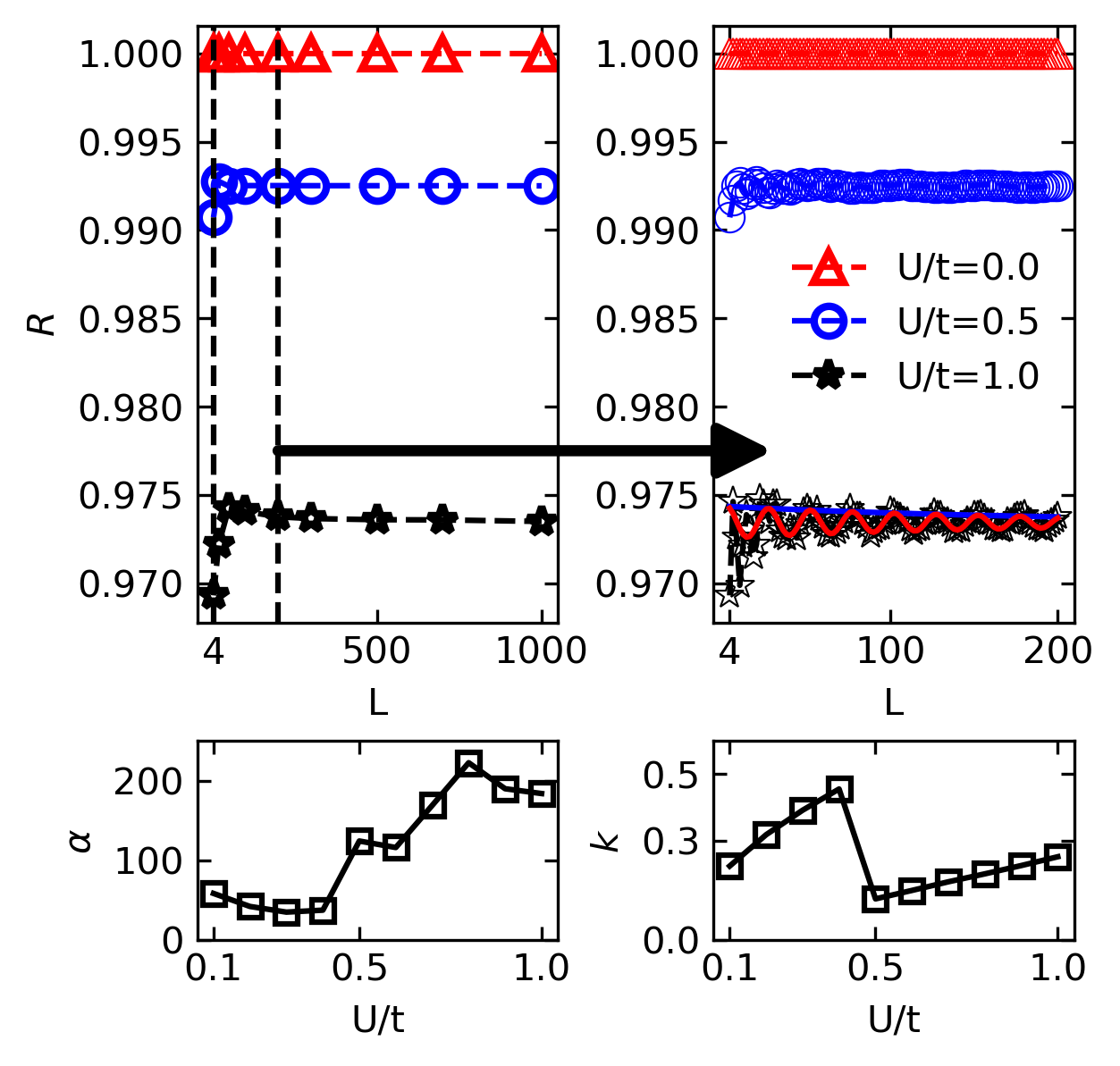}
    \caption{(Top left panel) Rectification ratio $R$ of the inhomogeneous BHM as a function of system size $L$ from the 3rd Q with Hartree approximation for $U/t=0, 0.5, 1$. (Top right panel) The details of the oscillations and an example (the red curve) of fitting to Eq.~\eqref{eq:dampedsin}. (Bottom row) The decay characteristic length $\alpha$ and the wave vector $k$ from fitting $R$.
    }
    \label{fig:rectvsites}
\end{figure}

For the 3rd Q with Hartree approximation, we present a more detailed analysis of the rectification against system size $L$. The diode effect measured by $R$ exhibits an initial steep linear rise followed by a plateau as $L$ increases, as shown in Fig.~\ref{fig:rectvsites}. Interestingly, there are oscillations similar to a damped harmonic oscillator as $R$ settles into the plateau. We therefore fit a damped sinusoidal function of the form
\begin{equation}
    R=A\exp{(-\frac{L}{\alpha})}\sin{(2\pi kL+\phi)} + C,
    \label{eq:dampedsin}
\end{equation}
where $A$ is the amplitude, $\alpha$ the characteristic length of decay, $k$ the wave vector, $\phi$ the phase shift, and $C$ the overall shift. From the fitting, we found that the wave vector  and characteristic lengths of oscillations as a function of interaction exhibit non-monotonic behavior, as shown in  Fig.~\ref{fig:rectvsites}. Around $U/t\approx 0.5$, there is a transition from rapid oscillations to smooth and longer wave length as the system changes from the weakly interacting regime to the intermediate interacting regime.  Moreover, the dependence of the rectification on the wave vector suggests that the rectification may be related to mismatched wave packets of different wave vectors within the inhomogeneous system. Therefore, the rectification in the presence of the inhomogeneous interactions is likely due to the suppression of wave packets in the system, as the two sides of the system block each other with the mismatch. The scalability of the 3rd Q with Hartree approximation to large system size thus allows us to reveal such an interesting phenomenon in large systems closer to the thermodynamic limit.

\subsection{Bosonic SSH-Hubbard model}
Here we investigate the rectification of the SSH-Hubbard model with alternating hopping coefficient $t_1$ and $t_2$ and uniform onsite coupling constant $U$ when the hopping coefficients $t_1$ and $t_2$ are swapped.
In the following, we consider even lattice sites, so the last hopping term is the same as the first. 
The hopping coefficients of the forward configuration has the following profile [$t_{2},t_{1},\cdots,t_{1},t_{2}$], and the reverse configuration has [$t_{1},t_{2},\cdots,t_{2},t_{1}$]. Explicitly, the BHM Hamiltonian~\eqref{eq:H} becomes $H_{SSH-H}=-\sum_{i=1}^{L/2-1}(t_a c^{\dagger}_{2i-1}c_{2i}+t_bc^{\dagger}_{2i}c_{2i+1}+H.c.)+U\sum_i n_i n_i$ with $(t_a,t_b)=(t_2,t_1)$ ($(t_a,t_b)=(t_1,t_2)$) for the forward (reverse) configuration.
Similar to the study of interaction induced diode effect, we also use Eq.~\eqref{eq:rect} to define the rectification as the ratio of the steady-state currents from the forward and reverse configurations. Different from the diode effect due to inhomogeneous interactions studied previously, here the two reservoirs are not swapped.

\begin{figure}[t]
    \centering
    \includegraphics[width=0.9\columnwidth]{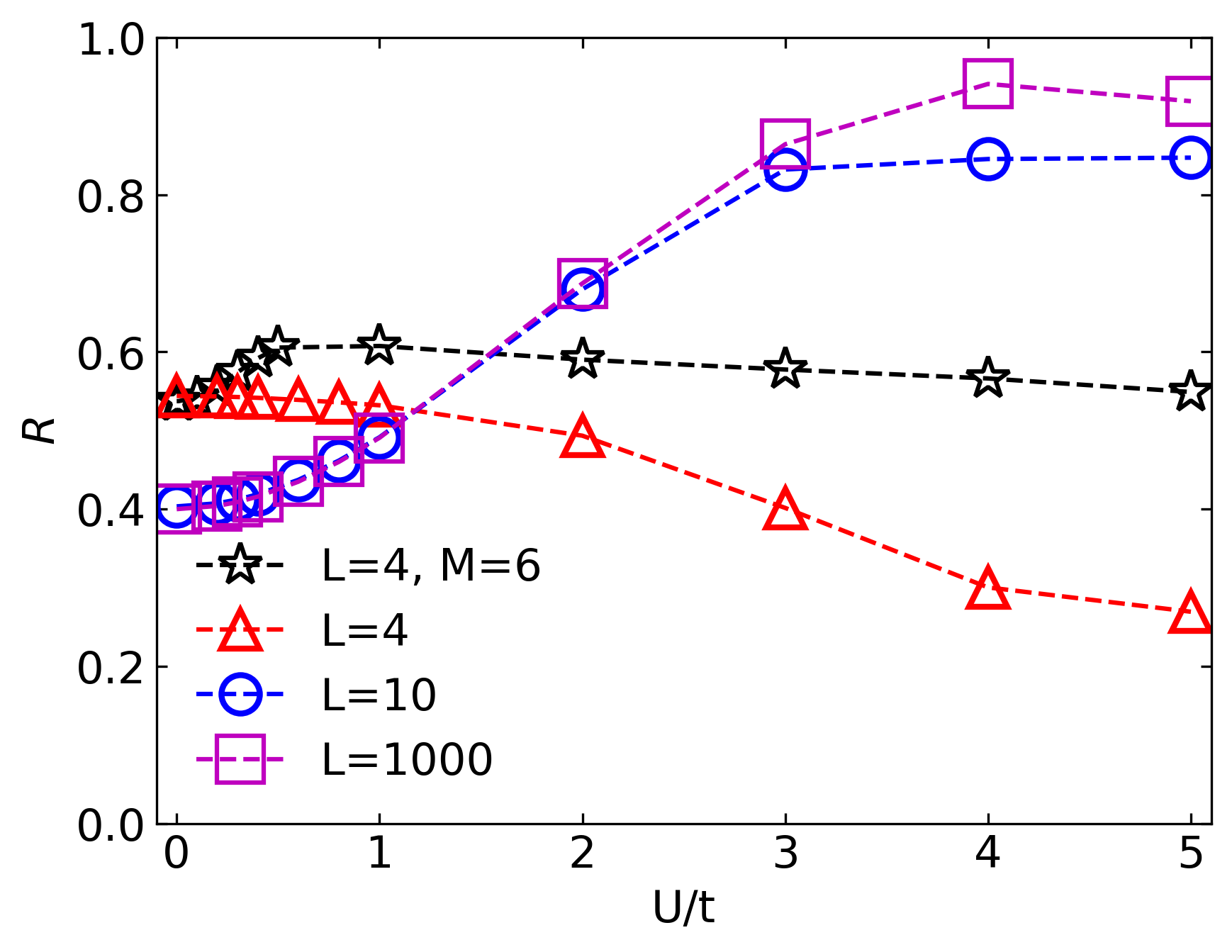}
    \caption{Rectification ratio $R$ of the SSH-Hubbard model as a function of $U$. The black stars show the QuTiP simulation with $L=4$ and $M=6$ while the triangles, circles, and squares shows the 3rd Q with Hartree approximation with $L=4,10,1000$. Here $\gamma_\mathcal{L}=\gamma_\mathcal{R}=\gamma=t_1$, $t_2/t_1=0.5$, $N_\mathcal{L}=0$, and $N_\mathcal{R}=1$.
    }
    \label{fig:sshvinter}
\end{figure}

Fig.~\ref{fig:sshvinter} shows the rectification ratio $R$ of the SSH-Hubbard model from the 3rd Q with Hartree approximation with $L=4, 10, 1000$ as a function of $U/t_1$. The number densities of the reservoirs are $N_\mathcal{L}=0$ and $N_\mathcal{R}=1$, respectively. We also show the small-system result from the QuTiP simulation with $L=4$ and $M=6$. For the small size with $L=4$, the results from QuTiP and 3rd Q with Hartree approximation in the regime with $U/t_1 <1$ are relatively close, again establishing the reliability of the latter. The deviation when $U/t_1 >1$ only offers qualitative indications because the Hartree approximation may not hold there. When we compare the $L=4$, $L=10$, and $L=1000$ results from the 3rd Q with Hartree approximation, they both show increasing $R$ with $U/t_1$ when $U/t_1 < 1$. However, their behavior differ in the strongly interacting regime when $U/t_1 >1$. Moreover, the value of $R$ decreases with $L$ when $U/t_1 < 1$ but increases with $L$ when $U/t_1 >1$. In the weakly interacting regime, the overlap between $L=10$ and $L=1000$ further demonstrates convergence towards the thermodynamic limit. We also caution that rectification due to a swap of the alternating hopping coefficients can survive in the noninteracting SSH model as shown in Fig.~\ref{fig:sshvinter}, and a possible explanation will be given later.

\begin{figure}[t]
    \centering
    \includegraphics[width=0.9\columnwidth]{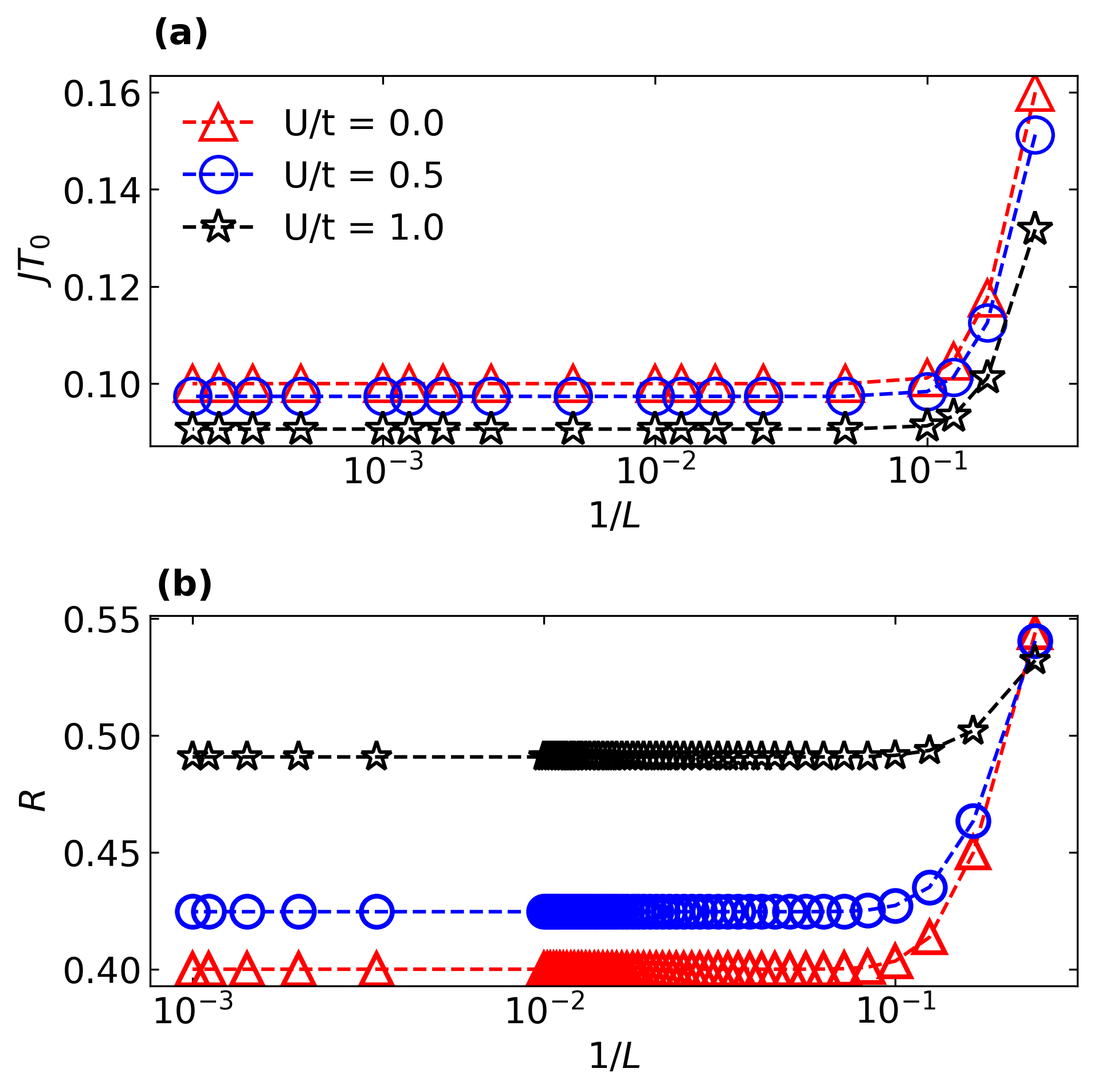}
    \caption{Finite scaling analyses (up to L=5000) of the forward current (a) and rectification ratio (b) of the SSH-Hubbard model as a function of $1/L$ according to the 3rd Q with Hartree approximation for $U/t_1=0,  0.5, 1$. Here  $\gamma_\mathcal{L}=\gamma_\mathcal{R}=\gamma=t_1$, $t_2/t_1=0.5$, $N_\mathcal{L}=0$, and $N_\mathcal{R}=1$.
    }
    \label{fig:sshvsite}
\end{figure}

To further investigate the size dependence, we show the finite-size scaling analyses of the steady-state current and the rectification ratio from the 3rd Q with Hartree approximation for $U/t_1 \le 1$ in Fig.~\ref{fig:sshvsite}. One can see that as $L$ increases, the results saturates into the thermodynamic-limit ($L\rightarrow\infty$) values according to the 3rd Q with Hartree approximation. The monotonic decay of $R$ as $L$ increases of the SSH-Hubbard model differs from that of the inhomogeneous BHM with damped oscillations shown in Fig.~\ref{fig:rectvsites}. This implies the alternating hopping coefficients of the SSH-Hubbard model do not induce observable wave packets of specific wave vectors, in contrast to the BHM with inhomogeneous interactions. Moreover, Fig.~\ref{fig:sshvsite} shows that $R$ reaches a plateau as $L\rightarrow\infty$.
By comparing the values of $R$ when $L$ is large, one can see that the diode effect of the SSH-Hubbard model due to the swapping of $t_1$ and $t_2$ becomes weaker as the onsite interaction increases in the regime $U/t_1 \le 1$.
The scattering effect from $U$ suppresses the steady-state current regardless of the configurations. Since the interactions are uniform, it pushes the system towards the regime without rectification and thus increases $R$.

\begin{figure}[t]
    \centering
    \includegraphics[width=0.9\columnwidth]{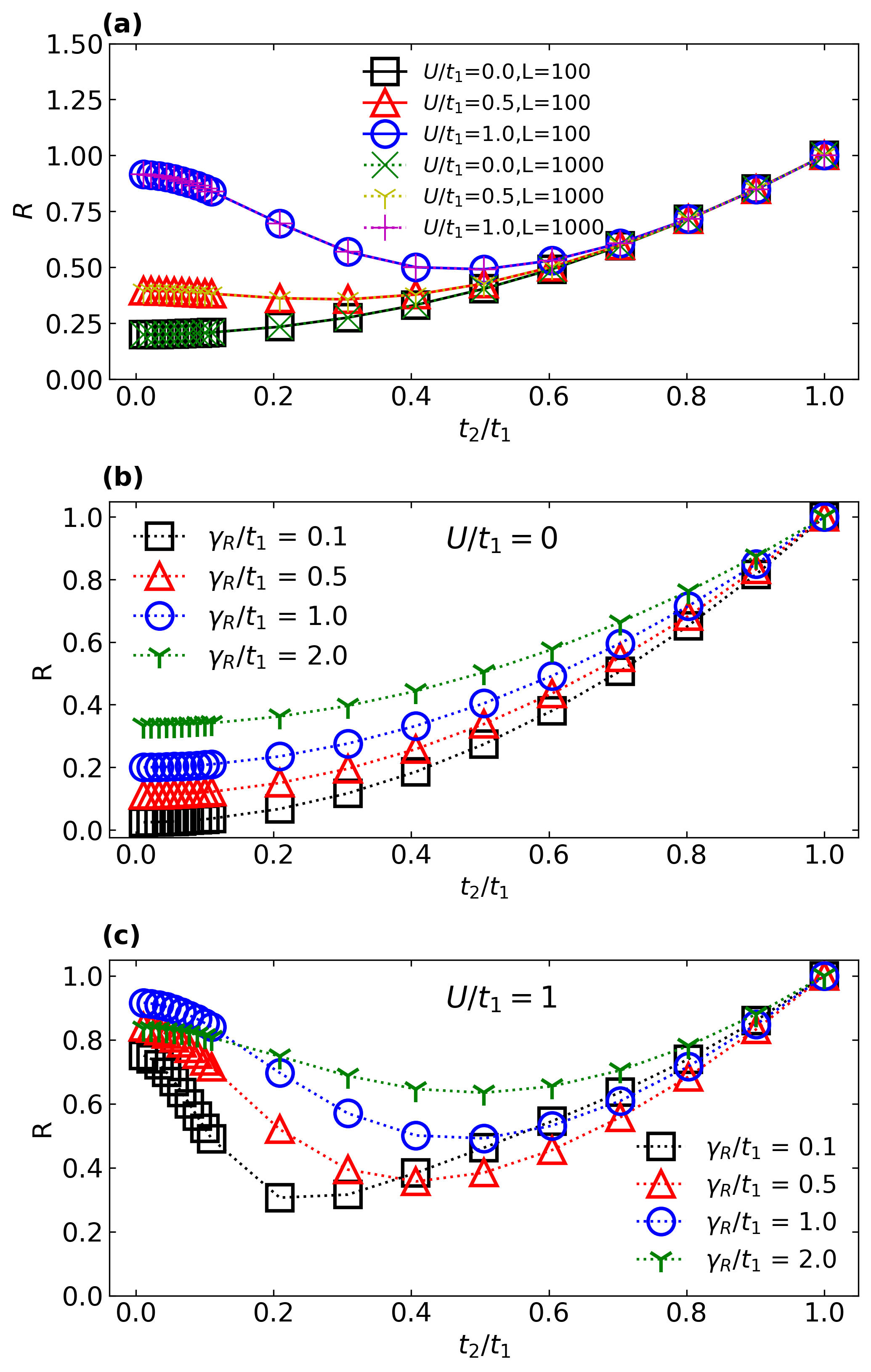}
    \caption{Rectification ratio $R$ as a function of $t_2/t_1$ of the SSH-Hubbard model from the 3rd Q with Hartree approximation. Panel (a) shows $U/t_1 = 0, 0.5, 1$ and fixed $\gamma_\mathcal{L}=\gamma_\mathcal{R}=\gamma=t_1$, Panels (b) and (c) shows $\gamma/t_1 = 0.1, 0.5, 1.0, 2.0$ with $U/t_1=0$ (b) and $U/t_1=1$ (c). Here $N_\mathcal{L}=0$ and $N_\mathcal{R}=1$ and $L=100,1000$ in (a) and $L=100$ in (b) and (c). 
    }
    \label{fig:sshvband}
\end{figure}

Next, Fig.~\ref{fig:sshvband} (a) shows $R$ of the SSH-Hubbard model as a function of $t_2/t_1$ for $L=100$ and $L=1000$ from the 3rd Q with Hartree approximation for selected values of $U/t_1 \le 1$. For the noninteracting case ($U=0$), $R$ decreases with $t_2/t_1$ monotonically. However, the rectification is suppressed by the presence of the onsite interactions, as one can see that the value of $R$ increases with $U/t_1$. When $U/t_1>0$, the rectification is no longer monotonic and shows a dip as $t_2/t_1$ varies. One can see that the $U/t_1=1$ case shows a minimal $R$ around $t_2/t_1=0.5$ and $R\approx 1$ as $t_2/t_1\rightarrow 0$. 

We note that the variation of $R$ with $t_2/t_1$ is not a bulk behavior of the system because a band calculation of the noninteracting ($U=0$) case of the SSH model in the thermodynamic limit shows that the two bands are $ E_{\pm}(k)=\pm\sqrt{t_1^2+t_2^2+2t_1t_2\cos(ka)}$, where $a$ is the lattice constant and $0\le k<2\pi$ is the crystal momentum in the first Brillouin zone. Therefore, the bands are invariant with respect to a swap of $t_1$ and $t_2$ and cannot contribute to the rectification. This indicates the rectification of the SSH-Hubbard model when the hopping coefficients are swapped is from the system-reservoir coupling.

As discussed in Ref.~\cite{he2023particle} for the fermionic SSH model, localized edge states from topological systems can couple to the reservoirs and affect quantum transport. In the present case of bosonic transport, the two ends of the SSH-Hubbard model have both $t_1$ or both $t_2$ connected to the reservoirs. The different couplings between the edges of the system and the reservoirs in the forward and reverse configurations due to the swap of $t_{1,2}$ thus lead to the rectification when we compare the currents through different configurations. This argument is also supported by the aforementioned observation that $U$ in general enhances $R$ as shown in Fig.~\ref{fig:sshvband} (a) and the argument that interaction induces scattering effects regardless of the hopping coefficients, thereby reducing any difference from swapping the configurations.

To further confirm the argument about system-reservoir coupling, we show in Fig.~\ref{fig:sshvband} (b) the rectification ratio as a function of $t_2/t_1$ for the noninteracting ($U=0$) SSH model with different values of the system-reservoir coupling $\gamma$. Indeed, $R$ increases with $\gamma$ because the system-reservoir coupling determines how efficient particles can be exchanged. A higher value of $\gamma$ implies more frequent exchanges between the system and reservoirs regardless of the hopping patterns, so the results are less sensitive to the ratio $t_2/t_1$. Moreover, the noninteracting SSH model approaches perfect rectification ($R=0$) as $t_2/t_1\rightarrow 0$ when $\gamma\rightarrow 0$, but in that limit the current also approaches zero in both configurations due to the vanishing system-reservoir coupling.
A similar increase of $R$ with $\gamma$ for the case with $U/t_1=1$ is shown in Fig.~\ref{fig:sshvband} (c), so the argument works for both noninteracting and interacting systems.

\section{Discussion}\label{Sec:Discussion}
The 3rd Q with Hartree approximation demonstrates its utility as a manageable and efficient approach for investigating qualitative behavior of bosonic transport. The infinite Fock space built in the formalism allows the method to faithfully follow the Bose-Einstein statistics without truncation. This is an advantage over exact constructions which truncates the Fock space, such as the QuTiP simulation or other generalizations~\cite{pepino2009atomtronic,pepino2010open,Lai2016,bychek2020open,palak2022geometry}. We mention that Hartree approximation for interacting bosons has also been implemented in many-body wavefunction calculations~\cite{Manthe17,10.21468/SciPostPhysCore.6.4.073}. However, Hartree approximation stems from the weakly-interacting regime and limits the range of its applicability. From our analysis, the 3rd Q with Hartree approximation is suitable to serve as a quick estimation tool for the rapid development of atomic devices with increasing particles and lattice sites~\cite{caliga2017experimental,caliga2016transport,gati2006realization,vochezer2018light,aidelsburger2011experimental,tao2024high,gyger2024continuous} until other resource-efficient methods for bosonic transport in strongly interacting systems become available. The steady-state currents presented here are particularly useful since measurements of local currents by local density correlations in bosonic systems realized by cold atoms have become available in experiments~\cite{kessler2014single,impertro2024local}.

The three examples studied here are realizable in cold-atom or condensed-matter experiments. For example,  Ref.~\cite{PhysRevResearch.5.023010} demonstrates the 1D BHM with uniform or tilted parameters by loading atoms into optical lattices. It was also used to measure energy ensemble averages in optical lattices~\cite{nakamura2019experimental}. The atomtronic battery has been realized via a double-well ~\cite{caliga2017experimental}, and diode-like behavior has been studied in a similar setup with an adjoined  well~\cite{caliga2016transport}. In addition to atomtronic diode effect~\cite{pepino2009atomtronic,pepino2010open}, diode effect may also be engineered with transmission line resonators and superconducting qubits~\cite{zhao2023engineering}. Nevertheless, our examples of the diode effect in the BHM with inhomogeneous interactions or alternating hopping coefficients show that many tunable parameters can be used to induce asymmetric currents in quantum bosonic systems as the configuration changes. The appearance of diode effect in electronic and bosonic systems shows that it transcends spin statistics and will continue to be an interesting research topic in condensed-matter and atomic physics.  

Meanwhile, Ref.~\cite{MEI201558} summarizes some studies of the SSH model and BHM in cold-atom systems. The bosonic SSH model has been realized in Ref.~\cite{Atala2013} for the measurement of the Zak phase and Ref.~\cite{Lohse2016} for realizing the Thouless pump. Ref.~\cite{doi:10.1126/science.aav9105} realized the SSH model with hard-core bosons by Rydberg atoms. Furthermore, Ref.~\cite{PhysRevResearch.5.L032035} generates a momentum-space lattice resembling the SSH model for interacting bosons and provides an alternative platform for studying the SSH-Hubbard model. Ref.~\cite{besedin2021topological} studies photonic SSH model with attractive Hubbard interactions using superconducting transmon qubits. A slightly modified system of a Bose Hubbard ladder known as the Harper–Hofstadter model has also been realized using cold atoms~\cite{tai2017microscopy} and superconducting processors \cite{ye2019propagation}. Given the relatively large boson numbers and lattice size in those experiments, the 3rd Q with Hartree approximation will be a handy tool for checking qualitative behavior and searching for interesting phenomena.

The calculation of the steady-state transport properties of the 3rd Q with Hartree approximation only scales polynomially with the total system size, as one can see from Eq.~\eqref{eq:slyfast} and its discussion. Moreover, the evaluation only requires the construction of the matrices corresponding to the mean-field Hamiltonian and Lindblad operators and the consistent solution to the steady-state equation. Thus, although here we focus on quantum transport and analyze only 1D examples, the 3rd Q with Hartree approximation can be applied straightforwardly to higher-dimensional weakly interacting many-body bosonic systems towards the thermodynamic limit by using manageable resources and time. Moreover, mean-field theories typically work better in higher dimensions~\cite{ChaikinBook}, so the 3rd Q with Hartree approximation may offer improved quantitative results as the dimensionality increases. We mention that a tensor-based study of 2D BHM investigates a periodic lattice of size $5\times 4$ with up to 7 bosonic states per site~\cite{Kaneko2022}, leaving plenty of room for future research.

The 3rd Q with Hartree approximation also complements other many-body methods. For example, multi-configurational Hartree (or Hartree-Fock) approximations~\cite{PhysRevA.77.033613,10.1063/1.4821350,PhysRevA.93.063601,10.1007/978-3-319-47066-5_6,e23040392,Molignini24} mainly focus on finite-time dynamics of ground-state properties while the 3rd Q with Hartree approximation can reveal the steady-state behavior in the long-time limit of general pure or mixed states. Meanwhile, phase-space methods~\cite{anton2013bosonic,kordas2015nonequilibrium,LEE1995147,LajosDiosi_2002,POLKOVNIKOV20101790,10.1063/1.5046663,10.1002/qute.202100016} are built on kinetic equations or path-integral formalisms whereas the 3rd Q with Hartree approximation is based on the Lindblad equation common in the open quantum system literature.

\section{Conclusion}\label{Sec:Conclusion}
We have presented the 3rd Q with Hartree approximation, which provides a resource-efficient and scalable method for studying steady-state transport properties of weakly interacting bosonic systems described by the Lindblad equation. Importantly, the infinite Fock space of bosons has been maintained in the formalism. By examining three examples from the BHM, the 3rd Q with Hartree approximation captures the qualitative behavior and provides an upper bound of the steady-state current when compared with small-scale simulations with capped boson numbers. Nevertheless, the 3rd Q with Hartree approximation enables us to examine the scaling with system size towards the thermodynamic limit in the studies of rectification due to inhomogeneous interactions or alternating hopping coefficients. The results reveal the dependence of the rectification on the wave vector and system-reservoir coupling. With the rapid progress in cold-atom experiments and quantum simulations relevant to quantum transport of interacting bosons~\cite{Amico2022atomtronic.94.041001,Polo_2024}, the 3rd Q with Hartree approximation will serve as a useful theoretical tool for analyses and predictions of many-body bosonic systems.

\begin{acknowledgments}
We thank Dr. Palak Dugar for helping set up the initial investigation. This work was supported
by the National Science Foundation under Grant No.
PHY-2310656.
\end{acknowledgments}

\appendix

\section{Small-system comparisons}\label{app:a}
\begin{figure}[t]
    \centering
    \includegraphics[width=0.9\columnwidth]{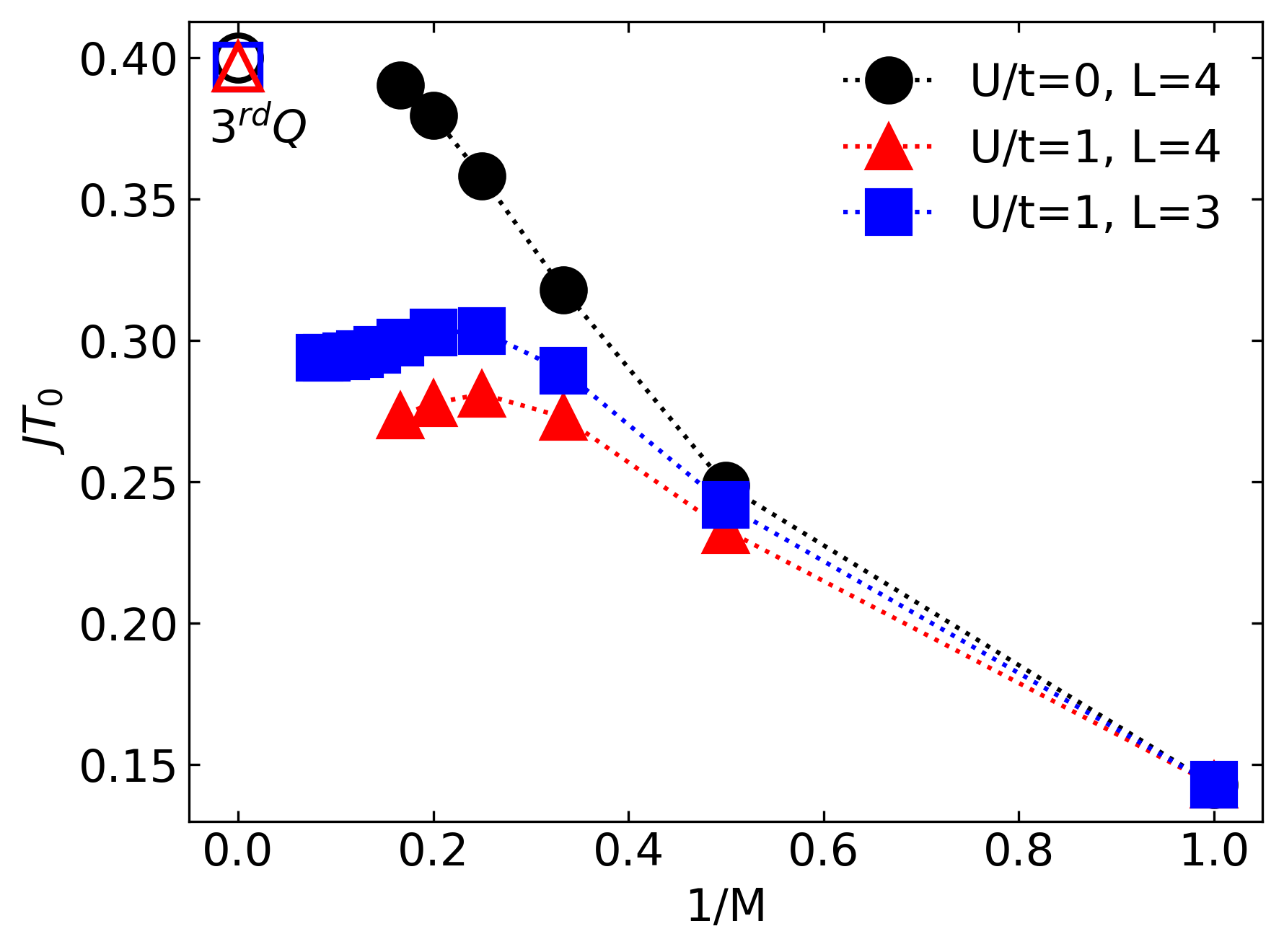}
    \caption{Boson-number scaling of the steady-state currents for small interacting and non-interacting uniform BHM. Here $M$ is the maximal boson numbers on each site. The results from the 3rd Q with Hartree approximation are shown by the empty symbols at $1/M=0$ while the QuTiP simulation results are shown by the solid symbols. Here $\gamma/t=1$, $N_\mathcal{L}=0$, and $N_\mathcal{R}=1$.
    }
    \label{fig:steady_vbose_inter}
\end{figure}

\begin{figure}[t]
    \centering
    \includegraphics[width=0.9\columnwidth]{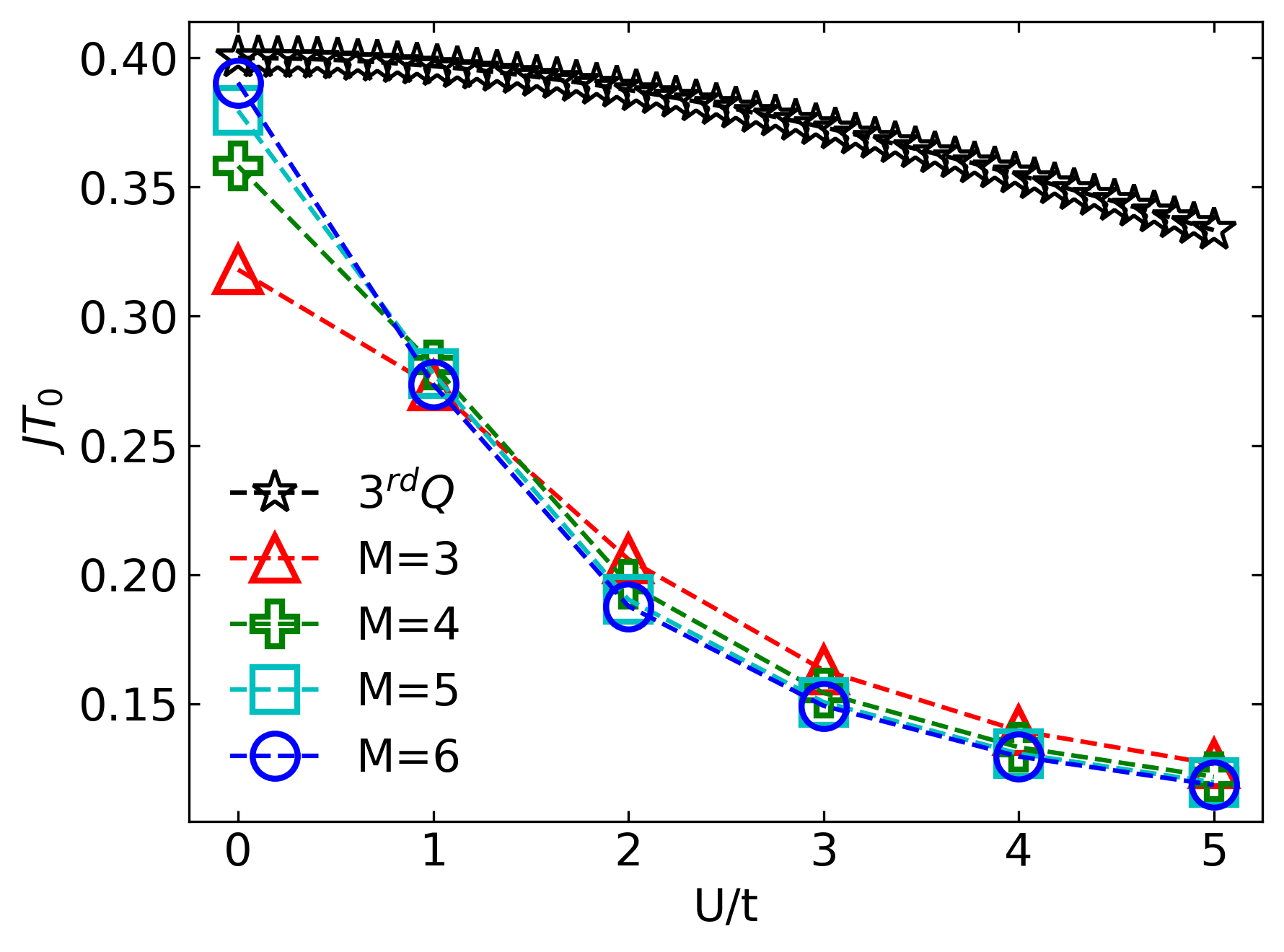}
    \caption{Steady-state currents from the 3rd Q with Hartree approximation and QuTiP simulations with different values of $M$ as a function of $U/t$ of the uniform BHM described by the Lindblad equation. Here $L=4$, $\gamma/t=1$, $N_\mathcal{L}=0$, and $N_\mathcal{R}=1$. 
    }
    \label{fig:steady_vinter}
\end{figure}

\begin{figure}[t]
    \centering
    \includegraphics[width=0.9\columnwidth]{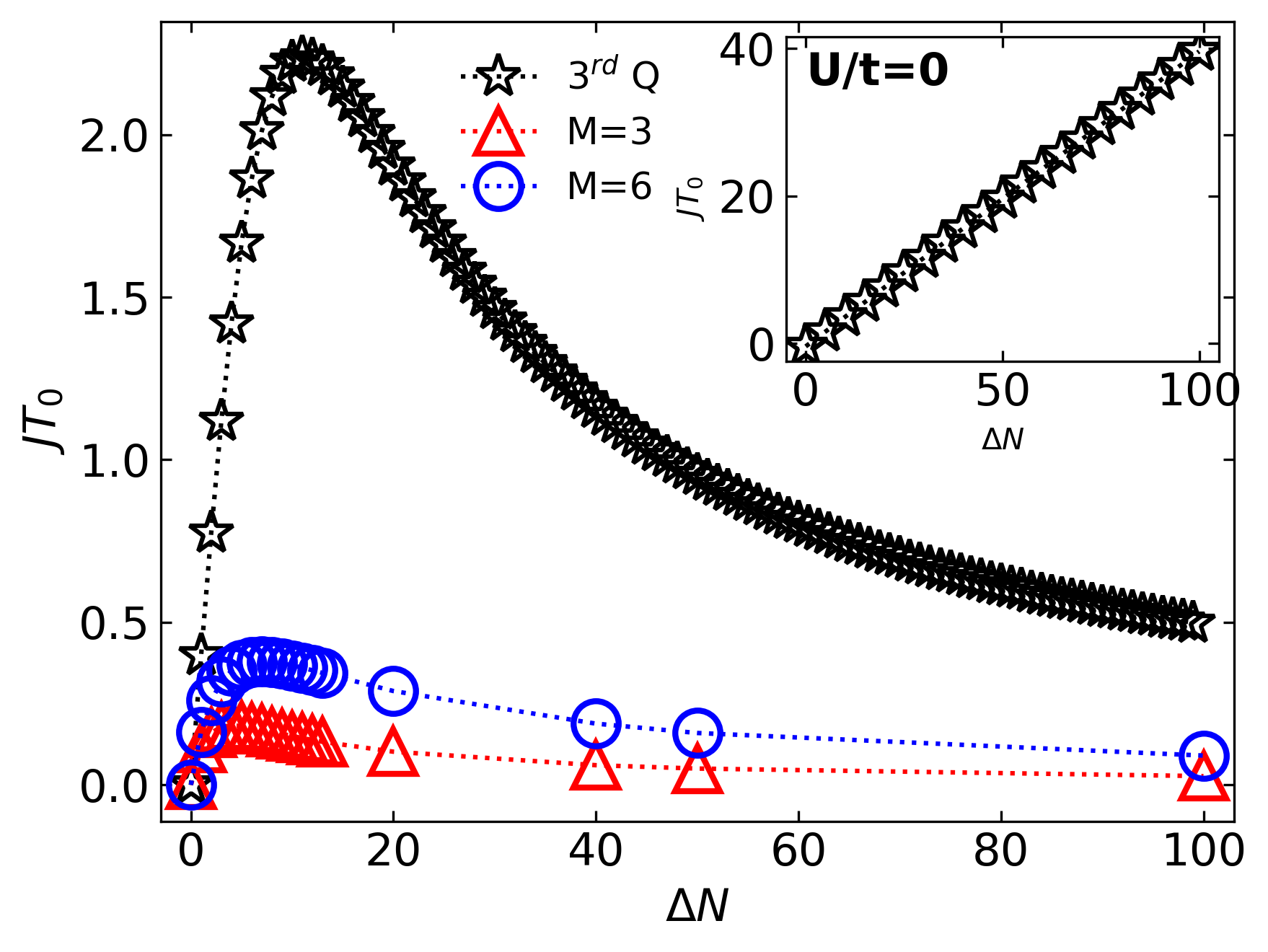}
    \caption{Steady-state currents from the 3rd Q with Hartree approximation and QuTiP simulations with $M=3, 6$ as a function of $\Delta N= N_\mathcal{R}-N_\mathcal{L}$ for the uniform BHM. Here $\gamma/t=1$, $N_\mathcal{L}=1$, $U/t=1$, and $L=4$. The inset shows the $U/t=0$ case with a linear relation.}
    \label{fig:steady_vresvr}
\end{figure}

Here We check the behavior of the 3rd Q with Hartree approximation against the exact solution from the QuTiP simulations for small ($L\le 4$) systems when the latter is still manageable and show the results in Figs.~\ref{fig:steady_vbose_inter}, \ref{fig:steady_vinter}, and \ref{fig:steady_vresvr}. When the system size is small ($L\le 4$), the results from the two methods are compared directly in Fig.~\ref{fig:steady_vbose_inter}. As the interaction changes from $U/t=0$ to $U/t=1$ or the site number changes from $L=4$ to $L=3$, the steady-state current from the 3rd Q with Hartree approximation exhibits very little changes, which are shown by the overlapped symbols at $1/M=0$ in Fig.~\ref{fig:steady_vbose_inter} (a). Meanwhile, the steady-state current of the $L=4$ noninteracting system ($U/t=0$) from the QuTiP simulation increases with $1/M$ and approaches the value from the 3rd Q at $M\rightarrow\infty$. For the interacting systems with $U/t=1$, however, the steady-state currents from the QuTiP simulations are not monotonic with respect to $1/M$ and may go under the corresponding 3rd Q values as $M$ increases.

For a small 1D BHM ($L=4$), Fig.~\ref{fig:steady_vinter} shows that the steady-state current decreases as $U/t$ increases according to both QuTiP simulations with different values of $M$ and 3rd Q with Hartree approximation. However, the QuTiP simulations shows faster decays of the steady-state current as $U/t$ increases while the 3rd Q with Hartree approximation seems to provide an over-estimation.

Next, we investigate the dependence of the steady-state current on $\Delta N= N_\mathcal{R}-N_\mathcal{L}$ by setting $N_\mathcal{L} = 1$ and varying the right number density $N_\mathcal{R}$ to drive the current. For a noninteracting ($U/t=0$) small uniform 1D BHM ($L=4$), the current is linearly dependent with respect to $\Delta N$, as shown in the inset of Fig.~\ref{fig:steady_vresvr}.

However, in the presence of the onsite interactions, the dependence of the steady-state current on $\Delta N$ becomes non-linear, as shown in Fig.~\ref{fig:steady_vresvr} for the 3rd Q with Hartree approximation and QuTiP simulations. The non-monotonic dependence on $\Delta N$ in interacting systems has been discussed by using the truncated Wigner approximation in a system of three quantum dots coupled to two Markovian reservoirs and a two-site system with two reservoirs~\cite{anton2013bosonic,kordas2015nonequilibrium}. Importantly, the 3rd Q with Hartree approximation correctly capture the dome structure for the 1D BHM with $U/t=1$ as $\Delta N$ increases, so the method should be viewed as a qualitative approach which provides an upper-bound of the exact values in the $M\rightarrow\infty$ limit.

\begin{thebibliography}{100}%
	\makeatletter
	\providecommand \@ifxundefined [1]{%
		\@ifx{#1\undefined}
	}%
	\providecommand \@ifnum [1]{%
		\ifnum #1\expandafter \@firstoftwo
		\else \expandafter \@secondoftwo
		\fi
	}%
	\providecommand \@ifx [1]{%
		\ifx #1\expandafter \@firstoftwo
		\else \expandafter \@secondoftwo
		\fi
	}%
	\providecommand \natexlab [1]{#1}%
	\providecommand \enquote  [1]{``#1''}%
	\providecommand \bibnamefont  [1]{#1}%
	\providecommand \bibfnamefont [1]{#1}%
	\providecommand \citenamefont [1]{#1}%
	\providecommand \href@noop [0]{\@secondoftwo}%
	\providecommand \href [0]{\begingroup \@sanitize@url \@href}%
	\providecommand \@href[1]{\@@startlink{#1}\@@href}%
	\providecommand \@@href[1]{\endgroup#1\@@endlink}%
	\providecommand \@sanitize@url [0]{\catcode `\\12\catcode `\$12\catcode
		`\&12\catcode `\#12\catcode `\^12\catcode `\_12\catcode `\%12\relax}%
	\providecommand \@@startlink[1]{}%
	\providecommand \@@endlink[0]{}%
	\providecommand \url  [0]{\begingroup\@sanitize@url \@url }%
	\providecommand \@url [1]{\endgroup\@href {#1}{\urlprefix }}%
	\providecommand \urlprefix  [0]{URL }%
	\providecommand \Eprint [0]{\href }%
	\providecommand \doibase [0]{https://doi.org/}%
	\providecommand \selectlanguage [0]{\@gobble}%
	\providecommand \bibinfo  [0]{\@secondoftwo}%
	\providecommand \bibfield  [0]{\@secondoftwo}%
	\providecommand \translation [1]{[#1]}%
	\providecommand \BibitemOpen [0]{}%
	\providecommand \bibitemStop [0]{}%
	\providecommand \bibitemNoStop [0]{.\EOS\space}%
	\providecommand \EOS [0]{\spacefactor3000\relax}%
	\providecommand \BibitemShut  [1]{\csname bibitem#1\endcsname}%
	\let\auto@bib@innerbib\@empty
	\bibitem [{\citenamefont {Pethick}\ and\ \citenamefont
		{Smith}(2002)}]{pethick2002bose}%
	\BibitemOpen
	\bibfield  {author} {\bibinfo {author} {\bibfnamefont {C.}~\bibnamefont
			{Pethick}}\ and\ \bibinfo {author} {\bibfnamefont {H.}~\bibnamefont
			{Smith}},\ }\href {https://books.google.com/books?id=iBk0G3_5iIQC} {\emph
		{\bibinfo {title} {Bose-Einstein Condensation in Dilute Gases}}}\ (\bibinfo
	{publisher} {Cambridge University Press},\ \bibinfo {address} {Cambridge,
		UK},\ \bibinfo {year} {2002})\BibitemShut {NoStop}%
	\bibitem [{\citenamefont {Ueda}(2010)}]{ueda2010fundamentals}%
	\BibitemOpen
	\bibfield  {author} {\bibinfo {author} {\bibfnamefont {M.}~\bibnamefont
			{Ueda}},\ }\href {https://books.google.com/books?id=YNc7DQAAQBAJ} {\emph
		{\bibinfo {title} {Fundamentals And New Frontiers Of Bose-einstein
				Condensation}}}\ (\bibinfo  {publisher} {World Scientific Publishing
		Company},\ \bibinfo {address} {Singapore},\ \bibinfo {year}
	{2010})\BibitemShut {NoStop}%
	\bibitem [{\citenamefont {Cosco}\ \emph {et~al.}(2018)\citenamefont {Cosco},
		\citenamefont {Borrelli}, \citenamefont {Mendoza-Arenas}, \citenamefont
		{Plastina}, \citenamefont {Jaksch},\ and\ \citenamefont
		{Maniscalco}}]{cosco2018bosehubbard}%
	\BibitemOpen
	\bibfield  {author} {\bibinfo {author} {\bibfnamefont {F.}~\bibnamefont
			{Cosco}}, \bibinfo {author} {\bibfnamefont {M.}~\bibnamefont {Borrelli}},
		\bibinfo {author} {\bibfnamefont {J.~J.}\ \bibnamefont {Mendoza-Arenas}},
		\bibinfo {author} {\bibfnamefont {F.}~\bibnamefont {Plastina}}, \bibinfo
		{author} {\bibfnamefont {D.}~\bibnamefont {Jaksch}},\ and\ \bibinfo {author}
		{\bibfnamefont {S.}~\bibnamefont {Maniscalco}},\ }\bibfield  {title}
	{\bibinfo {title} {Bose-hubbard lattice as a controllable environment for
			open quantum systems},\ }\href {https://doi.org/10.1103/PhysRevA.97.040101}
	{\bibfield  {journal} {\bibinfo  {journal} {Phys. Rev. A}\ }\textbf {\bibinfo
			{volume} {97}},\ \bibinfo {pages} {040101(R)} (\bibinfo {year}
		{2018})}\BibitemShut {NoStop}%
	\bibitem [{\citenamefont {Lewenstein}\ \emph {et~al.}(2007)\citenamefont
		{Lewenstein}, \citenamefont {Sanpera}, \citenamefont {Ahufinger},
		\citenamefont {Damski}, \citenamefont {Sen(De)},\ and\ \citenamefont
		{Sen}}]{maciej2007ultracold}%
	\BibitemOpen
	\bibfield  {author} {\bibinfo {author} {\bibfnamefont {M.}~\bibnamefont
			{Lewenstein}}, \bibinfo {author} {\bibfnamefont {A.}~\bibnamefont {Sanpera}},
		\bibinfo {author} {\bibfnamefont {V.}~\bibnamefont {Ahufinger}}, \bibinfo
		{author} {\bibfnamefont {B.}~\bibnamefont {Damski}}, \bibinfo {author}
		{\bibfnamefont {A.}~\bibnamefont {Sen(De)}},\ and\ \bibinfo {author}
		{\bibfnamefont {U.}~\bibnamefont {Sen}},\ }\bibfield  {title} {\bibinfo
		{title} {Ultracold atomic gases in optical lattices: mimicking condensed
			matter physics and beyond},\ }\href@noop {} {\bibfield  {journal} {\bibinfo
			{journal} {Adv. Phys.}\ }\textbf {\bibinfo {volume} {56}},\ \bibinfo {pages}
		{243} (\bibinfo {year} {2007})}\BibitemShut {NoStop}%
	\bibitem [{\citenamefont {Di~Ventra}(2008)}]{di2008electrical}%
	\BibitemOpen
	\bibfield  {author} {\bibinfo {author} {\bibfnamefont {M.}~\bibnamefont
			{Di~Ventra}},\ }\href {https://books.google.com/books?id=hLyryP7zZmsC} {\emph
		{\bibinfo {title} {Electrical Transport in Nanoscale Systems}}}\ (\bibinfo
	{publisher} {Cambridge University Press},\ \bibinfo {address} {Cambridge,
		UK},\ \bibinfo {year} {2008})\BibitemShut {NoStop}%
	\bibitem [{\citenamefont {Yuli V.~Nazarov}\ and\ \citenamefont
		{Blanter}(2009)}]{Nazarov_book}%
	\BibitemOpen
	\bibfield  {author} {\bibinfo {author} {\bibfnamefont {Y.~V.}\ \bibnamefont
			{Yuli V.~Nazarov}}\ and\ \bibinfo {author} {\bibfnamefont {Y.~M.}\
			\bibnamefont {Blanter}},\ }\href@noop {} {\emph {\bibinfo {title} {Quantum
				Transport: Introduction to Nanoscience}}}\ (\bibinfo  {publisher} {Cambridge
		University Press},\ \bibinfo {address} {Cambridge, UK},\ \bibinfo {year}
	{2009})\BibitemShut {NoStop}%
	\bibitem [{\citenamefont {Chien}\ \emph {et~al.}(2015)\citenamefont {Chien},
		\citenamefont {Peotta},\ and\ \citenamefont {Di~Ventra}}]{ChienNatPhys}%
	\BibitemOpen
	\bibfield  {author} {\bibinfo {author} {\bibfnamefont {C.~C.}\ \bibnamefont
			{Chien}}, \bibinfo {author} {\bibfnamefont {S.}~\bibnamefont {Peotta}},\ and\
		\bibinfo {author} {\bibfnamefont {M.}~\bibnamefont {Di~Ventra}},\ }\bibfield
	{title} {\bibinfo {title} {Quantum transport in ultracold atoms},\
	}\href@noop {} {\bibfield  {journal} {\bibinfo  {journal} {Nat. Phys.}\
		}\textbf {\bibinfo {volume} {11}},\ \bibinfo {pages} {998} (\bibinfo {year}
		{2015})}\BibitemShut {NoStop}%
	\bibitem [{\citenamefont {Hirose}\ and\ \citenamefont
		{Kobayashi}(2014)}]{Hirose_book}%
	\BibitemOpen
	\bibfield  {author} {\bibinfo {author} {\bibfnamefont {K.}~\bibnamefont
			{Hirose}}\ and\ \bibinfo {author} {\bibfnamefont {N.}~\bibnamefont
			{Kobayashi}},\ }\href@noop {} {\emph {\bibinfo {title} {Quantum Transport
				Calculations for Nanosystems}}}\ (\bibinfo  {publisher} {Jenny Stanford
		Publishing},\ \bibinfo {address} {Singapore},\ \bibinfo {year}
	{2014})\BibitemShut {NoStop}%
	\bibitem [{\citenamefont {Waintal}\ \emph {et~al.}(2024)\citenamefont
		{Waintal}, \citenamefont {Wimmer}, \citenamefont {Akhmerov}, \citenamefont
		{Groth}, \citenamefont {Nikolic}, \citenamefont {Istas}, \citenamefont
		{Rosdahl},\ and\ \citenamefont {Varjas}}]{Waintal24}%
	\BibitemOpen
	\bibfield  {author} {\bibinfo {author} {\bibfnamefont {X.}~\bibnamefont
			{Waintal}}, \bibinfo {author} {\bibfnamefont {M.}~\bibnamefont {Wimmer}},
		\bibinfo {author} {\bibfnamefont {A.}~\bibnamefont {Akhmerov}}, \bibinfo
		{author} {\bibfnamefont {C.}~\bibnamefont {Groth}}, \bibinfo {author}
		{\bibfnamefont {B.~K.}\ \bibnamefont {Nikolic}}, \bibinfo {author}
		{\bibfnamefont {M.}~\bibnamefont {Istas}}, \bibinfo {author} {\bibfnamefont
			{T.~O.}\ \bibnamefont {Rosdahl}},\ and\ \bibinfo {author} {\bibfnamefont
			{D.}~\bibnamefont {Varjas}},\ }\href@noop {} {\bibinfo {title} {Computational
			quantum transport}} (\bibinfo {year} {2024}),\ \bibinfo {note} {arXiv:
		2407.16257}\BibitemShut {NoStop}%
	\bibitem [{\citenamefont {Breuer}\ and\ \citenamefont
		{Petruccione}(2002)}]{breuer2002theory}%
	\BibitemOpen
	\bibfield  {author} {\bibinfo {author} {\bibfnamefont {H.-P.}\ \bibnamefont
			{Breuer}}\ and\ \bibinfo {author} {\bibfnamefont {F.}~\bibnamefont
			{Petruccione}},\ }\href@noop {} {\emph {\bibinfo {title} {The theory of open
				quantum systems}}}\ (\bibinfo  {publisher} {Oxford University Press},\
	\bibinfo {address} {Oxford, UK},\ \bibinfo {year} {2002})\BibitemShut
	{NoStop}%
	\bibitem [{\citenamefont {Weiss}(2012)}]{weiss2012quantum}%
	\BibitemOpen
	\bibfield  {author} {\bibinfo {author} {\bibfnamefont {U.}~\bibnamefont
			{Weiss}},\ }\href@noop {} {\emph {\bibinfo {title} {Quantum dissipative
				systems}}}\ (\bibinfo  {publisher} {World scientific},\ \bibinfo {address}
	{Singapore},\ \bibinfo {year} {2012})\BibitemShut {NoStop}%
	\bibitem [{\citenamefont {Li}\ \emph {et~al.}(2018)\citenamefont {Li},
		\citenamefont {Hall},\ and\ \citenamefont {Wiseman}}]{Li2018markov}%
	\BibitemOpen
	\bibfield  {author} {\bibinfo {author} {\bibfnamefont {L.}~\bibnamefont
			{Li}}, \bibinfo {author} {\bibfnamefont {M.~J.}\ \bibnamefont {Hall}},\ and\
		\bibinfo {author} {\bibfnamefont {H.~M.}\ \bibnamefont {Wiseman}},\
	}\bibfield  {title} {\bibinfo {title} {Concepts of quantum non-markovianity:
			A hierarchy},\ }\href
	{https://doi.org/https://doi.org/10.1016/j.physrep.2018.07.001} {\bibfield
		{journal} {\bibinfo  {journal} {Phys. Rep.}\ }\textbf {\bibinfo {volume}
			{759}},\ \bibinfo {pages} {1} (\bibinfo {year} {2018})}\BibitemShut {NoStop}%
	\bibitem [{\citenamefont {Minganti}\ and\ \citenamefont
		{Biella}(2024)}]{Minganti24}%
	\BibitemOpen
	\bibfield  {author} {\bibinfo {author} {\bibfnamefont {F.}~\bibnamefont
			{Minganti}}\ and\ \bibinfo {author} {\bibfnamefont {A.}~\bibnamefont
			{Biella}},\ }\href@noop {} {\bibinfo {title} {Open quantum systems: A brief
			introduction}} (\bibinfo {year} {2024}),\ \bibinfo {note} {arXiv:
		2407.16855}\BibitemShut {NoStop}%
	\bibitem [{\citenamefont {Pepino}\ \emph {et~al.}(2009)\citenamefont {Pepino},
		\citenamefont {Cooper}, \citenamefont {Anderson},\ and\ \citenamefont
		{Holland}}]{pepino2009atomtronic}%
	\BibitemOpen
	\bibfield  {author} {\bibinfo {author} {\bibfnamefont {R.~A.}\ \bibnamefont
			{Pepino}}, \bibinfo {author} {\bibfnamefont {J.}~\bibnamefont {Cooper}},
		\bibinfo {author} {\bibfnamefont {D.~Z.}\ \bibnamefont {Anderson}},\ and\
		\bibinfo {author} {\bibfnamefont {M.~J.}\ \bibnamefont {Holland}},\
	}\bibfield  {title} {\bibinfo {title} {Atomtronic circuits of diodes and
			transistors},\ }\href {https://doi.org/10.1103/PhysRevLett.103.140405}
	{\bibfield  {journal} {\bibinfo  {journal} {Phys. Rev. Lett.}\ }\textbf
		{\bibinfo {volume} {103}},\ \bibinfo {pages} {140405} (\bibinfo {year}
		{2009})}\BibitemShut {NoStop}%
	\bibitem [{\citenamefont {Pepino}\ \emph {et~al.}(2010)\citenamefont {Pepino},
		\citenamefont {Cooper}, \citenamefont {Meiser}, \citenamefont {Anderson},\
		and\ \citenamefont {Holland}}]{pepino2010open}%
	\BibitemOpen
	\bibfield  {author} {\bibinfo {author} {\bibfnamefont {R.~A.}\ \bibnamefont
			{Pepino}}, \bibinfo {author} {\bibfnamefont {J.}~\bibnamefont {Cooper}},
		\bibinfo {author} {\bibfnamefont {D.}~\bibnamefont {Meiser}}, \bibinfo
		{author} {\bibfnamefont {D.~Z.}\ \bibnamefont {Anderson}},\ and\ \bibinfo
		{author} {\bibfnamefont {M.~J.}\ \bibnamefont {Holland}},\ }\bibfield
	{title} {\bibinfo {title} {Open quantum systems approach to atomtronics},\
	}\href {https://doi.org/10.1103/PhysRevA.82.013640} {\bibfield  {journal}
		{\bibinfo  {journal} {Phys. Rev. A}\ }\textbf {\bibinfo {volume} {82}},\
		\bibinfo {pages} {013640} (\bibinfo {year} {2010})}\BibitemShut {NoStop}%
	\bibitem [{\citenamefont {Ghosh}\ \emph {et~al.}(2021)\citenamefont {Ghosh},
		\citenamefont {Chanda}, \citenamefont {Mal},\ and\ \citenamefont
		{Sen(De)}}]{ghosh2021fast}%
	\BibitemOpen
	\bibfield  {author} {\bibinfo {author} {\bibfnamefont {S.}~\bibnamefont
			{Ghosh}}, \bibinfo {author} {\bibfnamefont {T.}~\bibnamefont {Chanda}},
		\bibinfo {author} {\bibfnamefont {S.}~\bibnamefont {Mal}},\ and\ \bibinfo
		{author} {\bibfnamefont {A.}~\bibnamefont {Sen(De)}},\ }\bibfield  {title}
	{\bibinfo {title} {Fast charging of a quantum battery assisted by noise},\
	}\href {https://doi.org/10.1103/PhysRevA.104.032207} {\bibfield  {journal}
		{\bibinfo  {journal} {Phys. Rev. A}\ }\textbf {\bibinfo {volume} {104}},\
		\bibinfo {pages} {032207} (\bibinfo {year} {2021})}\BibitemShut {NoStop}%
	\bibitem [{\citenamefont {Amico}\ \emph {et~al.}(2022)\citenamefont {Amico},
		\citenamefont {Anderson}, \citenamefont {Boshier}, \citenamefont {Brantut},
		\citenamefont {Kwek}, \citenamefont {Minguzzi},\ and\ \citenamefont {von
			Klitzing}}]{Amico2022atomtronic.94.041001}%
	\BibitemOpen
	\bibfield  {author} {\bibinfo {author} {\bibfnamefont {L.}~\bibnamefont
			{Amico}}, \bibinfo {author} {\bibfnamefont {D.}~\bibnamefont {Anderson}},
		\bibinfo {author} {\bibfnamefont {M.}~\bibnamefont {Boshier}}, \bibinfo
		{author} {\bibfnamefont {J.-P.}\ \bibnamefont {Brantut}}, \bibinfo {author}
		{\bibfnamefont {L.-C.}\ \bibnamefont {Kwek}}, \bibinfo {author}
		{\bibfnamefont {A.}~\bibnamefont {Minguzzi}},\ and\ \bibinfo {author}
		{\bibfnamefont {W.}~\bibnamefont {von Klitzing}},\ }\bibfield  {title}
	{\bibinfo {title} {Colloquium: Atomtronic circuits: From many-body physics to
			quantum technologies},\ }\href {https://doi.org/10.1103/RevModPhys.94.041001}
	{\bibfield  {journal} {\bibinfo  {journal} {Rev. Mod. Phys.}\ }\textbf
		{\bibinfo {volume} {94}},\ \bibinfo {pages} {041001} (\bibinfo {year}
		{2022})}\BibitemShut {NoStop}%
	\bibitem [{\citenamefont {Caliga}\ \emph {et~al.}(2017)\citenamefont {Caliga},
		\citenamefont {Straatsma},\ and\ \citenamefont
		{Anderson}}]{caliga2017experimental}%
	\BibitemOpen
	\bibfield  {author} {\bibinfo {author} {\bibfnamefont {S.~C.}\ \bibnamefont
			{Caliga}}, \bibinfo {author} {\bibfnamefont {C.~J.~E.}\ \bibnamefont
			{Straatsma}},\ and\ \bibinfo {author} {\bibfnamefont {D.~Z.}\ \bibnamefont
			{Anderson}},\ }\bibfield  {title} {\bibinfo {title} {Experimental
			demonstration of an atomtronic battery},\ }\href
	{https://doi.org/10.1088/1367-2630/aa56d8} {\bibfield  {journal} {\bibinfo
			{journal} {New J. Phys.}\ }\textbf {\bibinfo {volume} {19}},\ \bibinfo
		{pages} {013036} (\bibinfo {year} {2017})}\BibitemShut {NoStop}%
	\bibitem [{\citenamefont {Caliga}\ \emph {et~al.}(2016)\citenamefont {Caliga},
		\citenamefont {Straatsma},\ and\ \citenamefont
		{Anderson}}]{caliga2016transport}%
	\BibitemOpen
	\bibfield  {author} {\bibinfo {author} {\bibfnamefont {S.~C.}\ \bibnamefont
			{Caliga}}, \bibinfo {author} {\bibfnamefont {C.~J.~E.}\ \bibnamefont
			{Straatsma}},\ and\ \bibinfo {author} {\bibfnamefont {D.~Z.}\ \bibnamefont
			{Anderson}},\ }\bibfield  {title} {\bibinfo {title} {Transport dynamics of
			ultracold atoms in a triple-well transistor-like potential},\ }\href
	{https://doi.org/10.1088/1367-2630/18/2/025010} {\bibfield  {journal}
		{\bibinfo  {journal} {New J. Phys.}\ }\textbf {\bibinfo {volume} {18}},\
		\bibinfo {pages} {025010} (\bibinfo {year} {2016})}\BibitemShut {NoStop}%
	\bibitem [{\citenamefont {Gati}\ \emph {et~al.}(2006)\citenamefont {Gati},
		\citenamefont {Albiez}, \citenamefont {F{\"o}lling}, \citenamefont
		{Hemmerling},\ and\ \citenamefont {Oberthaler}}]{gati2006realization}%
	\BibitemOpen
	\bibfield  {author} {\bibinfo {author} {\bibfnamefont {R.}~\bibnamefont
			{Gati}}, \bibinfo {author} {\bibfnamefont {M.}~\bibnamefont {Albiez}},
		\bibinfo {author} {\bibfnamefont {J.}~\bibnamefont {F{\"o}lling}}, \bibinfo
		{author} {\bibfnamefont {B.}~\bibnamefont {Hemmerling}},\ and\ \bibinfo
		{author} {\bibfnamefont {M.~K.}\ \bibnamefont {Oberthaler}},\ }\bibfield
	{title} {\bibinfo {title} {Realization of a single josephson junction for
			bose--einstein condensates},\ }\href
	{https://doi.org/10.1007/s00340-005-2059-z} {\bibfield  {journal} {\bibinfo
			{journal} {Appl. Phys. B}\ }\textbf {\bibinfo {volume} {82}},\ \bibinfo
		{pages} {207} (\bibinfo {year} {2006})}\BibitemShut {NoStop}%
	\bibitem [{\citenamefont {Gangat}\ \emph {et~al.}(2018)\citenamefont {Gangat},
		\citenamefont {McCulloch},\ and\ \citenamefont {Kao}}]{gangat2018symmetry}%
	\BibitemOpen
	\bibfield  {author} {\bibinfo {author} {\bibfnamefont {A.~A.}\ \bibnamefont
			{Gangat}}, \bibinfo {author} {\bibfnamefont {I.~P.}\ \bibnamefont
			{McCulloch}},\ and\ \bibinfo {author} {\bibfnamefont {Y.-J.}\ \bibnamefont
			{Kao}},\ }\bibfield  {title} {\bibinfo {title} {Symmetry between repulsive
			and attractive interactions in driven-dissipative bose-hubbard systems},\
	}\href {https://doi.org/10.1038/s41598-018-21845-5} {\bibfield  {journal}
		{\bibinfo  {journal} {Sci. Rep.}\ }\textbf {\bibinfo {volume} {8}},\ \bibinfo
		{pages} {3698} (\bibinfo {year} {2018})}\BibitemShut {NoStop}%
	\bibitem [{\citenamefont {Lai}\ and\ \citenamefont {Chien}(2016)}]{Lai2016}%
	\BibitemOpen
	\bibfield  {author} {\bibinfo {author} {\bibfnamefont {C.-Y.}\ \bibnamefont
			{Lai}}\ and\ \bibinfo {author} {\bibfnamefont {C.-C.}\ \bibnamefont
			{Chien}},\ }\bibfield  {title} {\bibinfo {title} {Challenges and constraints
			of dynamically emerged source and sink in atomtronic circuits: From
			closed-system to open-system approaches},\ }\href
	{https://doi.org/10.1038/srep37256} {\bibfield  {journal} {\bibinfo
			{journal} {Sci. Rep.}\ }\textbf {\bibinfo {volume} {6}},\ \bibinfo {pages}
		{37256} (\bibinfo {year} {2016})}\BibitemShut {NoStop}%
	\bibitem [{\citenamefont {Vochezer}\ \emph {et~al.}(2018)\citenamefont
		{Vochezer}, \citenamefont {Kampschulte}, \citenamefont {Hammerer},\ and\
		\citenamefont {Treutlein}}]{vochezer2018light}%
	\BibitemOpen
	\bibfield  {author} {\bibinfo {author} {\bibfnamefont {A.}~\bibnamefont
			{Vochezer}}, \bibinfo {author} {\bibfnamefont {T.}~\bibnamefont
			{Kampschulte}}, \bibinfo {author} {\bibfnamefont {K.}~\bibnamefont
			{Hammerer}},\ and\ \bibinfo {author} {\bibfnamefont {P.}~\bibnamefont
			{Treutlein}},\ }\bibfield  {title} {\bibinfo {title} {Light-mediated
			collective atomic motion in an optical lattice coupled to a membrane},\
	}\href {https://doi.org/10.1103/PhysRevLett.120.073602} {\bibfield  {journal}
		{\bibinfo  {journal} {Phys. Rev. Lett.}\ }\textbf {\bibinfo {volume} {120}},\
		\bibinfo {pages} {073602} (\bibinfo {year} {2018})}\BibitemShut {NoStop}%
	\bibitem [{\citenamefont {Yang}\ \emph {et~al.}(2020)\citenamefont {Yang},
		\citenamefont {Sun}, \citenamefont {Ott}, \citenamefont {Wang}, \citenamefont
		{Zache}, \citenamefont {Halimeh}, \citenamefont {Yuan}, \citenamefont
		{Hauke},\ and\ \citenamefont {Pan}}]{yang2020observation}%
	\BibitemOpen
	\bibfield  {author} {\bibinfo {author} {\bibfnamefont {B.}~\bibnamefont
			{Yang}}, \bibinfo {author} {\bibfnamefont {H.}~\bibnamefont {Sun}}, \bibinfo
		{author} {\bibfnamefont {R.}~\bibnamefont {Ott}}, \bibinfo {author}
		{\bibfnamefont {H.-Y.}\ \bibnamefont {Wang}}, \bibinfo {author}
		{\bibfnamefont {T.~V.}\ \bibnamefont {Zache}}, \bibinfo {author}
		{\bibfnamefont {J.~C.}\ \bibnamefont {Halimeh}}, \bibinfo {author}
		{\bibfnamefont {Z.-S.}\ \bibnamefont {Yuan}}, \bibinfo {author}
		{\bibfnamefont {P.}~\bibnamefont {Hauke}},\ and\ \bibinfo {author}
		{\bibfnamefont {J.-W.}\ \bibnamefont {Pan}},\ }\bibfield  {title} {\bibinfo
		{title} {Observation of gauge invariance in a 71-site bose--hubbard quantum
			simulator},\ }\href {https://doi.org/10.1038/s41586-020-2910-8} {\bibfield
		{journal} {\bibinfo  {journal} {Nature}\ }\textbf {\bibinfo {volume} {587}},\
		\bibinfo {pages} {392} (\bibinfo {year} {2020})}\BibitemShut {NoStop}%
	\bibitem [{\citenamefont {Aidelsburger}\ \emph {et~al.}(2011)\citenamefont
		{Aidelsburger}, \citenamefont {Atala}, \citenamefont {Nascimb\`ene},
		\citenamefont {Trotzky}, \citenamefont {Chen},\ and\ \citenamefont
		{Bloch}}]{aidelsburger2011experimental}%
	\BibitemOpen
	\bibfield  {author} {\bibinfo {author} {\bibfnamefont {M.}~\bibnamefont
			{Aidelsburger}}, \bibinfo {author} {\bibfnamefont {M.}~\bibnamefont {Atala}},
		\bibinfo {author} {\bibfnamefont {S.}~\bibnamefont {Nascimb\`ene}}, \bibinfo
		{author} {\bibfnamefont {S.}~\bibnamefont {Trotzky}}, \bibinfo {author}
		{\bibfnamefont {Y.-A.}\ \bibnamefont {Chen}},\ and\ \bibinfo {author}
		{\bibfnamefont {I.}~\bibnamefont {Bloch}},\ }\bibfield  {title} {\bibinfo
		{title} {Experimental realization of strong effective magnetic fields in an
			optical lattice},\ }\href {https://doi.org/10.1103/PhysRevLett.107.255301}
	{\bibfield  {journal} {\bibinfo  {journal} {Phys. Rev. Lett.}\ }\textbf
		{\bibinfo {volume} {107}},\ \bibinfo {pages} {255301} (\bibinfo {year}
		{2011})}\BibitemShut {NoStop}%
	\bibitem [{\citenamefont {Tao}\ \emph {et~al.}(2024)\citenamefont {Tao},
		\citenamefont {Ammenwerth}, \citenamefont {Gyger}, \citenamefont {Bloch},\
		and\ \citenamefont {Zeiher}}]{tao2024high}%
	\BibitemOpen
	\bibfield  {author} {\bibinfo {author} {\bibfnamefont {R.}~\bibnamefont
			{Tao}}, \bibinfo {author} {\bibfnamefont {M.}~\bibnamefont {Ammenwerth}},
		\bibinfo {author} {\bibfnamefont {F.}~\bibnamefont {Gyger}}, \bibinfo
		{author} {\bibfnamefont {I.}~\bibnamefont {Bloch}},\ and\ \bibinfo {author}
		{\bibfnamefont {J.}~\bibnamefont {Zeiher}},\ }\bibfield  {title} {\bibinfo
		{title} {High-fidelity detection of large-scale atom arrays in an optical
			lattice},\ }\href {https://doi.org/10.1103/PhysRevLett.133.013401} {\bibfield
		{journal} {\bibinfo  {journal} {Phys. Rev. Lett.}\ }\textbf {\bibinfo
			{volume} {133}},\ \bibinfo {pages} {013401} (\bibinfo {year}
		{2024})}\BibitemShut {NoStop}%
	\bibitem [{\citenamefont {Gyger}\ \emph {et~al.}(2024)\citenamefont {Gyger},
		\citenamefont {Ammenwerth}, \citenamefont {Tao}, \citenamefont {Timme},
		\citenamefont {Snigirev}, \citenamefont {Bloch},\ and\ \citenamefont
		{Zeiher}}]{gyger2024continuous}%
	\BibitemOpen
	\bibfield  {author} {\bibinfo {author} {\bibfnamefont {F.}~\bibnamefont
			{Gyger}}, \bibinfo {author} {\bibfnamefont {M.}~\bibnamefont {Ammenwerth}},
		\bibinfo {author} {\bibfnamefont {R.}~\bibnamefont {Tao}}, \bibinfo {author}
		{\bibfnamefont {H.}~\bibnamefont {Timme}}, \bibinfo {author} {\bibfnamefont
			{S.}~\bibnamefont {Snigirev}}, \bibinfo {author} {\bibfnamefont
			{I.}~\bibnamefont {Bloch}},\ and\ \bibinfo {author} {\bibfnamefont
			{J.}~\bibnamefont {Zeiher}},\ }\bibfield  {title} {\bibinfo {title}
		{Continuous operation of large-scale atom arrays in optical lattices},\
	}\href {https://doi.org/10.1103/PhysRevResearch.6.033104} {\bibfield
		{journal} {\bibinfo  {journal} {Phys. Rev. Res.}\ }\textbf {\bibinfo {volume}
			{6}},\ \bibinfo {pages} {033104} (\bibinfo {year} {2024})}\BibitemShut
	{NoStop}%
	\bibitem [{\citenamefont {Muraev}\ \emph {et~al.}(2022)\citenamefont {Muraev},
		\citenamefont {Maksimov},\ and\ \citenamefont
		{Kolovsky}}]{PhysRevA.105.013307}%
	\BibitemOpen
	\bibfield  {author} {\bibinfo {author} {\bibfnamefont {P.~S.}\ \bibnamefont
			{Muraev}}, \bibinfo {author} {\bibfnamefont {D.~N.}\ \bibnamefont
			{Maksimov}},\ and\ \bibinfo {author} {\bibfnamefont {A.~R.}\ \bibnamefont
			{Kolovsky}},\ }\bibfield  {title} {\bibinfo {title} {Resonant transport of
			bosonic carriers through a quantum device},\ }\href
	{https://doi.org/10.1103/PhysRevA.105.013307} {\bibfield  {journal} {\bibinfo
			{journal} {Phys. Rev. A}\ }\textbf {\bibinfo {volume} {105}},\ \bibinfo
		{pages} {013307} (\bibinfo {year} {2022})}\BibitemShut {NoStop}%
	\bibitem [{\citenamefont {Trivedi}\ \emph {et~al.}(2023)\citenamefont
		{Trivedi}, \citenamefont {Gupta}, \citenamefont {Agarwalla}, \citenamefont
		{Dhar}, \citenamefont {Kulkarni}, \citenamefont {Kundu},\ and\ \citenamefont
		{Sabhapandit}}]{trivedi2023filling}%
	\BibitemOpen
	\bibfield  {author} {\bibinfo {author} {\bibfnamefont {A.}~\bibnamefont
			{Trivedi}}, \bibinfo {author} {\bibfnamefont {S.}~\bibnamefont {Gupta}},
		\bibinfo {author} {\bibfnamefont {B.~K.}\ \bibnamefont {Agarwalla}}, \bibinfo
		{author} {\bibfnamefont {A.}~\bibnamefont {Dhar}}, \bibinfo {author}
		{\bibfnamefont {M.}~\bibnamefont {Kulkarni}}, \bibinfo {author}
		{\bibfnamefont {A.}~\bibnamefont {Kundu}},\ and\ \bibinfo {author}
		{\bibfnamefont {S.}~\bibnamefont {Sabhapandit}},\ }\bibfield  {title}
	{\bibinfo {title} {Filling an empty lattice by local injection of quantum
			particles},\ }\href {https://doi.org/10.1103/PhysRevA.108.052204} {\bibfield
		{journal} {\bibinfo  {journal} {Phys. Rev. A}\ }\textbf {\bibinfo {volume}
			{108}},\ \bibinfo {pages} {052204} (\bibinfo {year} {2023})}\BibitemShut
	{NoStop}%
	\bibitem [{\citenamefont {Kolovsky}\ \emph {et~al.}(2018)\citenamefont
		{Kolovsky}, \citenamefont {Denis},\ and\ \citenamefont
		{Wimberger}}]{PhysRevA.98.043623}%
	\BibitemOpen
	\bibfield  {author} {\bibinfo {author} {\bibfnamefont {A.~R.}\ \bibnamefont
			{Kolovsky}}, \bibinfo {author} {\bibfnamefont {Z.}~\bibnamefont {Denis}},\
		and\ \bibinfo {author} {\bibfnamefont {S.}~\bibnamefont {Wimberger}},\
	}\bibfield  {title} {\bibinfo {title} {Landauer-b\"uttiker equation for
			bosonic carriers},\ }\href {https://doi.org/10.1103/PhysRevA.98.043623}
	{\bibfield  {journal} {\bibinfo  {journal} {Phys. Rev. A}\ }\textbf {\bibinfo
			{volume} {98}},\ \bibinfo {pages} {043623} (\bibinfo {year}
		{2018})}\BibitemShut {NoStop}%
	\bibitem [{\citenamefont {Santos}\ and\ \citenamefont
		{Landi}(2016)}]{PhysRevE.94.062143}%
	\BibitemOpen
	\bibfield  {author} {\bibinfo {author} {\bibfnamefont {J.~P.}\ \bibnamefont
			{Santos}}\ and\ \bibinfo {author} {\bibfnamefont {G.~T.}\ \bibnamefont
			{Landi}},\ }\bibfield  {title} {\bibinfo {title} {Microscopic theory of a
			nonequilibrium open bosonic chain},\ }\href
	{https://doi.org/10.1103/PhysRevE.94.062143} {\bibfield  {journal} {\bibinfo
			{journal} {Phys. Rev. E}\ }\textbf {\bibinfo {volume} {94}},\ \bibinfo
		{pages} {062143} (\bibinfo {year} {2016})}\BibitemShut {NoStop}%
	\bibitem [{\citenamefont {Prosen}\ and\ \citenamefont
		{Seligman}(2010)}]{prosen2010quantization}%
	\BibitemOpen
	\bibfield  {author} {\bibinfo {author} {\bibfnamefont {T.}~\bibnamefont
			{Prosen}}\ and\ \bibinfo {author} {\bibfnamefont {T.~H.}\ \bibnamefont
			{Seligman}},\ }\bibfield  {title} {\bibinfo {title} {Quantization over boson
			operator spaces},\ }\href@noop {} {\bibfield  {journal} {\bibinfo  {journal}
			{J. Phys. A.: Math. Theor.}\ }\textbf {\bibinfo {volume} {43}},\ \bibinfo
		{pages} {392004} (\bibinfo {year} {2010})}\BibitemShut {NoStop}%
	\bibitem [{\citenamefont {Lindblad}(1976)}]{lindblad1976semigroup}%
	\BibitemOpen
	\bibfield  {author} {\bibinfo {author} {\bibfnamefont {G.}~\bibnamefont
			{Lindblad}},\ }\bibfield  {title} {\bibinfo {title} {On the generators of
			quantum dynamical semigroups},\ }\href {https://doi.org/10.1007/BF01608499}
	{\bibfield  {journal} {\bibinfo  {journal} {Commun. Math. Phys.}\ }\textbf
		{\bibinfo {volume} {48}},\ \bibinfo {pages} {119} (\bibinfo {year}
		{1976})}\BibitemShut {NoStop}%
	\bibitem [{\citenamefont {Schlosshauer}(2007)}]{Schlosshauer_book}%
	\BibitemOpen
	\bibfield  {author} {\bibinfo {author} {\bibfnamefont {M.}~\bibnamefont
			{Schlosshauer}},\ }\href@noop {} {\emph {\bibinfo {title} {Decoherence and
				the Quantum-To-Classical Transition}}}\ (\bibinfo  {publisher} {Springer},\
	\bibinfo {address} {Heidelberg, Germany},\ \bibinfo {year}
	{2007})\BibitemShut {NoStop}%
	\bibitem [{\citenamefont {Zhang}\ and\ \citenamefont
		{Barthel}(2022)}]{PhysRevLett.129.120401}%
	\BibitemOpen
	\bibfield  {author} {\bibinfo {author} {\bibfnamefont {Y.}~\bibnamefont
			{Zhang}}\ and\ \bibinfo {author} {\bibfnamefont {T.}~\bibnamefont
			{Barthel}},\ }\bibfield  {title} {\bibinfo {title} {Criticality and phase
			classification for quadratic open quantum many-body systems},\ }\href
	{https://doi.org/10.1103/PhysRevLett.129.120401} {\bibfield  {journal}
		{\bibinfo  {journal} {Phys. Rev. Lett.}\ }\textbf {\bibinfo {volume} {129}},\
		\bibinfo {pages} {120401} (\bibinfo {year} {2022})}\BibitemShut {NoStop}%
	\bibitem [{\citenamefont {Barthel}\ and\ \citenamefont
		{Zhang}(2022)}]{Barthel22}%
	\BibitemOpen
	\bibfield  {author} {\bibinfo {author} {\bibfnamefont {T.}~\bibnamefont
			{Barthel}}\ and\ \bibinfo {author} {\bibfnamefont {Y.}~\bibnamefont
			{Zhang}},\ }\bibfield  {title} {\bibinfo {title} {Solving quasi-free and
			quadratic lindblad master equations for open fermionic and bosonic systems},\
	}\href {https://doi.org/10.1088/1742-5468/ac8e5c} {\bibfield  {journal}
		{\bibinfo  {journal} {J. Stat. Mech.}\ ,\ \bibinfo {pages} {113101}}
		(\bibinfo {year} {2022})}\BibitemShut {NoStop}%
	\bibitem [{\citenamefont {He}\ and\ \citenamefont {Chien}(2022)}]{He_2022}%
	\BibitemOpen
	\bibfield  {author} {\bibinfo {author} {\bibfnamefont {Y.}~\bibnamefont
			{He}}\ and\ \bibinfo {author} {\bibfnamefont {C.-C.}\ \bibnamefont {Chien}},\
	}\bibfield  {title} {\bibinfo {title} {Topological classifications of
			quadratic bosonic excitations in closed and open systems with examples},\
	}\href {https://doi.org/10.1088/1361-648X/ac53da} {\bibfield  {journal}
		{\bibinfo  {journal} {J. Phys.: Condens. Matter}\ }\textbf {\bibinfo {volume}
			{34}},\ \bibinfo {pages} {175403} (\bibinfo {year} {2022})}\BibitemShut
	{NoStop}%
	\bibitem [{\citenamefont {Guo}\ and\ \citenamefont
		{Poletti}(2017)}]{PhysRevA.95.052107}%
	\BibitemOpen
	\bibfield  {author} {\bibinfo {author} {\bibfnamefont {C.}~\bibnamefont
			{Guo}}\ and\ \bibinfo {author} {\bibfnamefont {D.}~\bibnamefont {Poletti}},\
	}\bibfield  {title} {\bibinfo {title} {Solutions for bosonic and fermionic
			dissipative quadratic open systems},\ }\href
	{https://doi.org/10.1103/PhysRevA.95.052107} {\bibfield  {journal} {\bibinfo
			{journal} {Phys. Rev. A}\ }\textbf {\bibinfo {volume} {95}},\ \bibinfo
		{pages} {052107} (\bibinfo {year} {2017})}\BibitemShut {NoStop}%
	\bibitem [{\citenamefont {Mendoza-Arenas}\ and\ \citenamefont
		{Clark}(2024{\natexlab{a}})}]{PRXQuantum.5.010341}%
	\BibitemOpen
	\bibfield  {author} {\bibinfo {author} {\bibfnamefont {J.~J.}\ \bibnamefont
			{Mendoza-Arenas}}\ and\ \bibinfo {author} {\bibfnamefont {S.~R.}\
			\bibnamefont {Clark}},\ }\bibfield  {title} {\bibinfo {title} {Giant
			rectification in strongly interacting driven tilted systems},\ }\href
	{https://doi.org/10.1103/PRXQuantum.5.010341} {\bibfield  {journal} {\bibinfo
			{journal} {PRX Quantum}\ }\textbf {\bibinfo {volume} {5}},\ \bibinfo {pages}
		{010341} (\bibinfo {year} {2024}{\natexlab{a}})}\BibitemShut {NoStop}%
	\bibitem [{\citenamefont {Chien}\ \emph {et~al.}(2013)\citenamefont {Chien},
		\citenamefont {Gruss}, \citenamefont {Ventra},\ and\ \citenamefont
		{Zwolak}}]{chien2013interaction}%
	\BibitemOpen
	\bibfield  {author} {\bibinfo {author} {\bibfnamefont {C.-C.}\ \bibnamefont
			{Chien}}, \bibinfo {author} {\bibfnamefont {D.}~\bibnamefont {Gruss}},
		\bibinfo {author} {\bibfnamefont {M.~D.}\ \bibnamefont {Ventra}},\ and\
		\bibinfo {author} {\bibfnamefont {M.}~\bibnamefont {Zwolak}},\ }\bibfield
	{title} {\bibinfo {title} {Interaction-induced conducting–non-conducting
			transition of ultra-cold atoms in one-dimensional optical lattices},\ }\href
	{https://doi.org/10.1088/1367-2630/15/6/063026} {\bibfield  {journal}
		{\bibinfo  {journal} {New J. Phys.}\ }\textbf {\bibinfo {volume} {15}},\
		\bibinfo {pages} {063026} (\bibinfo {year} {2013})}\BibitemShut {NoStop}%
	\bibitem [{\citenamefont {Ramanan}\ \emph {et~al.}(2009)\citenamefont
		{Ramanan}, \citenamefont {Mishra}, \citenamefont {Luthra}, \citenamefont
		{Pai},\ and\ \citenamefont {Das}}]{Ramana2009signature}%
	\BibitemOpen
	\bibfield  {author} {\bibinfo {author} {\bibfnamefont {S.}~\bibnamefont
			{Ramanan}}, \bibinfo {author} {\bibfnamefont {T.}~\bibnamefont {Mishra}},
		\bibinfo {author} {\bibfnamefont {M.~S.}\ \bibnamefont {Luthra}}, \bibinfo
		{author} {\bibfnamefont {R.~V.}\ \bibnamefont {Pai}},\ and\ \bibinfo {author}
		{\bibfnamefont {B.~P.}\ \bibnamefont {Das}},\ }\bibfield  {title} {\bibinfo
		{title} {Signatures of the superfluid--to--mott-insulator transition in cold
			bosonic atoms in a one-dimensional optical lattice},\ }\href
	{https://doi.org/10.1103/PhysRevA.79.013625} {\bibfield  {journal} {\bibinfo
			{journal} {Phys. Rev. A}\ }\textbf {\bibinfo {volume} {79}},\ \bibinfo
		{pages} {013625} (\bibinfo {year} {2009})}\BibitemShut {NoStop}%
	\bibitem [{\citenamefont {Contessi}\ \emph {et~al.}(2021)\citenamefont
		{Contessi}, \citenamefont {Romito}, \citenamefont {Rizzi},\ and\
		\citenamefont {Recati}}]{contessi2021collisionless}%
	\BibitemOpen
	\bibfield  {author} {\bibinfo {author} {\bibfnamefont {D.}~\bibnamefont
			{Contessi}}, \bibinfo {author} {\bibfnamefont {D.}~\bibnamefont {Romito}},
		\bibinfo {author} {\bibfnamefont {M.}~\bibnamefont {Rizzi}},\ and\ \bibinfo
		{author} {\bibfnamefont {A.}~\bibnamefont {Recati}},\ }\bibfield  {title}
	{\bibinfo {title} {Collisionless drag for a one-dimensional two-component
			bose-hubbard model},\ }\href
	{https://doi.org/10.1103/PhysRevResearch.3.L022017} {\bibfield  {journal}
		{\bibinfo  {journal} {Phys. Rev. Res.}\ }\textbf {\bibinfo {volume} {3}},\
		\bibinfo {pages} {L022017} (\bibinfo {year} {2021})}\BibitemShut {NoStop}%
	\bibitem [{\citenamefont {{Gaude}}\ \emph {et~al.}(2022)\citenamefont
		{{Gaude}}, \citenamefont {{Das}},\ and\ \citenamefont
		{{Pai}}}]{cmf2022gaude}%
	\BibitemOpen
	\bibfield  {author} {\bibinfo {author} {\bibfnamefont {P.~P.}\ \bibnamefont
			{{Gaude}}}, \bibinfo {author} {\bibfnamefont {A.}~\bibnamefont {{Das}}},\
		and\ \bibinfo {author} {\bibfnamefont {R.~V.}\ \bibnamefont {{Pai}}},\
	}\bibfield  {title} {\bibinfo {title} {{Cluster mean field plus density
				matrix renormalization theory for the Bose Hubbard models}},\ }\href
	{https://doi.org/10.1088/1751-8121/ac71e7} {\bibfield  {journal} {\bibinfo
			{journal} {Journal of Physics A Mathematical General}\ }\textbf {\bibinfo
			{volume} {55}},\ \bibinfo {eid} {265004} (\bibinfo {year} {2022})},\ \Eprint
	{https://arxiv.org/abs/2201.01923} {arXiv:2201.01923 [cond-mat.quant-gas]}
	\BibitemShut {NoStop}%
	\bibitem [{\citenamefont {Fetter}\ and\ \citenamefont
		{Walecka}(1971)}]{fetter1971many}%
	\BibitemOpen
	\bibfield  {author} {\bibinfo {author} {\bibfnamefont {A.~L.}\ \bibnamefont
			{Fetter}}\ and\ \bibinfo {author} {\bibfnamefont {J.~D.}\ \bibnamefont
			{Walecka}},\ }\href@noop {} {\emph {\bibinfo {title} {Quantum Theory of
				Many-Particle Systems}}}\ (\bibinfo  {publisher} {McGraw-Hill},\ \bibinfo
	{address} {Boston},\ \bibinfo {year} {1971})\BibitemShut {NoStop}%
	\bibitem [{\citenamefont {Chern}\ \emph {et~al.}(2014)\citenamefont {Chern},
		\citenamefont {Chien},\ and\ \citenamefont
		{Di~Ventra}}]{chern2014dynamically}%
	\BibitemOpen
	\bibfield  {author} {\bibinfo {author} {\bibfnamefont {G.-W.}\ \bibnamefont
			{Chern}}, \bibinfo {author} {\bibfnamefont {C.-C.}\ \bibnamefont {Chien}},\
		and\ \bibinfo {author} {\bibfnamefont {M.}~\bibnamefont {Di~Ventra}},\
	}\bibfield  {title} {\bibinfo {title} {Dynamically generated flat-band phases
			in optical kagome lattices},\ }\href
	{https://doi.org/10.1103/PhysRevA.90.013609} {\bibfield  {journal} {\bibinfo
			{journal} {Phys. Rev. A}\ }\textbf {\bibinfo {volume} {90}},\ \bibinfo
		{pages} {013609} (\bibinfo {year} {2014})}\BibitemShut {NoStop}%
	\bibitem [{\citenamefont {Metcalf}\ \emph {et~al.}(2016)\citenamefont
		{Metcalf}, \citenamefont {Chern}, \citenamefont {Ventra},\ and\ \citenamefont
		{Chien}}]{metcalf2016matterwave}%
	\BibitemOpen
	\bibfield  {author} {\bibinfo {author} {\bibfnamefont {M.}~\bibnamefont
			{Metcalf}}, \bibinfo {author} {\bibfnamefont {G.-W.}\ \bibnamefont {Chern}},
		\bibinfo {author} {\bibfnamefont {M.~D.}\ \bibnamefont {Ventra}},\ and\
		\bibinfo {author} {\bibfnamefont {C.-C.}\ \bibnamefont {Chien}},\ }\bibfield
	{title} {\bibinfo {title} {Matter-wave propagation in optical lattices:
			geometrical and flat-band effects},\ }\href
	{https://doi.org/10.1088/0953-4075/49/7/075301} {\bibfield  {journal}
		{\bibinfo  {journal} {J. Phys. B: At. Mol. Opt. Phys.}\ }\textbf {\bibinfo
			{volume} {49}},\ \bibinfo {pages} {075301} (\bibinfo {year}
		{2016})}\BibitemShut {NoStop}%
	\bibitem [{\citenamefont {Parajuli}\ \emph {et~al.}(2019)\citenamefont
		{Parajuli}, \citenamefont {Pecak},\ and\ \citenamefont
		{Chien}}]{parajuli2019mass}%
	\BibitemOpen
	\bibfield  {author} {\bibinfo {author} {\bibfnamefont {B.}~\bibnamefont
			{Parajuli}}, \bibinfo {author} {\bibfnamefont {D.}~\bibnamefont {Pecak}},\
		and\ \bibinfo {author} {\bibfnamefont {C.-C.}\ \bibnamefont {Chien}},\
	}\bibfield  {title} {\bibinfo {title} {Mass-imbalance-induced structures of
			binary atomic mixtures in box potentials},\ }\href
	{https://doi.org/10.1103/PhysRevA.100.063623} {\bibfield  {journal} {\bibinfo
			{journal} {Phys. Rev. A}\ }\textbf {\bibinfo {volume} {100}},\ \bibinfo
		{pages} {063623} (\bibinfo {year} {2019})}\BibitemShut {NoStop}%
	\bibitem [{\citenamefont {Parajuli}\ \emph {et~al.}(2023)\citenamefont
		{Parajuli}, \citenamefont {Pecak},\ and\ \citenamefont
		{Chien}}]{parajuli2023atomic}%
	\BibitemOpen
	\bibfield  {author} {\bibinfo {author} {\bibfnamefont {B.}~\bibnamefont
			{Parajuli}}, \bibinfo {author} {\bibfnamefont {D.}~\bibnamefont {Pecak}},\
		and\ \bibinfo {author} {\bibfnamefont {C.-C.}\ \bibnamefont {Chien}},\
	}\bibfield  {title} {\bibinfo {title} {Atomic boson-fermion mixtures in
			one-dimensional box potentials: Few-body and mean-field many-body analyses},\
	}\href {https://doi.org/10.1103/PhysRevA.107.023308} {\bibfield  {journal}
		{\bibinfo  {journal} {Phys. Rev. A}\ }\textbf {\bibinfo {volume} {107}},\
		\bibinfo {pages} {023308} (\bibinfo {year} {2023})}\BibitemShut {NoStop}%
	\bibitem [{\citenamefont {Alon}\ \emph {et~al.}(2008)\citenamefont {Alon},
		\citenamefont {Streltsov},\ and\ \citenamefont
		{Cederbaum}}]{PhysRevA.77.033613}%
	\BibitemOpen
	\bibfield  {author} {\bibinfo {author} {\bibfnamefont {O.~E.}\ \bibnamefont
			{Alon}}, \bibinfo {author} {\bibfnamefont {A.~I.}\ \bibnamefont
			{Streltsov}},\ and\ \bibinfo {author} {\bibfnamefont {L.~S.}\ \bibnamefont
			{Cederbaum}},\ }\bibfield  {title} {\bibinfo {title} {Multiconfigurational
			time-dependent hartree method for bosons: Many-body dynamics of bosonic
			systems},\ }\href {https://doi.org/10.1103/PhysRevA.77.033613} {\bibfield
		{journal} {\bibinfo  {journal} {Phys. Rev. A}\ }\textbf {\bibinfo {volume}
			{77}},\ \bibinfo {pages} {033613} (\bibinfo {year} {2008})}\BibitemShut
	{NoStop}%
	\bibitem [{\citenamefont {Cao}\ \emph {et~al.}(2013)\citenamefont {Cao},
		\citenamefont {Krönke}, \citenamefont {Vendrell},\ and\ \citenamefont
		{Schmelcher}}]{10.1063/1.4821350}%
	\BibitemOpen
	\bibfield  {author} {\bibinfo {author} {\bibfnamefont {L.}~\bibnamefont
			{Cao}}, \bibinfo {author} {\bibfnamefont {S.}~\bibnamefont {Krönke}},
		\bibinfo {author} {\bibfnamefont {O.}~\bibnamefont {Vendrell}},\ and\
		\bibinfo {author} {\bibfnamefont {P.}~\bibnamefont {Schmelcher}},\ }\bibfield
	{title} {\bibinfo {title} {The multi-layer multi-configuration
			time-dependent hartree method for bosons: Theory, implementation, and
			applications},\ }\href {https://doi.org/10.1063/1.4821350} {\bibfield
		{journal} {\bibinfo  {journal} {The Journal of Chemical Physics}\ }\textbf
		{\bibinfo {volume} {139}},\ \bibinfo {pages} {134103} (\bibinfo {year}
		{2013})},\ \Eprint
	{https://arxiv.org/abs/https://pubs.aip.org/aip/jcp/article-pdf/doi/10.1063/1.4821350/15466111/134103\_1\_online.pdf}
	{https://pubs.aip.org/aip/jcp/article-pdf/doi/10.1063/1.4821350/15466111/134103\_1\_online.pdf}
	\BibitemShut {NoStop}%
	\bibitem [{\citenamefont {Lode}(2016)}]{PhysRevA.93.063601}%
	\BibitemOpen
	\bibfield  {author} {\bibinfo {author} {\bibfnamefont {A.~U.~J.}\
			\bibnamefont {Lode}},\ }\bibfield  {title} {\bibinfo {title}
		{Multiconfigurational time-dependent hartree method for bosons with internal
			degrees of freedom: Theory and composite fragmentation of multicomponent
			bose-einstein condensates},\ }\href
	{https://doi.org/10.1103/PhysRevA.93.063601} {\bibfield  {journal} {\bibinfo
			{journal} {Phys. Rev. A}\ }\textbf {\bibinfo {volume} {93}},\ \bibinfo
		{pages} {063601} (\bibinfo {year} {2016})}\BibitemShut {NoStop}%
	\bibitem [{\citenamefont {Alon}\ \emph {et~al.}(2016)\citenamefont {Alon},
		\citenamefont {Beinke}, \citenamefont {Cederbaum}, \citenamefont {Edmonds},
		\citenamefont {Fasshauer}, \citenamefont {Kasevich}, \citenamefont {Klaiman},
		\citenamefont {Lode}, \citenamefont {Parker}, \citenamefont {Sakmann},
		\citenamefont {Tsatsos},\ and\ \citenamefont
		{Streltsov}}]{10.1007/978-3-319-47066-5_6}%
	\BibitemOpen
	\bibfield  {author} {\bibinfo {author} {\bibfnamefont {O.~E.}\ \bibnamefont
			{Alon}}, \bibinfo {author} {\bibfnamefont {R.}~\bibnamefont {Beinke}},
		\bibinfo {author} {\bibfnamefont {L.~S.}\ \bibnamefont {Cederbaum}}, \bibinfo
		{author} {\bibfnamefont {M.~J.}\ \bibnamefont {Edmonds}}, \bibinfo {author}
		{\bibfnamefont {E.}~\bibnamefont {Fasshauer}}, \bibinfo {author}
		{\bibfnamefont {M.~A.}\ \bibnamefont {Kasevich}}, \bibinfo {author}
		{\bibfnamefont {S.}~\bibnamefont {Klaiman}}, \bibinfo {author} {\bibfnamefont
			{A.~U.~J.}\ \bibnamefont {Lode}}, \bibinfo {author} {\bibfnamefont {N.~G.}\
			\bibnamefont {Parker}}, \bibinfo {author} {\bibfnamefont {K.}~\bibnamefont
			{Sakmann}}, \bibinfo {author} {\bibfnamefont {M.~C.}\ \bibnamefont
			{Tsatsos}},\ and\ \bibinfo {author} {\bibfnamefont {A.~I.}\ \bibnamefont
			{Streltsov}},\ }\bibfield  {title} {\bibinfo {title} {Vorticity, variance,
			and the vigor of many-body phenomena in ultracold quantum systems: Mctdhb and
			mctdh-x},\ }in\ \href@noop {} {\emph {\bibinfo {booktitle} {High Performance
				Computing in Science and Engineering {\textasciiacute}16}}},\ \bibinfo
	{editor} {edited by\ \bibinfo {editor} {\bibfnamefont {W.~E.}\ \bibnamefont
			{Nagel}}, \bibinfo {editor} {\bibfnamefont {D.~H.}\ \bibnamefont
			{Kr{\"o}ner}},\ and\ \bibinfo {editor} {\bibfnamefont {M.~M.}\ \bibnamefont
			{Resch}}}\ (\bibinfo  {publisher} {Springer International Publishing},\
	\bibinfo {address} {Cham},\ \bibinfo {year} {2016})\ pp.\ \bibinfo {pages}
	{79--96}\BibitemShut {NoStop}%
	\bibitem [{\citenamefont {Lode}\ \emph {et~al.}(2021)\citenamefont {Lode},
		\citenamefont {Dutta},\ and\ \citenamefont {Lévêque}}]{e23040392}%
	\BibitemOpen
	\bibfield  {author} {\bibinfo {author} {\bibfnamefont {A.~U.~J.}\
			\bibnamefont {Lode}}, \bibinfo {author} {\bibfnamefont {S.}~\bibnamefont
			{Dutta}},\ and\ \bibinfo {author} {\bibfnamefont {C.}~\bibnamefont
			{Lévêque}},\ }\bibfield  {title} {\bibinfo {title} {Dynamics of ultracold
			bosons in artificial gauge fields—angular momentum, fragmentation, and the
			variance of entropy},\ }\bibfield  {journal} {\bibinfo  {journal} {Entropy}\
	}\textbf {\bibinfo {volume} {23}},\ \href {https://doi.org/10.3390/e23040392}
	{10.3390/e23040392} (\bibinfo {year} {2021})\BibitemShut {NoStop}%
	\bibitem [{\citenamefont {Molignini}\ \emph {et~al.}(2024)\citenamefont
		{Molignini}, \citenamefont {Dutta},\ and\ \citenamefont
		{Fasshauer}}]{Molignini24}%
	\BibitemOpen
	\bibfield  {author} {\bibinfo {author} {\bibfnamefont {P.}~\bibnamefont
			{Molignini}}, \bibinfo {author} {\bibfnamefont {S.}~\bibnamefont {Dutta}},\
		and\ \bibinfo {author} {\bibfnamefont {E.}~\bibnamefont {Fasshauer}},\
	}\href@noop {} {\bibinfo {title} {Lecture notes: many-body quantum dynamics
			with mctdh-x}} (\bibinfo {year} {2024}),\ \bibinfo {note}
	{arxiv:2407.20317}\BibitemShut {NoStop}%
	\bibitem [{\citenamefont {Ivanov}\ \emph {et~al.}(2013)\citenamefont {Ivanov},
		\citenamefont {Kordas}, \citenamefont {Komnik},\ and\ \citenamefont
		{Wimberger}}]{anton2013bosonic}%
	\BibitemOpen
	\bibfield  {author} {\bibinfo {author} {\bibfnamefont {A.}~\bibnamefont
			{Ivanov}}, \bibinfo {author} {\bibfnamefont {G.}~\bibnamefont {Kordas}},
		\bibinfo {author} {\bibfnamefont {A.}~\bibnamefont {Komnik}},\ and\ \bibinfo
		{author} {\bibfnamefont {S.}~\bibnamefont {Wimberger}},\ }\bibfield  {title}
	{\bibinfo {title} {Bosonic transport through a chain of quantum dots},\
	}\href {https://doi.org/10.1140/epjb/e2013-40417-4} {\bibfield  {journal}
		{\bibinfo  {journal} {Eur. Phys. J. B}\ }\textbf {\bibinfo {volume} {86}},\
		\bibinfo {pages} {345} (\bibinfo {year} {2013})}\BibitemShut {NoStop}%
	\bibitem [{\citenamefont {Kordas}\ \emph {et~al.}(2015)\citenamefont {Kordas},
		\citenamefont {Witthaut},\ and\ \citenamefont
		{Wimberger}}]{kordas2015nonequilibrium}%
	\BibitemOpen
	\bibfield  {author} {\bibinfo {author} {\bibfnamefont {G.}~\bibnamefont
			{Kordas}}, \bibinfo {author} {\bibfnamefont {D.}~\bibnamefont {Witthaut}},\
		and\ \bibinfo {author} {\bibfnamefont {S.}~\bibnamefont {Wimberger}},\
	}\bibfield  {title} {\bibinfo {title} {Non-equilibrium dynamics in
			dissipative bose-hubbard chains},\ }\href
	{https://doi.org/https://doi.org/10.1002/andp.201400189} {\bibfield
		{journal} {\bibinfo  {journal} {Ann. Phys. (Berlin)}\ }\textbf {\bibinfo
			{volume} {527}},\ \bibinfo {pages} {619} (\bibinfo {year}
		{2015})}\BibitemShut {NoStop}%
	\bibitem [{\citenamefont {Lee}(1995)}]{LEE1995147}%
	\BibitemOpen
	\bibfield  {author} {\bibinfo {author} {\bibfnamefont {H.-W.}\ \bibnamefont
			{Lee}},\ }\bibfield  {title} {\bibinfo {title} {Theory and application of the
			quantum phase-space distribution functions},\ }\href
	{https://doi.org/https://doi.org/10.1016/0370-1573(95)00007-4} {\bibfield
		{journal} {\bibinfo  {journal} {Physics Reports}\ }\textbf {\bibinfo {volume}
			{259}},\ \bibinfo {pages} {147} (\bibinfo {year} {1995})}\BibitemShut
	{NoStop}%
	\bibitem [{\citenamefont {Diósi}\ and\ \citenamefont
		{Kiefer}(2002)}]{LajosDiosi_2002}%
	\BibitemOpen
	\bibfield  {author} {\bibinfo {author} {\bibfnamefont {L.}~\bibnamefont
			{Diósi}}\ and\ \bibinfo {author} {\bibfnamefont {C.}~\bibnamefont
			{Kiefer}},\ }\bibfield  {title} {\bibinfo {title} {Exact positivity of the
			wigner and p-functions of a markovian open system},\ }\href
	{https://doi.org/10.1088/0305-4470/35/11/312} {\bibfield  {journal} {\bibinfo
			{journal} {Journal of Physics A: Mathematical and General}\ }\textbf
		{\bibinfo {volume} {35}},\ \bibinfo {pages} {2675} (\bibinfo {year}
		{2002})}\BibitemShut {NoStop}%
	\bibitem [{\citenamefont {Polkovnikov}(2010)}]{POLKOVNIKOV20101790}%
	\BibitemOpen
	\bibfield  {author} {\bibinfo {author} {\bibfnamefont {A.}~\bibnamefont
			{Polkovnikov}},\ }\bibfield  {title} {\bibinfo {title} {Phase space
			representation of quantum dynamics},\ }\href
	{https://doi.org/https://doi.org/10.1016/j.aop.2010.02.006} {\bibfield
		{journal} {\bibinfo  {journal} {Annals of Physics}\ }\textbf {\bibinfo
			{volume} {325}},\ \bibinfo {pages} {1790} (\bibinfo {year}
		{2010})}\BibitemShut {NoStop}%
	\bibitem [{\citenamefont {Weinbub}\ and\ \citenamefont
		{Ferry}(2018)}]{10.1063/1.5046663}%
	\BibitemOpen
	\bibfield  {author} {\bibinfo {author} {\bibfnamefont {J.}~\bibnamefont
			{Weinbub}}\ and\ \bibinfo {author} {\bibfnamefont {D.~K.}\ \bibnamefont
			{Ferry}},\ }\bibfield  {title} {\bibinfo {title} {Recent advances in wigner
			function approaches},\ }\href {https://doi.org/10.1063/1.5046663} {\bibfield
		{journal} {\bibinfo  {journal} {Applied Physics Reviews}\ }\textbf {\bibinfo
			{volume} {5}},\ \bibinfo {pages} {041104} (\bibinfo {year} {2018})},\ \Eprint
	{https://arxiv.org/abs/https://pubs.aip.org/aip/apr/article-pdf/doi/10.1063/1.5046663/19740878/041104\_1\_online.pdf}
	{https://pubs.aip.org/aip/apr/article-pdf/doi/10.1063/1.5046663/19740878/041104\_1\_online.pdf}
	\BibitemShut {NoStop}%
	\bibitem [{\citenamefont {Rundle}\ and\ \citenamefont
		{Everitt}(2021)}]{10.1002/qute.202100016}%
	\BibitemOpen
	\bibfield  {author} {\bibinfo {author} {\bibfnamefont {R.~P.}\ \bibnamefont
			{Rundle}}\ and\ \bibinfo {author} {\bibfnamefont {M.~J.}\ \bibnamefont
			{Everitt}},\ }\bibfield  {title} {\bibinfo {title} {Overview of the phase
			space formulation of quantum mechanics with application to quantum
			technologies},\ }\href
	{https://doi.org/https://doi.org/10.1002/qute.202100016} {\bibfield
		{journal} {\bibinfo  {journal} {Advanced Quantum Technologies}\ }\textbf
		{\bibinfo {volume} {4}},\ \bibinfo {pages} {2100016} (\bibinfo {year}
		{2021})},\ \Eprint
	{https://arxiv.org/abs/https://onlinelibrary.wiley.com/doi/pdf/10.1002/qute.202100016}
	{https://onlinelibrary.wiley.com/doi/pdf/10.1002/qute.202100016} \BibitemShut
	{NoStop}%
	\bibitem [{\citenamefont {Su}\ \emph {et~al.}(1979)\citenamefont {Su},
		\citenamefont {Schrieffer},\ and\ \citenamefont {Heeger}}]{SSH}%
	\BibitemOpen
	\bibfield  {author} {\bibinfo {author} {\bibfnamefont {W.~P.}\ \bibnamefont
			{Su}}, \bibinfo {author} {\bibfnamefont {J.~R.}\ \bibnamefont {Schrieffer}},\
		and\ \bibinfo {author} {\bibfnamefont {A.~J.}\ \bibnamefont {Heeger}},\
	}\bibfield  {title} {\bibinfo {title} {Solitons in polyacetylene},\ }\href
	{https://doi.org/10.1103/PhysRevLett.42.1698} {\bibfield  {journal} {\bibinfo
			{journal} {Phys. Rev. Lett.}\ }\textbf {\bibinfo {volume} {42}},\ \bibinfo
		{pages} {1698} (\bibinfo {year} {1979})}\BibitemShut {NoStop}%
	\bibitem [{\citenamefont {Asb{\'{o}}th}\ \emph {et~al.}(2016)\citenamefont
		{Asb{\'{o}}th}, \citenamefont {Oroszl{\'{a}}ny},\ and\ \citenamefont
		{P{\'{a}}lyi}}]{Asboth2016}%
	\BibitemOpen
	\bibfield  {author} {\bibinfo {author} {\bibfnamefont {J.~K.}\ \bibnamefont
			{Asb{\'{o}}th}}, \bibinfo {author} {\bibfnamefont {L.}~\bibnamefont
			{Oroszl{\'{a}}ny}},\ and\ \bibinfo {author} {\bibfnamefont {A.}~\bibnamefont
			{P{\'{a}}lyi}},\ }\href@noop {} {\emph {\bibinfo {title} {A Short Course on
				Topological Insulators: Band-structure topology and edge states in one and
				two dimensions}}}\ (\bibinfo  {publisher} {Springer},\ \bibinfo {address}
	{Berlin, Germany},\ \bibinfo {year} {2016})\BibitemShut {NoStop}%
	\bibitem [{\citenamefont {Di~Liberto}\ \emph {et~al.}(2017)\citenamefont
		{Di~Liberto}, \citenamefont {Recati}, \citenamefont {Carusotto},\ and\
		\citenamefont {Menotti}}]{diliberto2017twobound}%
	\BibitemOpen
	\bibfield  {author} {\bibinfo {author} {\bibfnamefont {M.}~\bibnamefont
			{Di~Liberto}}, \bibinfo {author} {\bibfnamefont {A.}~\bibnamefont {Recati}},
		\bibinfo {author} {\bibfnamefont {I.}~\bibnamefont {Carusotto}},\ and\
		\bibinfo {author} {\bibfnamefont {C.}~\bibnamefont {Menotti}},\ }\bibfield
	{title} {\bibinfo {title} {Two-body bound and edge states in the extended ssh
			bose-hubbard model},\ }\href {https://doi.org/10.1140/epjst/e2016-60388-y}
	{\bibfield  {journal} {\bibinfo  {journal} {Eur. Phys. J. Spec. Top.}\
		}\textbf {\bibinfo {volume} {226}},\ \bibinfo {pages} {2751} (\bibinfo {year}
		{2017})}\BibitemShut {NoStop}%
	\bibitem [{\citenamefont {Azcona}\ and\ \citenamefont
		{Downing}(2021)}]{azcona2021doublons}%
	\BibitemOpen
	\bibfield  {author} {\bibinfo {author} {\bibfnamefont {P.~M.}\ \bibnamefont
			{Azcona}}\ and\ \bibinfo {author} {\bibfnamefont {C.~A.}\ \bibnamefont
			{Downing}},\ }\bibfield  {title} {\bibinfo {title} {Doublons, topology and
			interactions in a one-dimensional lattice},\ }\href
	{https://doi.org/10.1038/s41598-021-91778-z} {\bibfield  {journal} {\bibinfo
			{journal} {Sci. Rep.}\ }\textbf {\bibinfo {volume} {11}},\ \bibinfo {pages}
		{12540} (\bibinfo {year} {2021})}\BibitemShut {NoStop}%
	\bibitem [{\citenamefont {Le}\ \emph {et~al.}(2020)\citenamefont {Le},
		\citenamefont {Fisher}, \citenamefont {Curson},\ and\ \citenamefont
		{Ginossar}}]{Le2020}%
	\BibitemOpen
	\bibfield  {author} {\bibinfo {author} {\bibfnamefont {N.~H.}\ \bibnamefont
			{Le}}, \bibinfo {author} {\bibfnamefont {A.~J.}\ \bibnamefont {Fisher}},
		\bibinfo {author} {\bibfnamefont {N.~J.}\ \bibnamefont {Curson}},\ and\
		\bibinfo {author} {\bibfnamefont {E.}~\bibnamefont {Ginossar}},\ }\bibfield
	{title} {\bibinfo {title} {Topological phases of a dimerized fermi--hubbard
			model for semiconductor nano-lattices},\ }\href
	{https://doi.org/10.1038/s41534-020-0253-9} {\bibfield  {journal} {\bibinfo
			{journal} {npj Quantum Inf.}\ }\textbf {\bibinfo {volume} {6}},\ \bibinfo
		{pages} {24} (\bibinfo {year} {2020})}\BibitemShut {NoStop}%
	\bibitem [{\citenamefont {Feng}\ \emph {et~al.}(2022)\citenamefont {Feng},
		\citenamefont {Xing}, \citenamefont {Poletti}, \citenamefont {Scalettar},\
		and\ \citenamefont {Batrouni}}]{PhysRevB.106.L081114}%
	\BibitemOpen
	\bibfield  {author} {\bibinfo {author} {\bibfnamefont {C.}~\bibnamefont
			{Feng}}, \bibinfo {author} {\bibfnamefont {B.}~\bibnamefont {Xing}}, \bibinfo
		{author} {\bibfnamefont {D.}~\bibnamefont {Poletti}}, \bibinfo {author}
		{\bibfnamefont {R.}~\bibnamefont {Scalettar}},\ and\ \bibinfo {author}
		{\bibfnamefont {G.}~\bibnamefont {Batrouni}},\ }\bibfield  {title} {\bibinfo
		{title} {Phase diagram of the su-schrieffer-heeger-hubbard model on a square
			lattice},\ }\href {https://doi.org/10.1103/PhysRevB.106.L081114} {\bibfield
		{journal} {\bibinfo  {journal} {Phys. Rev. B}\ }\textbf {\bibinfo {volume}
			{106}},\ \bibinfo {pages} {L081114} (\bibinfo {year} {2022})}\BibitemShut
	{NoStop}%
	\bibitem [{\citenamefont {Wichterich}\ \emph {et~al.}(2007)\citenamefont
		{Wichterich}, \citenamefont {Henrich}, \citenamefont {Breuer}, \citenamefont
		{Gemmer},\ and\ \citenamefont {Michel}}]{wichterich2007modeling}%
	\BibitemOpen
	\bibfield  {author} {\bibinfo {author} {\bibfnamefont {H.}~\bibnamefont
			{Wichterich}}, \bibinfo {author} {\bibfnamefont {M.~J.}\ \bibnamefont
			{Henrich}}, \bibinfo {author} {\bibfnamefont {H.-P.}\ \bibnamefont {Breuer}},
		\bibinfo {author} {\bibfnamefont {J.}~\bibnamefont {Gemmer}},\ and\ \bibinfo
		{author} {\bibfnamefont {M.}~\bibnamefont {Michel}},\ }\bibfield  {title}
	{\bibinfo {title} {Modeling heat transport through completely positive
			maps},\ }\href@noop {} {\bibfield  {journal} {\bibinfo  {journal} {Phys. Rev.
				E}\ }\textbf {\bibinfo {volume} {76}},\ \bibinfo {pages} {031115} (\bibinfo
		{year} {2007})}\BibitemShut {NoStop}%
	\bibitem [{\citenamefont {Purkayastha}\ \emph {et~al.}(2016)\citenamefont
		{Purkayastha}, \citenamefont {Dhar},\ and\ \citenamefont
		{Kulkarni}}]{purkayastha2016out}%
	\BibitemOpen
	\bibfield  {author} {\bibinfo {author} {\bibfnamefont {A.}~\bibnamefont
			{Purkayastha}}, \bibinfo {author} {\bibfnamefont {A.}~\bibnamefont {Dhar}},\
		and\ \bibinfo {author} {\bibfnamefont {M.}~\bibnamefont {Kulkarni}},\
	}\bibfield  {title} {\bibinfo {title} {Out-of-equilibrium open quantum
			systems: A comparison of approximate quantum master equation approaches with
			exact results},\ }\href@noop {} {\bibfield  {journal} {\bibinfo  {journal}
			{Phys. Rev. A}\ }\textbf {\bibinfo {volume} {93}},\ \bibinfo {pages} {062114}
		(\bibinfo {year} {2016})}\BibitemShut {NoStop}%
	\bibitem [{\citenamefont {Lidar}(2019)}]{lidar2019lecture}%
	\BibitemOpen
	\bibfield  {author} {\bibinfo {author} {\bibfnamefont {D.~A.}\ \bibnamefont
			{Lidar}},\ }\bibfield  {title} {\bibinfo {title} {Lecture notes on the theory
			of open quantum systems},\ }\href@noop {} {\bibfield  {journal} {\bibinfo
			{journal} {arXiv preprint arXiv:1902.00967}\ } (\bibinfo {year}
		{2019})}\BibitemShut {NoStop}%
	\bibitem [{\citenamefont {Lai}\ \emph {et~al.}(2018)\citenamefont {Lai},
		\citenamefont {Di~Ventra}, \citenamefont {Scheibner},\ and\ \citenamefont
		{Chien}}]{lai2018tunable}%
	\BibitemOpen
	\bibfield  {author} {\bibinfo {author} {\bibfnamefont {C.-Y.}\ \bibnamefont
			{Lai}}, \bibinfo {author} {\bibfnamefont {M.}~\bibnamefont {Di~Ventra}},
		\bibinfo {author} {\bibfnamefont {M.}~\bibnamefont {Scheibner}},\ and\
		\bibinfo {author} {\bibfnamefont {C.-C.}\ \bibnamefont {Chien}},\ }\bibfield
	{title} {\bibinfo {title} {Tunable current circulation in triangular
			quantum-dot metastructures},\ }\href@noop {} {\bibfield  {journal} {\bibinfo
			{journal} {Europhys. Lett.}\ }\textbf {\bibinfo {volume} {123}},\ \bibinfo
		{pages} {47002} (\bibinfo {year} {2018})}\BibitemShut {NoStop}%
	\bibitem [{\citenamefont {Dugar}\ and\ \citenamefont
		{Chien}(2022)}]{palak2022geometry}%
	\BibitemOpen
	\bibfield  {author} {\bibinfo {author} {\bibfnamefont {P.}~\bibnamefont
			{Dugar}}\ and\ \bibinfo {author} {\bibfnamefont {C.-C.}\ \bibnamefont
			{Chien}},\ }\bibfield  {title} {\bibinfo {title} {Geometry-based circulation
			of local thermal current in quantum harmonic and bose-hubbard systems},\
	}\href {https://doi.org/10.1103/PhysRevE.105.064111} {\bibfield  {journal}
		{\bibinfo  {journal} {Phys. Rev. E}\ }\textbf {\bibinfo {volume} {105}},\
		\bibinfo {pages} {064111} (\bibinfo {year} {2022})}\BibitemShut {NoStop}%
	\bibitem [{\citenamefont {Johansson}\ \emph {et~al.}(2013)\citenamefont
		{Johansson}, \citenamefont {Nation},\ and\ \citenamefont
		{Nori}}]{johansson2013qutip}%
	\BibitemOpen
	\bibfield  {author} {\bibinfo {author} {\bibfnamefont {J.}~\bibnamefont
			{Johansson}}, \bibinfo {author} {\bibfnamefont {P.}~\bibnamefont {Nation}},\
		and\ \bibinfo {author} {\bibfnamefont {F.}~\bibnamefont {Nori}},\ }\bibfield
	{title} {\bibinfo {title} {Qutip 2: A python framework for the dynamics of
			open quantum systems},\ }\href
	{https://doi.org/https://doi.org/10.1016/j.cpc.2012.11.019} {\bibfield
		{journal} {\bibinfo  {journal} {Comput. Phys. Commun.}\ }\textbf {\bibinfo
			{volume} {184}},\ \bibinfo {pages} {1234} (\bibinfo {year}
		{2013})}\BibitemShut {NoStop}%
	\bibitem [{\citenamefont {Dugar}\ \emph {et~al.}(2020)\citenamefont {Dugar},
		\citenamefont {Scheibner},\ and\ \citenamefont {Chien}}]{palak2020geometry}%
	\BibitemOpen
	\bibfield  {author} {\bibinfo {author} {\bibfnamefont {P.}~\bibnamefont
			{Dugar}}, \bibinfo {author} {\bibfnamefont {M.}~\bibnamefont {Scheibner}},\
		and\ \bibinfo {author} {\bibfnamefont {C.-C.}\ \bibnamefont {Chien}},\
	}\bibfield  {title} {\bibinfo {title} {Geometry-based circulation of local
			photonic transport in a triangular metastructure},\ }\href
	{https://doi.org/10.1103/PhysRevA.102.023704} {\bibfield  {journal} {\bibinfo
			{journal} {Phys. Rev. A}\ }\textbf {\bibinfo {volume} {102}},\ \bibinfo
		{pages} {023704} (\bibinfo {year} {2020})}\BibitemShut {NoStop}%
	\bibitem [{\citenamefont {Pi{\v{z}}orn}(2013)}]{pivzorn2013one}%
	\BibitemOpen
	\bibfield  {author} {\bibinfo {author} {\bibfnamefont {I.}~\bibnamefont
			{Pi{\v{z}}orn}},\ }\bibfield  {title} {\bibinfo {title} {One-dimensional
			bose-hubbard model far from equilibrium},\ }\href@noop {} {\bibfield
		{journal} {\bibinfo  {journal} {Phys. Rev. A}\ }\textbf {\bibinfo {volume}
			{88}},\ \bibinfo {pages} {043635} (\bibinfo {year} {2013})}\BibitemShut
	{NoStop}%
	\bibitem [{\citenamefont {de~Gennes}(1966)}]{gennes1966superconductivity}%
	\BibitemOpen
	\bibfield  {author} {\bibinfo {author} {\bibfnamefont {P.~G.}\ \bibnamefont
			{de~Gennes}},\ }\href@noop {} {\emph {\bibinfo {title} {Superconductivity of
				Metals and Alloys}}}\ (\bibinfo  {publisher} {Benjamin},\ \bibinfo {address}
	{New York},\ \bibinfo {year} {1966})\BibitemShut {NoStop}%
	\bibitem [{\citenamefont {Zhu}(2016)}]{zhu2016bogoliubov}%
	\BibitemOpen
	\bibfield  {author} {\bibinfo {author} {\bibfnamefont {J.~X.}\ \bibnamefont
			{Zhu}},\ }\href {https://books.google.com/books?id=Tep6DAAAQBAJ} {\emph
		{\bibinfo {title} {Bogoliubov-de Gennes Method and Its Applications}}},\
	Lecture Notes in Physics\ (\bibinfo  {publisher} {Springer International
		Publishing},\ \bibinfo {year} {2016})\BibitemShut {NoStop}%
	\bibitem [{\citenamefont {Polo}\ \emph {et~al.}(2024)\citenamefont {Polo},
		\citenamefont {Chetcuti}, \citenamefont {Domanti}, \citenamefont {Kitson},
		\citenamefont {Osterloh}, \citenamefont {Perciavalle}, \citenamefont
		{Singh},\ and\ \citenamefont {Amico}}]{Polo_2024}%
	\BibitemOpen
	\bibfield  {author} {\bibinfo {author} {\bibfnamefont {J.}~\bibnamefont
			{Polo}}, \bibinfo {author} {\bibfnamefont {W.~J.}\ \bibnamefont {Chetcuti}},
		\bibinfo {author} {\bibfnamefont {E.~C.}\ \bibnamefont {Domanti}}, \bibinfo
		{author} {\bibfnamefont {P.}~\bibnamefont {Kitson}}, \bibinfo {author}
		{\bibfnamefont {A.}~\bibnamefont {Osterloh}}, \bibinfo {author}
		{\bibfnamefont {F.}~\bibnamefont {Perciavalle}}, \bibinfo {author}
		{\bibfnamefont {V.~P.}\ \bibnamefont {Singh}},\ and\ \bibinfo {author}
		{\bibfnamefont {L.}~\bibnamefont {Amico}},\ }\bibfield  {title} {\bibinfo
		{title} {Perspective on new implementations of atomtronic circuits},\ }\href
	{https://doi.org/10.1088/2058-9565/ad48b2} {\bibfield  {journal} {\bibinfo
			{journal} {Quantum Sci. Technol.}\ }\textbf {\bibinfo {volume} {9}},\
		\bibinfo {pages} {030501} (\bibinfo {year} {2024})}\BibitemShut {NoStop}%
	\bibitem [{\citenamefont {Mendoza-Arenas}\ and\ \citenamefont
		{Clark}(2024{\natexlab{b}})}]{mendoza2024giant}%
	\BibitemOpen
	\bibfield  {author} {\bibinfo {author} {\bibfnamefont {J.~J.}\ \bibnamefont
			{Mendoza-Arenas}}\ and\ \bibinfo {author} {\bibfnamefont {S.~R.}\
			\bibnamefont {Clark}},\ }\bibfield  {title} {\bibinfo {title} {Giant
			rectification in strongly interacting driven tilted systems},\ }\href
	{https://doi.org/10.1103/PRXQuantum.5.010341} {\bibfield  {journal} {\bibinfo
			{journal} {PRX Quantum}\ }\textbf {\bibinfo {volume} {5}},\ \bibinfo {pages}
		{010341} (\bibinfo {year} {2024}{\natexlab{b}})}\BibitemShut {NoStop}%
	\bibitem [{\citenamefont {Capozzi}\ \emph {et~al.}(2015)\citenamefont
		{Capozzi}, \citenamefont {Xia}, \citenamefont {Adak}, \citenamefont {Dell},
		\citenamefont {Liu}, \citenamefont {Taylor}, \citenamefont {Neaton},
		\citenamefont {Campos},\ and\ \citenamefont
		{Venkataraman}}]{capozzi2015single}%
	\BibitemOpen
	\bibfield  {author} {\bibinfo {author} {\bibfnamefont {B.}~\bibnamefont
			{Capozzi}}, \bibinfo {author} {\bibfnamefont {J.}~\bibnamefont {Xia}},
		\bibinfo {author} {\bibfnamefont {O.}~\bibnamefont {Adak}}, \bibinfo {author}
		{\bibfnamefont {E.~J.}\ \bibnamefont {Dell}}, \bibinfo {author}
		{\bibfnamefont {Z.-F.}\ \bibnamefont {Liu}}, \bibinfo {author} {\bibfnamefont
			{J.~C.}\ \bibnamefont {Taylor}}, \bibinfo {author} {\bibfnamefont {J.~B.}\
			\bibnamefont {Neaton}}, \bibinfo {author} {\bibfnamefont {L.~M.}\
			\bibnamefont {Campos}},\ and\ \bibinfo {author} {\bibfnamefont
			{L.}~\bibnamefont {Venkataraman}},\ }\bibfield  {title} {\bibinfo {title}
		{Single-molecule diodes with high rectification ratios through environmental
			control},\ }\href {https://doi.org/10.1038/nnano.2015.97} {\bibfield
		{journal} {\bibinfo  {journal} {Nat. Nanotech.}\ }\textbf {\bibinfo {volume}
			{10}},\ \bibinfo {pages} {522} (\bibinfo {year} {2015})}\BibitemShut
	{NoStop}%
	\bibitem [{\citenamefont {Kornilovitch}\ \emph {et~al.}(2002)\citenamefont
		{Kornilovitch}, \citenamefont {Bratkovsky},\ and\ \citenamefont
		{Williams}}]{kornilovitch2002current}%
	\BibitemOpen
	\bibfield  {author} {\bibinfo {author} {\bibfnamefont {P.~E.}\ \bibnamefont
			{Kornilovitch}}, \bibinfo {author} {\bibfnamefont {A.~M.}\ \bibnamefont
			{Bratkovsky}},\ and\ \bibinfo {author} {\bibfnamefont {R.~S.}\ \bibnamefont
			{Williams}},\ }\bibfield  {title} {\bibinfo {title} {Current rectification by
			molecules with asymmetric tunneling barriers},\ }\href
	{https://doi.org/10.1103/PhysRevB.66.165436} {\bibfield  {journal} {\bibinfo
			{journal} {Phys. Rev. B}\ }\textbf {\bibinfo {volume} {66}},\ \bibinfo
		{pages} {165436} (\bibinfo {year} {2002})}\BibitemShut {NoStop}%
	\bibitem [{\citenamefont {He}\ and\ \citenamefont
		{Chien}(2023)}]{he2023particle}%
	\BibitemOpen
	\bibfield  {author} {\bibinfo {author} {\bibfnamefont {Y.}~\bibnamefont
			{He}}\ and\ \bibinfo {author} {\bibfnamefont {C.-C.}\ \bibnamefont {Chien}},\
	}\bibfield  {title} {\bibinfo {title} {Particle and thermal transport through
			one dimensional topological systems via lindblad formalism},\ }\href
	{https://doi.org/https://doi.org/10.1016/j.physleta.2023.128826} {\bibfield
		{journal} {\bibinfo  {journal} {Phys. Lett. A}\ }\textbf {\bibinfo {volume}
			{473}},\ \bibinfo {pages} {128826} (\bibinfo {year} {2023})}\BibitemShut
	{NoStop}%
	\bibitem [{\citenamefont {Bychek}\ \emph {et~al.}(2020)\citenamefont {Bychek},
		\citenamefont {Muraev}, \citenamefont {Maksimov},\ and\ \citenamefont
		{Kolovsky}}]{bychek2020open}%
	\BibitemOpen
	\bibfield  {author} {\bibinfo {author} {\bibfnamefont {A.~A.}\ \bibnamefont
			{Bychek}}, \bibinfo {author} {\bibfnamefont {P.~S.}\ \bibnamefont {Muraev}},
		\bibinfo {author} {\bibfnamefont {D.~N.}\ \bibnamefont {Maksimov}},\ and\
		\bibinfo {author} {\bibfnamefont {A.~R.}\ \bibnamefont {Kolovsky}},\
	}\bibfield  {title} {\bibinfo {title} {Open bose-hubbard chain:
			Pseudoclassical approach},\ }\href
	{https://doi.org/10.1103/PhysRevE.101.012208} {\bibfield  {journal} {\bibinfo
			{journal} {Phys. Rev. E}\ }\textbf {\bibinfo {volume} {101}},\ \bibinfo
		{pages} {012208} (\bibinfo {year} {2020})}\BibitemShut {NoStop}%
	\bibitem [{\citenamefont {Manthe}(2017)}]{Manthe17}%
	\BibitemOpen
	\bibfield  {author} {\bibinfo {author} {\bibfnamefont {U.}~\bibnamefont
			{Manthe}},\ }\bibfield  {title} {\bibinfo {title} {Wavepacket dynamics and
			the multi-configurational time-dependent hartree approach},\ }\href
	{https://doi.org/10.1088/1361-648X/aa6e96} {\bibfield  {journal} {\bibinfo
			{journal} {J. Phys.: Condens. Matter}\ }\textbf {\bibinfo {volume} {29}},\
		\bibinfo {pages} {253001} (\bibinfo {year} {2017})}\BibitemShut {NoStop}%
	\bibitem [{\citenamefont {Roy}\ \emph {et~al.}(2023)\citenamefont {Roy},
		\citenamefont {Chakrabarti},\ and\ \citenamefont
		{Gammal}}]{10.21468/SciPostPhysCore.6.4.073}%
	\BibitemOpen
	\bibfield  {author} {\bibinfo {author} {\bibfnamefont {R.}~\bibnamefont
			{Roy}}, \bibinfo {author} {\bibfnamefont {B.}~\bibnamefont {Chakrabarti}},\
		and\ \bibinfo {author} {\bibfnamefont {A.}~\bibnamefont {Gammal}},\
	}\bibfield  {title} {\bibinfo {title} {{Out of equilibrium many-body
				expansion dynamics of strongly interacting bosons}},\ }\href
	{https://doi.org/10.21468/SciPostPhysCore.6.4.073} {\bibfield  {journal}
		{\bibinfo  {journal} {SciPost Phys. Core}\ }\textbf {\bibinfo {volume} {6}},\
		\bibinfo {pages} {073} (\bibinfo {year} {2023})}\BibitemShut {NoStop}%
	\bibitem [{\citenamefont {Ke\ss{}ler}\ and\ \citenamefont
		{Marquardt}(2014)}]{kessler2014single}%
	\BibitemOpen
	\bibfield  {author} {\bibinfo {author} {\bibfnamefont {S.}~\bibnamefont
			{Ke\ss{}ler}}\ and\ \bibinfo {author} {\bibfnamefont {F.}~\bibnamefont
			{Marquardt}},\ }\bibfield  {title} {\bibinfo {title} {Single-site-resolved
			measurement of the current statistics in optical lattices},\ }\href
	{https://doi.org/10.1103/PhysRevA.89.061601} {\bibfield  {journal} {\bibinfo
			{journal} {Phys. Rev. A}\ }\textbf {\bibinfo {volume} {89}},\ \bibinfo
		{pages} {061601(R)} (\bibinfo {year} {2014})}\BibitemShut {NoStop}%
	\bibitem [{\citenamefont {Impertro}\ \emph {et~al.}(2024)\citenamefont
		{Impertro}, \citenamefont {Karch}, \citenamefont {Wienand}, \citenamefont
		{Huh}, \citenamefont {Schweizer}, \citenamefont {Bloch},\ and\ \citenamefont
		{Aidelsburger}}]{impertro2024local}%
	\BibitemOpen
	\bibfield  {author} {\bibinfo {author} {\bibfnamefont {A.}~\bibnamefont
			{Impertro}}, \bibinfo {author} {\bibfnamefont {S.}~\bibnamefont {Karch}},
		\bibinfo {author} {\bibfnamefont {J.~F.}\ \bibnamefont {Wienand}}, \bibinfo
		{author} {\bibfnamefont {S.~J.}\ \bibnamefont {Huh}}, \bibinfo {author}
		{\bibfnamefont {C.}~\bibnamefont {Schweizer}}, \bibinfo {author}
		{\bibfnamefont {I.}~\bibnamefont {Bloch}},\ and\ \bibinfo {author}
		{\bibfnamefont {M.}~\bibnamefont {Aidelsburger}},\ }\bibfield  {title}
	{\bibinfo {title} {Local readout and control of current and kinetic energy
			operators in optical lattices},\ }\href
	{https://doi.org/10.1103/PhysRevLett.133.063401} {\bibfield  {journal}
		{\bibinfo  {journal} {Phys. Rev. Lett.}\ }\textbf {\bibinfo {volume} {133}},\
		\bibinfo {pages} {063401} (\bibinfo {year} {2024})}\BibitemShut {NoStop}%
	\bibitem [{\citenamefont {Su}\ \emph {et~al.}(2023)\citenamefont {Su},
		\citenamefont {Sun}, \citenamefont {Hudomal}, \citenamefont {Desaules},
		\citenamefont {Zhou}, \citenamefont {Yang}, \citenamefont {Halimeh},
		\citenamefont {Yuan}, \citenamefont {Papi\ifmmode~\acute{c}\else
			\'{c}\fi{}},\ and\ \citenamefont {Pan}}]{PhysRevResearch.5.023010}%
	\BibitemOpen
	\bibfield  {author} {\bibinfo {author} {\bibfnamefont {G.-X.}\ \bibnamefont
			{Su}}, \bibinfo {author} {\bibfnamefont {H.}~\bibnamefont {Sun}}, \bibinfo
		{author} {\bibfnamefont {A.}~\bibnamefont {Hudomal}}, \bibinfo {author}
		{\bibfnamefont {J.-Y.}\ \bibnamefont {Desaules}}, \bibinfo {author}
		{\bibfnamefont {Z.-Y.}\ \bibnamefont {Zhou}}, \bibinfo {author}
		{\bibfnamefont {B.}~\bibnamefont {Yang}}, \bibinfo {author} {\bibfnamefont
			{J.~C.}\ \bibnamefont {Halimeh}}, \bibinfo {author} {\bibfnamefont {Z.-S.}\
			\bibnamefont {Yuan}}, \bibinfo {author} {\bibfnamefont {Z.}~\bibnamefont
			{Papi\ifmmode~\acute{c}\else \'{c}\fi{}}},\ and\ \bibinfo {author}
		{\bibfnamefont {J.-W.}\ \bibnamefont {Pan}},\ }\bibfield  {title} {\bibinfo
		{title} {Observation of many-body scarring in a bose-hubbard quantum
			simulator},\ }\href {https://doi.org/10.1103/PhysRevResearch.5.023010}
	{\bibfield  {journal} {\bibinfo  {journal} {Phys. Rev. Res.}\ }\textbf
		{\bibinfo {volume} {5}},\ \bibinfo {pages} {023010} (\bibinfo {year}
		{2023})}\BibitemShut {NoStop}%
	\bibitem [{\citenamefont {Nakamura}\ \emph {et~al.}(2019)\citenamefont
		{Nakamura}, \citenamefont {Takasu}, \citenamefont {Kobayashi}, \citenamefont
		{Asaka}, \citenamefont {Fukushima}, \citenamefont {Inaba}, \citenamefont
		{Yamashita},\ and\ \citenamefont {Takahashi}}]{nakamura2019experimental}%
	\BibitemOpen
	\bibfield  {author} {\bibinfo {author} {\bibfnamefont {Y.}~\bibnamefont
			{Nakamura}}, \bibinfo {author} {\bibfnamefont {Y.}~\bibnamefont {Takasu}},
		\bibinfo {author} {\bibfnamefont {J.}~\bibnamefont {Kobayashi}}, \bibinfo
		{author} {\bibfnamefont {H.}~\bibnamefont {Asaka}}, \bibinfo {author}
		{\bibfnamefont {Y.}~\bibnamefont {Fukushima}}, \bibinfo {author}
		{\bibfnamefont {K.}~\bibnamefont {Inaba}}, \bibinfo {author} {\bibfnamefont
			{M.}~\bibnamefont {Yamashita}},\ and\ \bibinfo {author} {\bibfnamefont
			{Y.}~\bibnamefont {Takahashi}},\ }\bibfield  {title} {\bibinfo {title}
		{Experimental determination of bose-hubbard energies},\ }\href
	{https://doi.org/10.1103/PhysRevA.99.033609} {\bibfield  {journal} {\bibinfo
			{journal} {Phys. Rev. A}\ }\textbf {\bibinfo {volume} {99}},\ \bibinfo
		{pages} {033609} (\bibinfo {year} {2019})}\BibitemShut {NoStop}%
	\bibitem [{\citenamefont {Zhao}\ \emph {et~al.}(2023)\citenamefont {Zhao},
		\citenamefont {Xing}, \citenamefont {Cao}, \citenamefont {Liu}, \citenamefont
		{Cui},\ and\ \citenamefont {Wang}}]{zhao2023engineering}%
	\BibitemOpen
	\bibfield  {author} {\bibinfo {author} {\bibfnamefont {X.}~\bibnamefont
			{Zhao}}, \bibinfo {author} {\bibfnamefont {Y.}~\bibnamefont {Xing}}, \bibinfo
		{author} {\bibfnamefont {J.}~\bibnamefont {Cao}}, \bibinfo {author}
		{\bibfnamefont {S.}~\bibnamefont {Liu}}, \bibinfo {author} {\bibfnamefont
			{W.-X.}\ \bibnamefont {Cui}},\ and\ \bibinfo {author} {\bibfnamefont {H.-F.}\
			\bibnamefont {Wang}},\ }\bibfield  {title} {\bibinfo {title} {Engineering
			quantum diode in one-dimensional time-varying superconducting circuits},\
	}\href {https://doi.org/10.1038/s41534-023-00729-1} {\bibfield  {journal}
		{\bibinfo  {journal} {npj Quantum Inf.}\ }\textbf {\bibinfo {volume} {9}},\
		\bibinfo {pages} {59} (\bibinfo {year} {2023})}\BibitemShut {NoStop}%
	\bibitem [{\citenamefont {Mei}\ \emph {et~al.}(2015)\citenamefont {Mei},
		\citenamefont {Zhang},\ and\ \citenamefont {Zhu}}]{MEI201558}%
	\BibitemOpen
	\bibfield  {author} {\bibinfo {author} {\bibfnamefont {F.}~\bibnamefont
			{Mei}}, \bibinfo {author} {\bibfnamefont {D.-W.}\ \bibnamefont {Zhang}},\
		and\ \bibinfo {author} {\bibfnamefont {S.-L.}\ \bibnamefont {Zhu}},\
	}\bibfield  {title} {\bibinfo {title} {Some topological states in
			one-dimensional cold atomic systems},\ }\href
	{https://doi.org/https://doi.org/10.1016/j.aop.2014.12.030} {\bibfield
		{journal} {\bibinfo  {journal} {Ann. Phys.}\ }\textbf {\bibinfo {volume}
			{358}},\ \bibinfo {pages} {58} (\bibinfo {year} {2015})}\BibitemShut
	{NoStop}%
	\bibitem [{\citenamefont {Atala}\ \emph {et~al.}(2013)\citenamefont {Atala},
		\citenamefont {Aidelsburger}, \citenamefont {Barreiro}, \citenamefont
		{Abanin}, \citenamefont {Kitagawa}, \citenamefont {Demler},\ and\
		\citenamefont {Bloch}}]{Atala2013}%
	\BibitemOpen
	\bibfield  {author} {\bibinfo {author} {\bibfnamefont {M.}~\bibnamefont
			{Atala}}, \bibinfo {author} {\bibfnamefont {M.}~\bibnamefont {Aidelsburger}},
		\bibinfo {author} {\bibfnamefont {J.~T.}\ \bibnamefont {Barreiro}}, \bibinfo
		{author} {\bibfnamefont {D.}~\bibnamefont {Abanin}}, \bibinfo {author}
		{\bibfnamefont {T.}~\bibnamefont {Kitagawa}}, \bibinfo {author}
		{\bibfnamefont {E.}~\bibnamefont {Demler}},\ and\ \bibinfo {author}
		{\bibfnamefont {I.}~\bibnamefont {Bloch}},\ }\bibfield  {title} {\bibinfo
		{title} {Direct measurement of the zak phase in topological bloch bands},\
	}\href {https://doi.org/10.1038/nphys2790} {\bibfield  {journal} {\bibinfo
			{journal} {Nat. Phys.}\ }\textbf {\bibinfo {volume} {9}},\ \bibinfo {pages}
		{795} (\bibinfo {year} {2013})}\BibitemShut {NoStop}%
	\bibitem [{\citenamefont {Lohse}\ \emph {et~al.}(2016)\citenamefont {Lohse},
		\citenamefont {Schweizer}, \citenamefont {Zilberberg}, \citenamefont
		{Aidelsburger},\ and\ \citenamefont {Bloch}}]{Lohse2016}%
	\BibitemOpen
	\bibfield  {author} {\bibinfo {author} {\bibfnamefont {M.}~\bibnamefont
			{Lohse}}, \bibinfo {author} {\bibfnamefont {C.}~\bibnamefont {Schweizer}},
		\bibinfo {author} {\bibfnamefont {O.}~\bibnamefont {Zilberberg}}, \bibinfo
		{author} {\bibfnamefont {M.}~\bibnamefont {Aidelsburger}},\ and\ \bibinfo
		{author} {\bibfnamefont {I.}~\bibnamefont {Bloch}},\ }\bibfield  {title}
	{\bibinfo {title} {A thouless quantum pump with ultracold bosonic atoms in an
			optical superlattice},\ }\href {https://doi.org/10.1038/nphys3584} {\bibfield
		{journal} {\bibinfo  {journal} {Nat. Phys.}\ }\textbf {\bibinfo {volume}
			{12}},\ \bibinfo {pages} {350} (\bibinfo {year} {2016})}\BibitemShut
	{NoStop}%
	\bibitem [{\citenamefont {de~Léséleuc}\ \emph {et~al.}(2019)\citenamefont
		{de~Léséleuc}, \citenamefont {Lienhard}, \citenamefont {Scholl},
		\citenamefont {Barredo}, \citenamefont {Weber}, \citenamefont {Lang},
		\citenamefont {Büchler}, \citenamefont {Lahaye},\ and\ \citenamefont
		{Browaeys}}]{doi:10.1126/science.aav9105}%
	\BibitemOpen
	\bibfield  {author} {\bibinfo {author} {\bibfnamefont {S.}~\bibnamefont
			{de~Léséleuc}}, \bibinfo {author} {\bibfnamefont {V.}~\bibnamefont
			{Lienhard}}, \bibinfo {author} {\bibfnamefont {P.}~\bibnamefont {Scholl}},
		\bibinfo {author} {\bibfnamefont {D.}~\bibnamefont {Barredo}}, \bibinfo
		{author} {\bibfnamefont {S.}~\bibnamefont {Weber}}, \bibinfo {author}
		{\bibfnamefont {N.}~\bibnamefont {Lang}}, \bibinfo {author} {\bibfnamefont
			{H.~P.}\ \bibnamefont {Büchler}}, \bibinfo {author} {\bibfnamefont
			{T.}~\bibnamefont {Lahaye}},\ and\ \bibinfo {author} {\bibfnamefont
			{A.}~\bibnamefont {Browaeys}},\ }\bibfield  {title} {\bibinfo {title}
		{Observation of a symmetry-protected topological phase of interacting bosons
			with rydberg atoms},\ }\href {https://doi.org/10.1126/science.aav9105}
	{\bibfield  {journal} {\bibinfo  {journal} {Science}\ }\textbf {\bibinfo
			{volume} {365}},\ \bibinfo {pages} {775} (\bibinfo {year} {2019})},\ \Eprint
	{https://arxiv.org/abs/https://www.science.org/doi/pdf/10.1126/science.aav9105}
	{https://www.science.org/doi/pdf/10.1126/science.aav9105} \BibitemShut
	{NoStop}%
	\bibitem [{\citenamefont {Li}\ \emph {et~al.}(2023)\citenamefont {Li},
		\citenamefont {Wang}, \citenamefont {Zhao}, \citenamefont {Du}, \citenamefont
		{Zhang}, \citenamefont {Hu}, \citenamefont {Mei}, \citenamefont {Xiao},
		\citenamefont {Ma},\ and\ \citenamefont {Jia}}]{PhysRevResearch.5.L032035}%
	\BibitemOpen
	\bibfield  {author} {\bibinfo {author} {\bibfnamefont {Y.}~\bibnamefont
			{Li}}, \bibinfo {author} {\bibfnamefont {Y.}~\bibnamefont {Wang}}, \bibinfo
		{author} {\bibfnamefont {H.}~\bibnamefont {Zhao}}, \bibinfo {author}
		{\bibfnamefont {H.}~\bibnamefont {Du}}, \bibinfo {author} {\bibfnamefont
			{J.}~\bibnamefont {Zhang}}, \bibinfo {author} {\bibfnamefont
			{Y.}~\bibnamefont {Hu}}, \bibinfo {author} {\bibfnamefont {F.}~\bibnamefont
			{Mei}}, \bibinfo {author} {\bibfnamefont {L.}~\bibnamefont {Xiao}}, \bibinfo
		{author} {\bibfnamefont {J.}~\bibnamefont {Ma}},\ and\ \bibinfo {author}
		{\bibfnamefont {S.}~\bibnamefont {Jia}},\ }\bibfield  {title} {\bibinfo
		{title} {Interaction-induced breakdown of chiral dynamics in the
			su-schrieffer-heeger model},\ }\href
	{https://doi.org/10.1103/PhysRevResearch.5.L032035} {\bibfield  {journal}
		{\bibinfo  {journal} {Phys. Rev. Res.}\ }\textbf {\bibinfo {volume} {5}},\
		\bibinfo {pages} {L032035} (\bibinfo {year} {2023})}\BibitemShut {NoStop}%
	\bibitem [{\citenamefont {Besedin}\ \emph {et~al.}(2021)\citenamefont
		{Besedin}, \citenamefont {Gorlach}, \citenamefont {Abramov}, \citenamefont
		{Tsitsilin}, \citenamefont {Moskalenko}, \citenamefont {Dobronosova},
		\citenamefont {Moskalev}, \citenamefont {Matanin}, \citenamefont {Smirnov},
		\citenamefont {Rodionov}, \citenamefont {Poddubny},\ and\ \citenamefont
		{Ustinov}}]{besedin2021topological}%
	\BibitemOpen
	\bibfield  {author} {\bibinfo {author} {\bibfnamefont {I.~S.}\ \bibnamefont
			{Besedin}}, \bibinfo {author} {\bibfnamefont {M.~A.}\ \bibnamefont
			{Gorlach}}, \bibinfo {author} {\bibfnamefont {N.~N.}\ \bibnamefont
			{Abramov}}, \bibinfo {author} {\bibfnamefont {I.}~\bibnamefont {Tsitsilin}},
		\bibinfo {author} {\bibfnamefont {I.~N.}\ \bibnamefont {Moskalenko}},
		\bibinfo {author} {\bibfnamefont {A.~A.}\ \bibnamefont {Dobronosova}},
		\bibinfo {author} {\bibfnamefont {D.~O.}\ \bibnamefont {Moskalev}}, \bibinfo
		{author} {\bibfnamefont {A.~R.}\ \bibnamefont {Matanin}}, \bibinfo {author}
		{\bibfnamefont {N.~S.}\ \bibnamefont {Smirnov}}, \bibinfo {author}
		{\bibfnamefont {I.~A.}\ \bibnamefont {Rodionov}}, \bibinfo {author}
		{\bibfnamefont {A.~N.}\ \bibnamefont {Poddubny}},\ and\ \bibinfo {author}
		{\bibfnamefont {A.~V.}\ \bibnamefont {Ustinov}},\ }\bibfield  {title}
	{\bibinfo {title} {Topological excitations and bound photon pairs in a
			superconducting quantum metamaterial},\ }\href
	{https://doi.org/10.1103/PhysRevB.103.224520} {\bibfield  {journal} {\bibinfo
			{journal} {Phys. Rev. B}\ }\textbf {\bibinfo {volume} {103}},\ \bibinfo
		{pages} {224520} (\bibinfo {year} {2021})}\BibitemShut {NoStop}%
	\bibitem [{\citenamefont {Tai}\ \emph {et~al.}(2017)\citenamefont {Tai},
		\citenamefont {Lukin}, \citenamefont {Rispoli}, \citenamefont {Schittko},
		\citenamefont {Menke}, \citenamefont {Borgnia}, \citenamefont {Preiss},
		\citenamefont {Grusdt}, \citenamefont {Kaufman},\ and\ \citenamefont
		{Greiner}}]{tai2017microscopy}%
	\BibitemOpen
	\bibfield  {author} {\bibinfo {author} {\bibfnamefont {M.~E.}\ \bibnamefont
			{Tai}}, \bibinfo {author} {\bibfnamefont {A.}~\bibnamefont {Lukin}}, \bibinfo
		{author} {\bibfnamefont {M.}~\bibnamefont {Rispoli}}, \bibinfo {author}
		{\bibfnamefont {R.}~\bibnamefont {Schittko}}, \bibinfo {author}
		{\bibfnamefont {T.}~\bibnamefont {Menke}}, \bibinfo {author} {\bibfnamefont
			{D.}~\bibnamefont {Borgnia}}, \bibinfo {author} {\bibfnamefont {P.~M.}\
			\bibnamefont {Preiss}}, \bibinfo {author} {\bibfnamefont {F.}~\bibnamefont
			{Grusdt}}, \bibinfo {author} {\bibfnamefont {A.~M.}\ \bibnamefont
			{Kaufman}},\ and\ \bibinfo {author} {\bibfnamefont {M.}~\bibnamefont
			{Greiner}},\ }\bibfield  {title} {\bibinfo {title} {Microscopy of the
			interacting harper--hofstadter model in the two-body limit},\ }\href
	{https://doi.org/10.1038/nature22811} {\bibfield  {journal} {\bibinfo
			{journal} {Nature}\ }\textbf {\bibinfo {volume} {546}},\ \bibinfo {pages}
		{519} (\bibinfo {year} {2017})}\BibitemShut {NoStop}%
	\bibitem [{\citenamefont {Ye}\ \emph {et~al.}(2019)\citenamefont {Ye},
		\citenamefont {Ge}, \citenamefont {Wu}, \citenamefont {Wang}, \citenamefont
		{Gong}, \citenamefont {Zhang}, \citenamefont {Zhu}, \citenamefont {Yang},
		\citenamefont {Li}, \citenamefont {Liang}, \citenamefont {Lin}, \citenamefont
		{Xu}, \citenamefont {Guo}, \citenamefont {Sun}, \citenamefont {Cheng},
		\citenamefont {Ma}, \citenamefont {Meng}, \citenamefont {Deng}, \citenamefont
		{Rong}, \citenamefont {Lu}, \citenamefont {Peng}, \citenamefont {Fan},
		\citenamefont {Zhu},\ and\ \citenamefont {Pan}}]{ye2019propagation}%
	\BibitemOpen
	\bibfield  {author} {\bibinfo {author} {\bibfnamefont {Y.}~\bibnamefont
			{Ye}}, \bibinfo {author} {\bibfnamefont {Z.-Y.}\ \bibnamefont {Ge}}, \bibinfo
		{author} {\bibfnamefont {Y.}~\bibnamefont {Wu}}, \bibinfo {author}
		{\bibfnamefont {S.}~\bibnamefont {Wang}}, \bibinfo {author} {\bibfnamefont
			{M.}~\bibnamefont {Gong}}, \bibinfo {author} {\bibfnamefont {Y.-R.}\
			\bibnamefont {Zhang}}, \bibinfo {author} {\bibfnamefont {Q.}~\bibnamefont
			{Zhu}}, \bibinfo {author} {\bibfnamefont {R.}~\bibnamefont {Yang}}, \bibinfo
		{author} {\bibfnamefont {S.}~\bibnamefont {Li}}, \bibinfo {author}
		{\bibfnamefont {F.}~\bibnamefont {Liang}}, \bibinfo {author} {\bibfnamefont
			{J.}~\bibnamefont {Lin}}, \bibinfo {author} {\bibfnamefont {Y.}~\bibnamefont
			{Xu}}, \bibinfo {author} {\bibfnamefont {C.}~\bibnamefont {Guo}}, \bibinfo
		{author} {\bibfnamefont {L.}~\bibnamefont {Sun}}, \bibinfo {author}
		{\bibfnamefont {C.}~\bibnamefont {Cheng}}, \bibinfo {author} {\bibfnamefont
			{N.}~\bibnamefont {Ma}}, \bibinfo {author} {\bibfnamefont {Z.~Y.}\
			\bibnamefont {Meng}}, \bibinfo {author} {\bibfnamefont {H.}~\bibnamefont
			{Deng}}, \bibinfo {author} {\bibfnamefont {H.}~\bibnamefont {Rong}}, \bibinfo
		{author} {\bibfnamefont {C.-Y.}\ \bibnamefont {Lu}}, \bibinfo {author}
		{\bibfnamefont {C.-Z.}\ \bibnamefont {Peng}}, \bibinfo {author}
		{\bibfnamefont {H.}~\bibnamefont {Fan}}, \bibinfo {author} {\bibfnamefont
			{X.}~\bibnamefont {Zhu}},\ and\ \bibinfo {author} {\bibfnamefont {J.-W.}\
			\bibnamefont {Pan}},\ }\bibfield  {title} {\bibinfo {title} {Propagation and
			localization of collective excitations on a 24-qubit superconducting
			processor},\ }\href {https://doi.org/10.1103/PhysRevLett.123.050502}
	{\bibfield  {journal} {\bibinfo  {journal} {Phys. Rev. Lett.}\ }\textbf
		{\bibinfo {volume} {123}},\ \bibinfo {pages} {050502} (\bibinfo {year}
		{2019})}\BibitemShut {NoStop}%
	\bibitem [{\citenamefont {Chaikin}\ and\ \citenamefont
		{Lubensky}(2000)}]{ChaikinBook}%
	\BibitemOpen
	\bibfield  {author} {\bibinfo {author} {\bibfnamefont {P.~M.}\ \bibnamefont
			{Chaikin}}\ and\ \bibinfo {author} {\bibfnamefont {T.~C.}\ \bibnamefont
			{Lubensky}},\ }\href@noop {} {\emph {\bibinfo {title} {Principles of
				Condensed Matter Physics}}}\ (\bibinfo  {publisher} {Cambridge University
		Press},\ \bibinfo {address} {Cambridge, UK},\ \bibinfo {year}
	{2000})\BibitemShut {NoStop}%
	\bibitem [{\citenamefont {Kaneko}\ and\ \citenamefont
		{Danshita}(2022)}]{Kaneko2022}%
	\BibitemOpen
	\bibfield  {author} {\bibinfo {author} {\bibfnamefont {R.}~\bibnamefont
			{Kaneko}}\ and\ \bibinfo {author} {\bibfnamefont {I.}~\bibnamefont
			{Danshita}},\ }\bibfield  {title} {\bibinfo {title} {Tensor-network study of
			correlation-spreading dynamics in the two-dimensional bose-hubbard model},\
	}\href {https://doi.org/10.1038/s42005-022-00848-9} {\bibfield  {journal}
		{\bibinfo  {journal} {Communications Physics}\ }\textbf {\bibinfo {volume}
			{5}},\ \bibinfo {pages} {65} (\bibinfo {year} {2022})}\BibitemShut {NoStop}%
\end{thebibliography}
%

\end{document}